\documentclass[sigconf,screen]{acmart}

\setcopyright{cc}
\setcctype{by}
\copyrightyear{2026}
\acmYear{}
\acmDOI{}
\acmConference[]{}{}{}
\acmISBN{}

\usepackage{cite}
\usepackage{amsmath,amsfonts}

\usepackage{hyperref}

\usepackage{algorithmic}
\usepackage{graphicx}
\usepackage{textcomp}
\usepackage{xcolor}
\usepackage{tikz}
\usepackage{tikz-cd}

\usepackage{mathpartir}
\usepackage{xparse}
\usepackage{mathtools}

\usepackage{amsthm}
\usepackage{thmtools}

\declaretheoremstyle[
  spaceabove=3pt,
  spacebelow=3pt,
  headfont=\normalfont\scshape,
  notefont=\mdseries, notebraces={(}{)},
  bodyfont=\normalfont,
  postheadspace=.5em,
  qed=$\triangle$
]{mythmstyle}
\declaretheorem[style=mythmstyle,numberwithin=section]{example}
\declaretheorem[style=mythmstyle,numberwithin=section]{definition}

\declaretheoremstyle[
  spaceabove=3pt,
  spacebelow=-3pt,
  headfont=\normalfont\scshape,
  notefont=\mdseries, notebraces={(}{)},
  bodyfont=\normalfont,
  postheadspace=.5em,
]{propstyle}
\declaretheorem[style=propstyle,numberwithin=section]{lemma}
\declaretheorem[style=propstyle,numberwithin=section]{theorem}

\AtBeginDocument{%

}

\newcommand{\sply}{\mathsf{sply}}
\newcommand{\gtyp}{\mathsf{gtyp}}
\newcommand{\ityp}{\mathsf{ityp}}
\newcommand{\ltyp}{\mathsf{ltyp}}
\newcommand{\isctxt}{\ \mathsf{ctxt}}
\newcommand{\issply}{\ \mathsf{sply}}
\newcommand{\isgtyp}{\ \mathsf{gtyp}}
\newcommand{\isityp}{\ \mathsf{ityp}}
\newcommand{\isltyp}{\ \mathsf{ltyp}}
\newcommand{\ltoi}[1]{{#1}^{\text{\large\textbullet}}}

\newcommand{\finltyp}{\mathsf{ltyp}_\mathsf{fin}}

\newcommand{\lino}[1]{{#1}^\circ}

\newcommand{\ctxtsep}{\mathrel{,}}
\newcommand{\splysep}{\mathrel{,}}
\newcommand{\di}{\diamond}

\newcommand{\finsply}[3]{%
  \mathop{\mathchoice{\textstyle\bigotimes}{\bigotimes}{\bigotimes}{\bigotimes}}_{\isof.{#1}:{#2}}{#3}%
}

\newcommand{\subst}[2]{{#1} [ {#2} ]}

\NewDocumentCommand \isof { o d<> d.: m o } {%
  \IfNoValueF{#1}{%
    #1 \vdash%
  }%
  \IfNoValueF{#2}{%
    #2 \Vdash%
  }%
  \IfNoValueTF{#3}{
      \IfNoValueTF{#5}{
        {#4}
      }{
        {#4} \mathrel{\hat{}} {#5}
      }
  }{%
      \IfNoValueTF{#5}{
        {#3} : {#4}%
      }{
        ({#3} : {#4}) \mathrel{\hat{}} {#5}
      }
  }
}%

\newcommand{\tyapp}[2]{{#1}({#2})}

\newcommand{\splyapp}[2]{{#1}({#2})}

\newcommand{\Eq}[3]{\mathsf{Id}_{#1}({#2},{#3})}

\newcommand{\Nat}{\mathbb{N}}
\newcommand{\zero}{\mathsf{0}}
\renewcommand{\succ}[1]{\mathsf{succ}~{#1}}
\newcommand{\elimNat}[3]{\mathsf{elim}_{\mathbb{N}}({#1},{#2},{#3})} 

\newcommand{\Empty}{\bot}
\newcommand{\Emptyo}{\lino{\bot\!}}

\newcommand{\elimEmptyo}[1]{\mathsf{elim}_{\Emptyo}({#1})}

\newcommand{\Unit}{\top}
\renewcommand{\tt}{\star}

\newcommand{\Bool}{\mathbb{B}}
\newcommand{\true}{\mathsf{true}}
\newcommand{\false}{\mathsf{false}}
\newcommand{\elimBool}[3]{\mathsf{if}~{#1}~\mathsf{then}~{#2}~\mathsf{else}~{#3}}

\newcommand{\Unito}{\lino{\Unit\!}}
\newcommand{\tto}{\tt}
\newcommand{\elimUnito}[2]{\mathsf{let}~\tt={#1}~\mathsf{in}~{#2}}

\newcommand{\Boolo}{\lino{\mathbb{B}\!}}

\newcommand{\booltonat}[1]{|{#1}|}

\newcommand{\BTree}[1]{\mathsf{BTree}~{#1}}

\NewDocumentCommand \Fun {o m m}{
    \IfNoValueTF{#1}{
        {#2} \to {#3}
  }{%
        (\isof.{#1}:{#2}) \to {#3}
    }%
}

\newcommand{\fun}[2]{\lambda {#1} . {#2}}
\newcommand{\app}[2]{{#1}~{#2}}

\NewDocumentCommand \Pair {o m m}{
  \IfNoValueTF{#1}{
    {#2} \times {#3}
  }{%
    (\isof.{#1}:{#2}) \times {#3}
  }%
}
\newcommand{\pair}[2]{({#1}\mathrel{,}{#2})}

\NewDocumentCommand \Pairelim {o o m m}{
    \IfNoValueTF{#1}{
        \mathsf{elim}_{\times}({#3} , {#4})
  }{%
        \mathsf{let}~\pair{#1}{#2}=#3~\mathsf{in}~#4
    }%
}

\NewDocumentCommand \W {o m m}{
  \IfNoValueTF{#1}{
    \mathcal{W}({#2},{#3})
  }{%
    \mathcal{W}(\isof.{#1}:{#2},{#3})
  }%
}
\renewcommand{\sup}[2]{\mathsf{sup}({#1},{#2})}
\newcommand{\elimW}[2]{\mathsf{elim_{\mathcal{W}}}({#1},{#2})} 

\newcommand{\expsply}[2]{{#1} \mathrel{\hat{}} {\underline{}{#2}}}

\NewDocumentCommand \Natfun {o m o m}{
    \IfNoValueTF{#1}{
        \IfNoValueTF{#3}{
          {#2} \multimap {#4}
        }{
          {#2} \mathrel{\hat{}} {#3} \multimap {#4}
        }
  }{%
        \IfNoValueTF{#3}{
          (\isof.{#1}:{#2}) \multimap {#4}
        }{
          (\isof.{#1}:{#2}) \mathrel{\hat{}} {#3} \multimap {#4}
        }
    }%
}

\NewDocumentCommand \natfun {m o m}{
  \lambda
  \IfNoValueTF{#2}{
    {#1} \to {#3}
  }{%
    {#1} \mathrel{\hat{}} {#2} . {#3}
  }%
}

\newcommand{\natapp}[2]{{#1}~{#2}}

\NewDocumentCommand \Bangfun {o m m}{
    \IfNoValueTF{#1}{
      ! {#2} \multimap {#3}
    }{%
      !(\isof.{#1}:{#2}) \multimap {#3}
    }%
}

\NewDocumentCommand \bangfun {m m}{
  \lambda
    ! {#1} . {#2}
}

\NewDocumentCommand \Natpair {o m o m}{
  \IfNoValueTF{#1}{
    \IfNoValueTF{#3}{
      {#2} \otimes {#4}
    }{
      {#2} \mathrel{\hat{}} {#3} \otimes {#4}
    }
  }{%
    \IfNoValueTF{#3}{
      (\isof.{#1}:{#2}) \otimes {#4}
    }{
      (\isof.{#1}:{#2}) \mathrel{\hat{}} {#3} \otimes {#4}
    }
  }%
}
\NewDocumentCommand \natpair {m o m}{
  \IfNoValueTF{#2}{
    ({#1} \mathrel{,} {#3})
  }{%
    ({#1} \mathrel{\hat{}} {#2} \mathrel{,} {#3})
  }%
}

\NewDocumentCommand \Natpairelim {o o m m}{ 
    \IfNoValueTF{#1}{
        \mathsf{elim}_{\times}({#3} , {#4})
  }{%
        \mathsf{let}~\pair{#1}{#2}=#3~\mathsf{in}~#4
    }%
}

\NewDocumentCommand \Wo {o m o m}{
  \IfNoValueTF{#1}{
    \IfNoValueTF{#3}{
      \lino{\mathcal{W}}({#2},{#4})
    }{%
      \lino{\mathcal{W}}(\expsply{#2}{#3},{#4})
    }%
  }{%
    \IfNoValueTF{#3}{
      \lino{\mathcal{W}}(\isof.{#1}:{#2},{#4})
    }{%
      \lino{\mathcal{W}}(\expsply{\isof.{#1}:{#2}}{#3},{#4})
    }%
  }%
}

\NewDocumentCommand \supo {m o m}{
  \IfNoValueTF{#2}{
    \mathsf{sup}({#1},{#3})
  }{%
    \mathsf{sup}(\expsply{#1}{#2},{#3})
  }%
}

\newcommand{\elimWo}[2]{\mathsf{elim}_{\lino{\mathcal{W}}}({#1},{#2})}

\newcommand{\extend}[2]{{#1}.{#2}}
\newcommand{\extendsubst}[2]{{#1}.{#2}}
\DeclareDocumentCommand \p { o } { \mathbf{p}\IfNoValueF{#1}{_{#1}} }
\DeclareDocumentCommand \q { o } { \mathbf{q}\IfNoValueF{#1}{_{#1}} }

\newcommand{\catfont}[1]{{\mathbf{#1}}}
\newcommand{\modelfont}[1]{{\mathcal #1}}

\newcommand{\objof}{:}
\newcommand{\ob}[1]{\mathit{ob}({#1})}
\NewDocumentCommand{\idmorphism}{o}{\IfNoValueTF{#1}{\mathsf{id}}{\mathsf{id}_{#1}}}
\renewcommand{\hom}[3]{\mathit{hom}_{#1}({#2},{#3}) }

\newcommand{\catelem}[2]{\int_{#1} {{#2}} }
\newcommand{\op}[1]{{#1}^{\mathit{op}}}
\newcommand{\psh}[1]{\catfont{Psh}({#1})}

\newcommand{\Cx}{\modelfont{Cx}}
\NewDocumentCommand{\Ty}{o}{\modelfont{T}\!y\IfNoValueTF{#1}{}{(#1)}}
\NewDocumentCommand{\ATy}{o}{\modelfont{T}\!y_0\IfNoValueTF{#1}{}{(#1)}}
\NewDocumentCommand{\Tm}{o o}{\modelfont{T}\!m\IfNoValueTF{#1}{}{(#1,#2)}}
\NewDocumentCommand{\ATm}{o}{\modelfont{T}\!m_0\IfNoValueTF{#1}{}{(#1)}}

\NewDocumentCommand{\Sp}{o}{\modelfont{Sp}\IfNoValueTF{#1}{}{(#1)}}

\newcommand{\agf}[1]{\AgdaFunction{#1}}

\newcommand{\RULEsplyityp}{
  \inferrule*[Right=$\sply$Emb]{ }{ \isof[\Gamma]{\sply \isityp} }
}
\newcommand{\RULEltypityp}{
  \inferrule*[Right=$\ltyp$Emb]{ }{ \isof[\Gamma]{\ltyp \isityp} }
}

\newcommand{\RULEltmsubst}{
  \inferrule*[Right=$\mathsf{ltm}$Sb]{
    \isof[\Gamma'].\gamma:\Gamma \\
    \isof< \Delta >.a:A
  }{ \isof< \subst{\Delta}{\gamma} >.{\subst{a}{\gamma}}:{ \subst{A}{\gamma} } }
}

\newcommand{\RULEsplyconv}{
  \inferrule*[Right=$\sply$Conv]{
    \isof<\Delta>{\mathcal{J}} \\
    \isof[\Gamma]{\Eq{\sply}{\Delta}{\Delta'}} }
  { \isof<\Delta' >{\mathcal{J}}
  }
}

\newcommand{\RULEemptysply}{
  \inferrule*[Right=$\di\sply$]{ }{ \isof[\Gamma]{\di \issply} }
}

\newcommand{\RULEbangsply}{
  \inferrule*[Right=$!\sply$]{
    \isof[\Gamma ]{ \Delta \issply }
  }{
    \isof[\Gamma]{ ! \Delta \issply }
  }
}

\newcommand{\RULEbangfunintro}{
  \inferrule*[Right=$!\!\!\multimap$I]{ 
    \isof[\Gamma]{\Delta \issply} \\
    \isof<\Delta \splysep !(\isof.x:A)  > . b(x) : {\tyapp{B}{x}}
  }{
    \isof<\Delta> . \bangfun{x}{b} : {\Bangfun[x]{A}{\tyapp{B}{x}}}
  }
}

\newcommand{\RULEbangfunapp}{
  \inferrule*[Right=$!\!\!\multimap$App]{
    \isof<\Delta_0> . f : {\Bangfun[x]{A}{\tyapp{B}{x}}} \\
    \isof<\Delta_1> . a : A
  }{
    \isof<\Delta_0 \splysep !\Delta_1>. \natapp{f}{a} : {\tyapp{B}{a}}
  }
}

\newcommand{\RULEvarsply}{%
  \inferrule*[Right=Var$\sply$]{%
    \isof[\Gamma]{A \isltyp}
  }{%
    \isof[\Gamma \ctxtsep \isof.x:{\ltoi{A}} ]{\isof.x:A \issply}
  }%
}

\newcommand{\RULEjoinsply}{
  \inferrule*[Right=$\ctxtsep\sply$]{
    \isof[\Gamma]{\Delta_0 \issply } \\
    \isof[\Gamma]{\Delta_1 \issply }
  }{ \isof[\Gamma]{\Delta_0 \splysep \Delta_1 \issply} }
}

\newcommand{\RULEltyp}{
  \inferrule*[Right=$\iota\gtyp$]{ \isof[\Gamma]{A \isgtyp} }{ \isof[\Gamma]{A \isltyp} }
}

\newcommand{\RULElvar}{%
  \inferrule*[Right=$\lino{\text{Var}}$]{%
    \isof[\Gamma]{A \isltyp}
  }{%
    \isof<\isof.x:A>{\isof.x:A}
  }%
}

\newcommand{\RULEexchange}{
  \inferrule*[Right=Exch]
  { \isof<\Delta_0 \splysep \Delta_1 >{\mathcal{J}} }
  { \isof<\Delta_1 \splysep \Delta_0 >{\mathcal{J}} }
}

\newcommand{\RULEunitointro}{
  \inferrule*[Right=$\Unito$I]{  }{ \isof[\di].{\tto}:{\Unito} }
}

\newcommand{\RULEunitoelim}{
  \inferrule*[Right=$\Unito$E]{
    \isof<\Delta_0>.a:{\Unito} \\
    \isof<\Delta_1>.c:{\tyapp{C}{\tto}}
  }{ \isof<\Delta_0 \splysep \Delta_1>.{ \elimUnito{a}{c} }:{\tyapp{C}{a}} }
}

\newcommand{\RULEemptyoelim}{
  \inferrule*[Right=$\Emptyo$E]{
    \isof<\Delta>.e:{\Emptyo}
  }{ \isof<\Delta>.{ \elimEmptyo{e} }:{ \tyapp{C}{e}} }
}

\newcommand{\RULEoboolintro}{
  \inferrule*[Right=$\Boolo$I]{  }{ \isof[\diamond].{\true, \false}:{\Boolo} }
}

\newcommand{\RULEoboolelim}{
  \inferrule*[Right=$\Boolo$E]
  {
    \isof[\Gamma].b:{\Bool} \\
    \isof<\Delta_0>.c_0:{\tyapp{C}{\true}} \\
    \isof<\Delta_1>.c_1:{\tyapp{C}{\false}}
  }{ \isof<\expsply{\Delta_0}{\booltonat{b}} \splysep \expsply{\Delta_1}{\booltonat{\neg b}}>.{\elimBool{b}{c_0}{c_1}}:{\tyapp{C}{b}} }
}

\newcommand{\RULEnatfunform}{
  \inferrule*[Right=$\multimap$F]{
    \isof[\Gamma]{A \isltyp} \\
    \isof[\Gamma \ctxtsep \isof.x:{\ltoi{A}}]{\tyapp{B}{x} \isltyp} \\
    \isof[\Gamma].m:{\Nat}
  }{
    \isof[\Gamma]{ \Natfun[x]{A}[m]{\tyapp{B}{x}} \isltyp }
  }
}

\newcommand{\RULEnatfunintro}{
  \inferrule*[Right=$\multimap$I]{ 
    \isof[\Gamma]{\Delta \issply} \\
    \isof<\Delta \splysep \isof.x:A[m] > . b : {\tyapp{B}{x}}
  }{
    \isof<\Delta> . \natfun{x}[m]{b} : {\Natfun[x]{A}[m]{\tyapp{B}{x}}}
  }
}

\newcommand{\RULEnatfunapp}{
  \inferrule*[Right=$\multimap$App]{
    \isof<\Delta_0> . f : {\Natfun[x]{A}[m]{\tyapp{B}{x}}} \\
    \isof<\Delta_1> . a : A
  }{
    \isof<\Delta_0 \splysep \expsply{\Delta_1}{m}>. \natapp{f}{a} : {\tyapp{B}{a}}
  }
}

\newcommand{\RULEnatfuncomp}{
  \inferrule*[Right=$\multimap\!\!\beta$]{
    \isof<\Delta_0 \splysep \isof.x:A[m] > . {b(x)} : {\tyapp{B}{x}} \\
    \isof<\Delta_1>.a:A
  }{
    \isof<\Delta_0 \splysep \expsply{\Delta_1}{m}> .
    {\natapp{(\natfun{x}[m]{b})}{a} \equiv \subst{b}{a/x} } : { \tyapp{B}{a} }
  }
}

\newcommand{\RULEnatpairform}{
  \inferrule*[Right=$\otimes$F]{
    \isof[\Gamma]{A \isltyp} \\
    \isof[\Gamma \ctxtsep \isof.x:{\ltoi{A}}]{\tyapp{B}{x} \isltyp} \\
    \isof[\Gamma].m:{\Nat}
  }{
    \isof[\Gamma]{ \Natpair[x]{A}[m]{\tyapp{B}{x}} \isltyp }
  }
}

\newcommand{\RULEnatpairintro}{
  \inferrule*[Right=$\otimes$I]{
    \isof<\Delta_0>.a:A \\
    \isof<\Delta_1>.b:{\tyapp{B}{a}} 
    }{ \isof<\expsply{\Delta_0}{m} \splysep \Delta_1 >.{\natpair{a}[m]{b}}:{\Natpair[x]{A}[m]{\tyapp{B}{x}}} }
}

\newcommand{\RULEnatpairelim}{
  \inferrule*[Right=$\otimes$E]{ 
    \isof<\Delta_0>.p:{\Natpair[x]{A}[m]{\tyapp{B}{x}}} \\
    \isof<\Delta_1 \splysep \isof.x:{A}[m] \splysep \isof.y:{\tyapp{B}{x}} >.c:{C\pair{x}{y}}
    }{ \isof<\Delta_0 \splysep \Delta_1 >.{\Natpairelim[x][y]{p}{c}}:{\tyapp{C}{p}} }
}

\newcommand{\RULEfinsply}{
  \inferrule*[Right=$\mathsf{fin}\sply$]{
    \isof[\Gamma]{B \ \finltyp} \\
    \isof[\Gamma \ctxtsep \isof.y:{\ltoi{B}} ]{ \Delta \issply }
  }{
    \isof[\Gamma]{ \finsply{y}{B}{\splyapp{\Delta}{y}} \issply }
  }
}

\newcommand{\RULEWoform}{
  \inferrule*[Right=$\lino{\mathcal{W}}$F]{ 
    \isof[\Gamma]{A \isltyp } \\
    \isof[\Gamma \ctxtsep \isof.x:{\ltoi{A}} ]{\tyapp{B}{x}\ \finltyp } \\
    \isof[\Gamma].m:{\Nat}
  }{
    \isof[\Gamma]{\Wo[x]{A}[m]{\tyapp{B}{x}} \isltyp}
  }
}

\newcommand{\RULEWointro}{
  \inferrule*[Right=$\lino{\mathcal{W}}$I]{
    \isof<\Delta_0>.a:{A} \\
    \isof[\Gamma \ctxtsep \isof.y:{\ltoi{\tyapp{B}{a}}} ]{ \Delta_1 \issply } \\
    \isof<\Delta_1>.{f}:{\Wo[x]{A}[m]{\tyapp{B}{x} }}
  }{
    \isof<\expsply{\Delta_0}{m} \splysep \finsply{y}{\tyapp{B}{a}}{\splyapp{\Delta_1}{y}} >.{\supo{a}[m]{f}}:{\Wo[x]{A}[m]{\tyapp{B}{x} }}
  }
}

\newcommand{\RULEWoelim}{
  \inferrule*[Right=$\!\!\lino{\mathcal{W}\!}$E]{
    \isof<\Delta_0>.p:{\Wo[x]{A}[m]{\tyapp{B}{x}} } \\
    \isof<\isof.{x}:{A}[m] \splysep \splyapp{\Delta_1}{x}
    \splysep \hspace{-.8em}\finsply{y}{\tyapp{B}{a}}{\hspace{-.6em}(\isof.{ \app{g}{y} }:{ \tyapp{C}{\app{f}{y}} })} >
    .{c(x,f,g)}: { \tyapp{C}{\sup{x}{f}} }
  }{
    \isof< \Delta_0 \splysep \elimW{ x \Delta . (\splyapp{\Delta_1}{x} \splysep
      \hspace{-.8em}\finsply{y}{\tyapp{B}{x}}{\hspace{-.6em} \splyapp{\Delta}{y} }  )}{p} >.{\elimWo{c}{p}}:{\tyapp{C}{p}}
  }
}

\newcommand{\RULEWocomp}{
  \inferrule*[]{ 
    \isof<\Delta_0>.a:{A} \\
    \isof[\Gamma \ctxtsep \isof.y:{\ltoi{\tyapp{B}{a}}}].{f_a(y)}:{\W{\ltoi{A}}{\ltoi{B}} }  \\
    \isof<\isof.{x}:{A}[m] \splysep \splyapp{\Delta_1}{x}
    \splysep \hspace{-.6em}\finsply{y}{\tyapp{B}{a}}{\hspace{-.6em}(\isof.{ \app{g}{y} }:{ \tyapp{C}{\app{f}{y}} })} >
    .{c(x,f,g)}: { \tyapp{C}{\sup{x}{f}} }
  }{
     \expsply{\Delta_0}{m} \splysep \splyapp{\Delta_1}{a} \splysep
      \hspace{-.8em}\finsply{y}{\tyapp{B}{a}}{\hspace{-.6em} \isof.{\elimW{c}{f_a(y)}}:{\tyapp{C}{f_a(y)}} } \Vdash\\
      {
      \elimWo{c}{\sup{a}{f_a}} \equiv c(a,{f_a},{y}.{\elimWo{c}{f_a(y)}})
      }:{\tyapp{C}{\sup{a}{f_a}}}
  }
}

\usepackage{code/latex/agda}
\usepackage{newunicodechar}
\newunicodechar{₀}{\ensuremath{_0}}
\newunicodechar{₁}{\ensuremath{_1}}
\newunicodechar{⊥}{\ensuremath{\bot}}
\newunicodechar{∷}{\ensuremath{\Colon}}
\newunicodechar{δ}{\ensuremath{\delta}}
\newunicodechar{Δ}{\ensuremath{\Delta}}
\newunicodechar{Π}{\ensuremath{\Pi}}
\newunicodechar{Σ}{\ensuremath{\Sigma}}
\newunicodechar{ℓ}{\ensuremath{\ell}}
\newunicodechar{⨾}{\ensuremath{\fcmp}}
\newunicodechar{∈}{\ensuremath{\in}}
\newunicodechar{λ}{\ensuremath{\lambda}}
\newunicodechar{Λ}{\ensuremath{\forall}}
\newunicodechar{⟨}{\ensuremath{\langle}}
\newunicodechar{⟦}{\ensuremath{\lBrack}}
\newunicodechar{←}{\ensuremath{\leftarrow}}
\newunicodechar{ℕ}{\ensuremath{\mathbb{N}}}
\newunicodechar{∶}{\ensuremath{\mathratio}}
\newunicodechar{◇}{\ensuremath{\mdlgwhtdiamond}}
\newunicodechar{⊸}{\ensuremath{\multimap}}
\newunicodechar{⊕}{\ensuremath{\oplus}}
\newunicodechar{⟩}{\ensuremath{\rangle}}
\newunicodechar{⟧}{\ensuremath{\rBrack}}
\newunicodechar{⇒}{\ensuremath{\Rightarrow}}
\newunicodechar{⇨}{\ensuremath{\rightwhitearrow}}
\newunicodechar{→}{\ensuremath{\to}}
\newunicodechar{⊤}{\ensuremath{\top}}
\newunicodechar{⊩}{\ensuremath{\Vdash}}
\newunicodechar{∘}{\ensuremath{\vysmwhtcircle}}
\newunicodechar{ᶠ}{\ensuremath{^f}}
\newunicodechar{ᵒ}{\ensuremath{^\circ}}
\newunicodechar{．}{\ensuremath{^{\scriptscriptstyle \medbullet}}}
\newunicodechar{⊗}{\ensuremath{\otimes}}

\begin{code}[hide]%
\>[0]\AgdaSymbol{\{-\#}\AgdaSpace{}%
\AgdaKeyword{OPTIONS}\AgdaSpace{}%
\AgdaPragma{--safe}\AgdaSpace{}%
\AgdaSymbol{\#-\}}\<%
\\
\>[0]\AgdaKeyword{module}\AgdaSpace{}%
\AgdaModule{Base}\AgdaSpace{}%
\AgdaKeyword{where}\<%
\\
\\[\AgdaEmptyExtraSkip]%
\>[0]\AgdaKeyword{open}\AgdaSpace{}%
\AgdaKeyword{import}\AgdaSpace{}%
\AgdaModule{Prelude}\AgdaSpace{}%
\AgdaKeyword{public}\<%
\\
\\[\AgdaEmptyExtraSkip]%
\\[\AgdaEmptyExtraSkip]%
\>[0]\AgdaKeyword{data}\AgdaSpace{}%
\AgdaDatatype{sply}\AgdaSpace{}%
\AgdaSymbol{(}\AgdaBound{ℓ}\AgdaSpace{}%
\AgdaSymbol{:}\AgdaSpace{}%
\AgdaPostulate{Level}\AgdaSymbol{)}\AgdaSpace{}%
\AgdaSymbol{:}\AgdaSpace{}%
\AgdaPrimitive{ityp}\AgdaSpace{}%
\AgdaSymbol{(}\AgdaPrimitive{ℓ-suc}\AgdaSpace{}%
\AgdaBound{ℓ}\AgdaSymbol{)}\AgdaSpace{}%
\AgdaKeyword{where}\<%
\\
\>[0][@{}l@{\AgdaIndent{0}}]%
\>[2]\AgdaInductiveConstructor{◇}\AgdaSpace{}%
\AgdaSymbol{:}\AgdaSpace{}%
\AgdaDatatype{sply}\AgdaSpace{}%
\AgdaBound{ℓ}\<%
\\
\>[2]\AgdaInductiveConstructor{ι}\AgdaSpace{}%
\AgdaSymbol{:}\AgdaSpace{}%
\AgdaSymbol{\{}\AgdaBound{A}\AgdaSpace{}%
\AgdaSymbol{:}\AgdaSpace{}%
\AgdaPrimitive{ityp}\AgdaSpace{}%
\AgdaBound{ℓ}\AgdaSymbol{\}}\AgdaSpace{}%
\AgdaSymbol{(}\AgdaBound{a}\AgdaSpace{}%
\AgdaSymbol{:}\AgdaSpace{}%
\AgdaBound{A}\AgdaSymbol{)}\AgdaSpace{}%
\AgdaSymbol{→}\AgdaSpace{}%
\AgdaDatatype{sply}\AgdaSpace{}%
\AgdaBound{ℓ}\<%
\\
\>[2]\AgdaOperator{\AgdaInductiveConstructor{\AgdaUnderscore{}⨾\AgdaUnderscore{}}}\AgdaSpace{}%
\AgdaSymbol{:}\AgdaSpace{}%
\AgdaDatatype{sply}\AgdaSpace{}%
\AgdaBound{ℓ}\AgdaSpace{}%
\AgdaSymbol{→}\AgdaSpace{}%
\AgdaDatatype{sply}\AgdaSpace{}%
\AgdaBound{ℓ}\AgdaSpace{}%
\AgdaSymbol{→}\AgdaSpace{}%
\AgdaDatatype{sply}\AgdaSpace{}%
\AgdaBound{ℓ}\<%
\\
\>[2]\AgdaOperator{\AgdaInductiveConstructor{[\AgdaUnderscore{},\AgdaUnderscore{}]}}\AgdaSpace{}%
\AgdaSymbol{:}\AgdaSpace{}%
\AgdaDatatype{sply}\AgdaSpace{}%
\AgdaBound{ℓ}\AgdaSpace{}%
\AgdaSymbol{→}\AgdaSpace{}%
\AgdaDatatype{sply}\AgdaSpace{}%
\AgdaBound{ℓ}\AgdaSpace{}%
\AgdaSymbol{→}\AgdaSpace{}%
\AgdaDatatype{sply}\AgdaSpace{}%
\AgdaBound{ℓ}\<%
\\
\>[2]\AgdaInductiveConstructor{Λ}\AgdaSpace{}%
\AgdaSymbol{:}\AgdaSpace{}%
\AgdaSymbol{(}\AgdaBound{A}\AgdaSpace{}%
\AgdaSymbol{:}\AgdaSpace{}%
\AgdaPrimitive{ityp}\AgdaSpace{}%
\AgdaBound{ℓ}\AgdaSymbol{)}\AgdaSpace{}%
\AgdaSymbol{→}\AgdaSpace{}%
\AgdaSymbol{(}\AgdaBound{Δ}\AgdaSpace{}%
\AgdaSymbol{:}\AgdaSpace{}%
\AgdaBound{A}\AgdaSpace{}%
\AgdaSymbol{→}\AgdaSpace{}%
\AgdaDatatype{sply}\AgdaSpace{}%
\AgdaBound{ℓ}\AgdaSymbol{)}\AgdaSpace{}%
\AgdaSymbol{→}\AgdaSpace{}%
\AgdaDatatype{sply}\AgdaSpace{}%
\AgdaBound{ℓ}\<%
\\
\>[2]\AgdaInductiveConstructor{!}\AgdaSpace{}%
\AgdaSymbol{:}\AgdaSpace{}%
\AgdaDatatype{sply}\AgdaSpace{}%
\AgdaBound{ℓ}\AgdaSpace{}%
\AgdaSymbol{→}\AgdaSpace{}%
\AgdaDatatype{sply}\AgdaSpace{}%
\AgdaBound{ℓ}\<%
\\
\\[\AgdaEmptyExtraSkip]%
\>[0]\AgdaKeyword{infixr}\AgdaSpace{}%
\AgdaNumber{-12}\AgdaSpace{}%
\AgdaOperator{\AgdaInductiveConstructor{\AgdaUnderscore{}⨾\AgdaUnderscore{}}}\<%
\\
\>[0]\AgdaKeyword{infixr}\AgdaSpace{}%
\AgdaNumber{30}\AgdaSpace{}%
\AgdaOperator{\AgdaInductiveConstructor{[\AgdaUnderscore{},\AgdaUnderscore{}]}}\<%
\\
\\[\AgdaEmptyExtraSkip]%
\>[0]\AgdaKeyword{module}\AgdaSpace{}%
\AgdaModule{\AgdaUnderscore{}}\AgdaSpace{}%
\AgdaSymbol{\{}\AgdaBound{ℓ}\AgdaSpace{}%
\AgdaSymbol{:}\AgdaSpace{}%
\AgdaPostulate{Level}\AgdaSymbol{\}}\AgdaSpace{}%
\AgdaKeyword{where}\<%
\\
\>[0][@{}l@{\AgdaIndent{0}}]%
\>[2]\AgdaKeyword{infixl}\AgdaSpace{}%
\AgdaNumber{-100}\AgdaSpace{}%
\AgdaOperator{\AgdaDatatype{\AgdaUnderscore{}⊩\AgdaUnderscore{}}}\<%
\\
\>[2]\AgdaKeyword{infixr}\AgdaSpace{}%
\AgdaNumber{11}\AgdaSpace{}%
\AgdaOperator{\AgdaInductiveConstructor{\AgdaUnderscore{}⨾ᶠ\AgdaUnderscore{}}}\<%
\\
\>[2]\AgdaKeyword{infixl}\AgdaSpace{}%
\AgdaNumber{10}\AgdaSpace{}%
\AgdaOperator{\AgdaInductiveConstructor{\AgdaUnderscore{}∘\AgdaUnderscore{}}}\<%
\\
\\[\AgdaEmptyExtraSkip]%
\>[2]\AgdaKeyword{data}\AgdaSpace{}%
\AgdaOperator{\AgdaDatatype{\AgdaUnderscore{}⊩\AgdaUnderscore{}}}\AgdaSpace{}%
\AgdaSymbol{:}\AgdaSpace{}%
\AgdaDatatype{sply}\AgdaSpace{}%
\AgdaBound{ℓ}\AgdaSpace{}%
\AgdaSymbol{→}\AgdaSpace{}%
\AgdaDatatype{sply}\AgdaSpace{}%
\AgdaBound{ℓ}\AgdaSpace{}%
\AgdaSymbol{→}\AgdaSpace{}%
\AgdaPrimitive{ityp}\AgdaSpace{}%
\AgdaSymbol{(}\AgdaPrimitive{ℓ-suc}\AgdaSpace{}%
\AgdaBound{ℓ}\AgdaSymbol{)}\AgdaSpace{}%
\AgdaKeyword{where}\<%
\\
\>[2][@{}l@{\AgdaIndent{0}}]%
\>[4]\AgdaInductiveConstructor{id}\AgdaSpace{}%
\AgdaSymbol{:}\AgdaSpace{}%
\AgdaSymbol{∀}\AgdaSpace{}%
\AgdaBound{Δ}\AgdaSpace{}%
\AgdaSymbol{→}\AgdaSpace{}%
\AgdaBound{Δ}\AgdaSpace{}%
\AgdaOperator{\AgdaDatatype{⊩}}\AgdaSpace{}%
\AgdaBound{Δ}\<%
\\
\>[4]\AgdaOperator{\AgdaInductiveConstructor{\AgdaUnderscore{}∘\AgdaUnderscore{}}}\AgdaSpace{}%
\AgdaSymbol{:}\AgdaSpace{}%
\AgdaSymbol{∀}\AgdaSpace{}%
\AgdaSymbol{\{}\AgdaBound{Δ₀}\AgdaSpace{}%
\AgdaBound{Δ₁}\AgdaSpace{}%
\AgdaBound{Δ₂}\AgdaSymbol{\}}\AgdaSpace{}%
\AgdaSymbol{→}\AgdaSpace{}%
\AgdaBound{Δ₁}\AgdaSpace{}%
\AgdaOperator{\AgdaDatatype{⊩}}\AgdaSpace{}%
\AgdaBound{Δ₂}\AgdaSpace{}%
\AgdaSymbol{→}\AgdaSpace{}%
\AgdaBound{Δ₀}\AgdaSpace{}%
\AgdaOperator{\AgdaDatatype{⊩}}\AgdaSpace{}%
\AgdaBound{Δ₁}\AgdaSpace{}%
\AgdaSymbol{→}\AgdaSpace{}%
\AgdaBound{Δ₀}\AgdaSpace{}%
\AgdaOperator{\AgdaDatatype{⊩}}\AgdaSpace{}%
\AgdaBound{Δ₂}\<%
\\
\>[4]\AgdaOperator{\AgdaInductiveConstructor{\AgdaUnderscore{}⨾ᶠ\AgdaUnderscore{}}}\AgdaSpace{}%
\AgdaSymbol{:}\AgdaSpace{}%
\AgdaSymbol{∀}\AgdaSpace{}%
\AgdaSymbol{\{}\AgdaBound{Δ₀}\AgdaSpace{}%
\AgdaBound{Δ₁}\AgdaSpace{}%
\AgdaBound{Δ₂}\AgdaSpace{}%
\AgdaBound{Δ₃}\AgdaSymbol{\}}\AgdaSpace{}%
\AgdaSymbol{→}\AgdaSpace{}%
\AgdaBound{Δ₀}\AgdaSpace{}%
\AgdaOperator{\AgdaDatatype{⊩}}\AgdaSpace{}%
\AgdaBound{Δ₁}\AgdaSpace{}%
\AgdaSymbol{→}\AgdaSpace{}%
\AgdaBound{Δ₂}\AgdaSpace{}%
\AgdaOperator{\AgdaDatatype{⊩}}\AgdaSpace{}%
\AgdaBound{Δ₃}\AgdaSpace{}%
\AgdaSymbol{→}\AgdaSpace{}%
\AgdaBound{Δ₀}\AgdaSpace{}%
\AgdaOperator{\AgdaInductiveConstructor{⨾}}\AgdaSpace{}%
\AgdaBound{Δ₂}\AgdaSpace{}%
\AgdaOperator{\AgdaDatatype{⊩}}\AgdaSpace{}%
\AgdaBound{Δ₁}\AgdaSpace{}%
\AgdaOperator{\AgdaInductiveConstructor{⨾}}\AgdaSpace{}%
\AgdaBound{Δ₃}\<%
\\
\\[\AgdaEmptyExtraSkip]%
\>[4]\AgdaInductiveConstructor{unitr}\AgdaSpace{}%
\AgdaSymbol{:}\AgdaSpace{}%
\AgdaSymbol{∀}\AgdaSpace{}%
\AgdaBound{Δ}\AgdaSpace{}%
\AgdaSymbol{→}\AgdaSpace{}%
\AgdaBound{Δ}\AgdaSpace{}%
\AgdaOperator{\AgdaInductiveConstructor{⨾}}\AgdaSpace{}%
\AgdaInductiveConstructor{◇}\AgdaSpace{}%
\AgdaOperator{\AgdaDatatype{⊩}}\AgdaSpace{}%
\AgdaBound{Δ}\<%
\\
\>[4]\AgdaInductiveConstructor{unitr'}\AgdaSpace{}%
\AgdaSymbol{:}\AgdaSpace{}%
\AgdaSymbol{∀}\AgdaSpace{}%
\AgdaBound{Δ}\AgdaSpace{}%
\AgdaSymbol{→}\AgdaSpace{}%
\AgdaBound{Δ}\AgdaSpace{}%
\AgdaOperator{\AgdaDatatype{⊩}}\AgdaSpace{}%
\AgdaBound{Δ}\AgdaSpace{}%
\AgdaOperator{\AgdaInductiveConstructor{⨾}}\AgdaSpace{}%
\AgdaInductiveConstructor{◇}\<%
\\
\>[4]\AgdaInductiveConstructor{swap}\AgdaSpace{}%
\AgdaSymbol{:}\AgdaSpace{}%
\AgdaSymbol{∀}\AgdaSpace{}%
\AgdaBound{Δ₀}\AgdaSpace{}%
\AgdaBound{Δ₁}\AgdaSpace{}%
\AgdaSymbol{→}\AgdaSpace{}%
\AgdaBound{Δ₀}\AgdaSpace{}%
\AgdaOperator{\AgdaInductiveConstructor{⨾}}\AgdaSpace{}%
\AgdaBound{Δ₁}\AgdaSpace{}%
\AgdaOperator{\AgdaDatatype{⊩}}\AgdaSpace{}%
\AgdaBound{Δ₁}\AgdaSpace{}%
\AgdaOperator{\AgdaInductiveConstructor{⨾}}\AgdaSpace{}%
\AgdaBound{Δ₀}\<%
\\
\>[4]\AgdaInductiveConstructor{assoc}\AgdaSpace{}%
\AgdaSymbol{:}\AgdaSpace{}%
\AgdaSymbol{∀}\AgdaSpace{}%
\AgdaBound{Δ₀}\AgdaSpace{}%
\AgdaBound{Δ₁}\AgdaSpace{}%
\AgdaBound{Δ₂}\AgdaSpace{}%
\AgdaSymbol{→}\AgdaSpace{}%
\AgdaBound{Δ₀}\AgdaSpace{}%
\AgdaOperator{\AgdaInductiveConstructor{⨾}}\AgdaSpace{}%
\AgdaSymbol{(}\AgdaBound{Δ₁}\AgdaSpace{}%
\AgdaOperator{\AgdaInductiveConstructor{⨾}}\AgdaSpace{}%
\AgdaBound{Δ₂}\AgdaSymbol{)}\AgdaSpace{}%
\AgdaOperator{\AgdaDatatype{⊩}}\AgdaSpace{}%
\AgdaSymbol{(}\AgdaBound{Δ₀}\AgdaSpace{}%
\AgdaOperator{\AgdaInductiveConstructor{⨾}}\AgdaSpace{}%
\AgdaBound{Δ₁}\AgdaSymbol{)}\AgdaSpace{}%
\AgdaOperator{\AgdaInductiveConstructor{⨾}}\AgdaSpace{}%
\AgdaBound{Δ₂}\<%
\\
\\[\AgdaEmptyExtraSkip]%
\>[4]\AgdaInductiveConstructor{curry}\AgdaSpace{}%
\AgdaSymbol{:}\AgdaSpace{}%
\AgdaSymbol{∀}\AgdaSpace{}%
\AgdaSymbol{\{}\AgdaBound{Δ₀}\AgdaSpace{}%
\AgdaBound{Δ₁}\AgdaSpace{}%
\AgdaBound{Δ₂}\AgdaSymbol{\}}\AgdaSpace{}%
\AgdaSymbol{→}%
\>[28]\AgdaSymbol{(}\AgdaBound{Δ₀}\AgdaSpace{}%
\AgdaOperator{\AgdaInductiveConstructor{⨾}}\AgdaSpace{}%
\AgdaBound{Δ₁}\AgdaSymbol{)}\AgdaSpace{}%
\AgdaOperator{\AgdaDatatype{⊩}}\AgdaSpace{}%
\AgdaBound{Δ₂}\AgdaSpace{}%
\AgdaSymbol{→}\AgdaSpace{}%
\AgdaBound{Δ₀}\AgdaSpace{}%
\AgdaOperator{\AgdaDatatype{⊩}}\AgdaSpace{}%
\AgdaOperator{\AgdaInductiveConstructor{[}}\AgdaSpace{}%
\AgdaBound{Δ₁}\AgdaSpace{}%
\AgdaOperator{\AgdaInductiveConstructor{,}}\AgdaSpace{}%
\AgdaBound{Δ₂}\AgdaSpace{}%
\AgdaOperator{\AgdaInductiveConstructor{]}}\<%
\\
\>[4]\AgdaInductiveConstructor{uncurry}\AgdaSpace{}%
\AgdaSymbol{:}\AgdaSpace{}%
\AgdaSymbol{∀}\AgdaSpace{}%
\AgdaSymbol{\{}\AgdaBound{Δ₀}\AgdaSpace{}%
\AgdaBound{Δ₁}\AgdaSpace{}%
\AgdaBound{Δ₂}\AgdaSymbol{\}}\AgdaSpace{}%
\AgdaSymbol{→}\AgdaSpace{}%
\AgdaBound{Δ₀}\AgdaSpace{}%
\AgdaOperator{\AgdaDatatype{⊩}}\AgdaSpace{}%
\AgdaOperator{\AgdaInductiveConstructor{[}}\AgdaSpace{}%
\AgdaBound{Δ₁}\AgdaSpace{}%
\AgdaOperator{\AgdaInductiveConstructor{,}}\AgdaSpace{}%
\AgdaBound{Δ₂}\AgdaSpace{}%
\AgdaOperator{\AgdaInductiveConstructor{]}}\AgdaSpace{}%
\AgdaSymbol{→}%
\>[49]\AgdaSymbol{(}\AgdaBound{Δ₀}\AgdaSpace{}%
\AgdaOperator{\AgdaInductiveConstructor{⨾}}\AgdaSpace{}%
\AgdaBound{Δ₁}\AgdaSymbol{)}\AgdaSpace{}%
\AgdaOperator{\AgdaDatatype{⊩}}%
\>[62]\AgdaBound{Δ₂}\<%
\\
\\[\AgdaEmptyExtraSkip]%
\>[4]\AgdaInductiveConstructor{bind}\AgdaSpace{}%
\AgdaSymbol{:}\AgdaSpace{}%
\AgdaSymbol{\{}\AgdaBound{Δ}\AgdaSpace{}%
\AgdaSymbol{:}\AgdaSpace{}%
\AgdaDatatype{sply}\AgdaSpace{}%
\AgdaBound{ℓ}\AgdaSymbol{\}}\AgdaSpace{}%
\AgdaSymbol{\{}\AgdaBound{(}\AgdaBound{A}\AgdaSpace{}%
\AgdaOperator{\AgdaInductiveConstructor{,}}\AgdaSpace{}%
\AgdaBound{Θ}\AgdaBound{)}\AgdaSpace{}%
\AgdaSymbol{:}\AgdaSpace{}%
\AgdaFunction{Σ⟨}\AgdaSpace{}%
\AgdaBound{A}\AgdaSpace{}%
\AgdaFunction{∶}\AgdaSpace{}%
\AgdaPrimitive{ityp}\AgdaSpace{}%
\AgdaBound{ℓ}\AgdaSpace{}%
\AgdaFunction{⟩}\AgdaSpace{}%
\AgdaSymbol{(}\AgdaBound{A}\AgdaSpace{}%
\AgdaSymbol{→}\AgdaSpace{}%
\AgdaDatatype{sply}\AgdaSpace{}%
\AgdaBound{ℓ}\AgdaSymbol{)\}}\<%
\\
\>[4][@{}l@{\AgdaIndent{0}}]%
\>[6]\AgdaSymbol{→}\AgdaSpace{}%
\AgdaSymbol{((}\AgdaBound{x}\AgdaSpace{}%
\AgdaSymbol{:}\AgdaSpace{}%
\AgdaBound{A}\AgdaSymbol{)}\AgdaSpace{}%
\AgdaSymbol{→}\AgdaSpace{}%
\AgdaBound{Δ}\AgdaSpace{}%
\AgdaOperator{\AgdaDatatype{⊩}}\AgdaSpace{}%
\AgdaBound{Θ}\AgdaSpace{}%
\AgdaBound{x}\AgdaSymbol{)}\AgdaSpace{}%
\AgdaSymbol{→}\AgdaSpace{}%
\AgdaSymbol{(}\AgdaBound{Δ}\AgdaSpace{}%
\AgdaOperator{\AgdaDatatype{⊩}}\AgdaSpace{}%
\AgdaInductiveConstructor{Λ}\AgdaSpace{}%
\AgdaBound{A}\AgdaSpace{}%
\AgdaBound{Θ}\AgdaSymbol{)}\<%
\\
\>[4]\AgdaInductiveConstructor{free}\AgdaSpace{}%
\AgdaSymbol{:}\AgdaSpace{}%
\AgdaSymbol{\{}\AgdaBound{Δ}\AgdaSpace{}%
\AgdaSymbol{:}\AgdaSpace{}%
\AgdaDatatype{sply}\AgdaSpace{}%
\AgdaBound{ℓ}\AgdaSymbol{\}}%
\>[25]\AgdaSymbol{\{}\AgdaBound{(}\AgdaBound{A}\AgdaSpace{}%
\AgdaOperator{\AgdaInductiveConstructor{,}}\AgdaSpace{}%
\AgdaBound{Θ}\AgdaBound{)}\AgdaSpace{}%
\AgdaSymbol{:}\AgdaSpace{}%
\AgdaFunction{Σ⟨}\AgdaSpace{}%
\AgdaBound{A}\AgdaSpace{}%
\AgdaFunction{∶}\AgdaSpace{}%
\AgdaPrimitive{ityp}\AgdaSpace{}%
\AgdaBound{ℓ}\AgdaSpace{}%
\AgdaFunction{⟩}\AgdaSpace{}%
\AgdaSymbol{(}\AgdaBound{A}\AgdaSpace{}%
\AgdaSymbol{→}\AgdaSpace{}%
\AgdaDatatype{sply}\AgdaSpace{}%
\AgdaBound{ℓ}\AgdaSymbol{)\}}\<%
\\
\>[4][@{}l@{\AgdaIndent{0}}]%
\>[6]\AgdaSymbol{→}\AgdaSpace{}%
\AgdaSymbol{(}\AgdaBound{Δ}\AgdaSpace{}%
\AgdaOperator{\AgdaDatatype{⊩}}\AgdaSpace{}%
\AgdaInductiveConstructor{Λ}\AgdaSpace{}%
\AgdaBound{A}\AgdaSpace{}%
\AgdaBound{Θ}\AgdaSymbol{)}\AgdaSpace{}%
\AgdaSymbol{→}\AgdaSpace{}%
\AgdaSymbol{((}\AgdaBound{x}\AgdaSpace{}%
\AgdaSymbol{:}\AgdaSpace{}%
\AgdaBound{A}\AgdaSymbol{)}\AgdaSpace{}%
\AgdaSymbol{→}\AgdaSpace{}%
\AgdaBound{Δ}\AgdaSpace{}%
\AgdaOperator{\AgdaDatatype{⊩}}\AgdaSpace{}%
\AgdaBound{Θ}\AgdaSpace{}%
\AgdaBound{x}\AgdaSymbol{)}\<%
\\
\\[\AgdaEmptyExtraSkip]%
\>[4]\AgdaInductiveConstructor{!ᶠ}\AgdaSpace{}%
\AgdaSymbol{:}\AgdaSpace{}%
\AgdaSymbol{∀}\AgdaSpace{}%
\AgdaSymbol{\{}\AgdaBound{Δ₀}\AgdaSpace{}%
\AgdaBound{Δ₁}\AgdaSymbol{\}}\AgdaSpace{}%
\AgdaSymbol{→}\AgdaSpace{}%
\AgdaBound{Δ₀}\AgdaSpace{}%
\AgdaOperator{\AgdaDatatype{⊩}}\AgdaSpace{}%
\AgdaBound{Δ₁}\AgdaSpace{}%
\AgdaSymbol{→}\AgdaSpace{}%
\AgdaInductiveConstructor{!}\AgdaSpace{}%
\AgdaBound{Δ₀}\AgdaSpace{}%
\AgdaOperator{\AgdaDatatype{⊩}}\AgdaSpace{}%
\AgdaInductiveConstructor{!}\AgdaSpace{}%
\AgdaBound{Δ₁}\<%
\\
\>[4]\AgdaInductiveConstructor{use}\AgdaSpace{}%
\AgdaSymbol{:}\AgdaSpace{}%
\AgdaSymbol{∀}\AgdaSpace{}%
\AgdaSymbol{\{}\AgdaBound{Δ}\AgdaSymbol{\}}\AgdaSpace{}%
\AgdaSymbol{→}\AgdaSpace{}%
\AgdaInductiveConstructor{!}\AgdaSpace{}%
\AgdaBound{Δ}\AgdaSpace{}%
\AgdaOperator{\AgdaDatatype{⊩}}\AgdaSpace{}%
\AgdaBound{Δ}\<%
\\
\>[4]\AgdaInductiveConstructor{mult}\AgdaSpace{}%
\AgdaSymbol{:}\AgdaSpace{}%
\AgdaSymbol{∀}\AgdaSpace{}%
\AgdaSymbol{\{}\AgdaBound{Δ}\AgdaSymbol{\}}\AgdaSpace{}%
\AgdaSymbol{→}\AgdaSpace{}%
\AgdaInductiveConstructor{!}\AgdaSpace{}%
\AgdaBound{Δ}\AgdaSpace{}%
\AgdaOperator{\AgdaDatatype{⊩}}\AgdaSpace{}%
\AgdaInductiveConstructor{!}\AgdaSpace{}%
\AgdaSymbol{(}\AgdaInductiveConstructor{!}\AgdaSpace{}%
\AgdaBound{Δ}\AgdaSymbol{)}\<%
\\
\>[4]\AgdaInductiveConstructor{coh◇}\AgdaSpace{}%
\AgdaSymbol{:}\AgdaSpace{}%
\AgdaInductiveConstructor{◇}\AgdaSpace{}%
\AgdaOperator{\AgdaDatatype{⊩}}\AgdaSpace{}%
\AgdaInductiveConstructor{!}\AgdaSpace{}%
\AgdaInductiveConstructor{◇}\<%
\\
\>[4]\AgdaInductiveConstructor{coh⨾}\AgdaSpace{}%
\AgdaSymbol{:}\AgdaSpace{}%
\AgdaSymbol{∀}\AgdaSpace{}%
\AgdaSymbol{\{}\AgdaBound{Δ₀}\AgdaSpace{}%
\AgdaBound{Δ₁}\AgdaSymbol{\}}\AgdaSpace{}%
\AgdaSymbol{→}\AgdaSpace{}%
\AgdaInductiveConstructor{!}\AgdaSpace{}%
\AgdaBound{Δ₀}\AgdaSpace{}%
\AgdaOperator{\AgdaInductiveConstructor{⨾}}\AgdaSpace{}%
\AgdaInductiveConstructor{!}\AgdaSpace{}%
\AgdaBound{Δ₁}\AgdaSpace{}%
\AgdaOperator{\AgdaDatatype{⊩}}\AgdaSpace{}%
\AgdaInductiveConstructor{!}\AgdaSpace{}%
\AgdaSymbol{(}\AgdaBound{Δ₀}\AgdaSpace{}%
\AgdaOperator{\AgdaInductiveConstructor{⨾}}\AgdaSpace{}%
\AgdaBound{Δ₁}\AgdaSymbol{)}\<%
\\
\>[4]\AgdaInductiveConstructor{coh⨾'}\AgdaSpace{}%
\AgdaSymbol{:}\AgdaSpace{}%
\AgdaSymbol{∀}\AgdaSpace{}%
\AgdaSymbol{\{}\AgdaBound{Δ₀}\AgdaSpace{}%
\AgdaBound{Δ₁}\AgdaSymbol{\}}\AgdaSpace{}%
\AgdaSymbol{→}\AgdaSpace{}%
\AgdaInductiveConstructor{!}\AgdaSpace{}%
\AgdaSymbol{(}\AgdaBound{Δ₀}\AgdaSpace{}%
\AgdaOperator{\AgdaInductiveConstructor{⨾}}\AgdaSpace{}%
\AgdaBound{Δ₁}\AgdaSymbol{)}\AgdaSpace{}%
\AgdaOperator{\AgdaDatatype{⊩}}\AgdaSpace{}%
\AgdaInductiveConstructor{!}\AgdaSpace{}%
\AgdaBound{Δ₀}\AgdaSpace{}%
\AgdaOperator{\AgdaInductiveConstructor{⨾}}\AgdaSpace{}%
\AgdaInductiveConstructor{!}\AgdaSpace{}%
\AgdaBound{Δ₁}\<%
\\
\\[\AgdaEmptyExtraSkip]%
\>[4]\AgdaInductiveConstructor{dupl}\AgdaSpace{}%
\AgdaSymbol{:}\AgdaSpace{}%
\AgdaSymbol{∀}\AgdaSpace{}%
\AgdaSymbol{\{}\AgdaBound{Δ}\AgdaSymbol{\}}\AgdaSpace{}%
\AgdaSymbol{→}\AgdaSpace{}%
\AgdaInductiveConstructor{!}\AgdaSpace{}%
\AgdaBound{Δ}\AgdaSpace{}%
\AgdaOperator{\AgdaDatatype{⊩}}\AgdaSpace{}%
\AgdaInductiveConstructor{!}\AgdaSpace{}%
\AgdaSymbol{(}\AgdaBound{Δ}\AgdaSpace{}%
\AgdaOperator{\AgdaInductiveConstructor{⨾}}\AgdaSpace{}%
\AgdaBound{Δ}\AgdaSymbol{)}\<%
\\
\>[4]\AgdaInductiveConstructor{erase}\AgdaSpace{}%
\AgdaSymbol{:}\AgdaSpace{}%
\AgdaSymbol{∀}\AgdaSpace{}%
\AgdaSymbol{\{}\AgdaBound{Δ}\AgdaSymbol{\}}\AgdaSpace{}%
\AgdaSymbol{→}\AgdaSpace{}%
\AgdaInductiveConstructor{!}\AgdaSpace{}%
\AgdaBound{Δ}\AgdaSpace{}%
\AgdaOperator{\AgdaDatatype{⊩}}\AgdaSpace{}%
\AgdaInductiveConstructor{◇}\<%
\\
\\[\AgdaEmptyExtraSkip]%
\\[\AgdaEmptyExtraSkip]%
\>[0]\AgdaKeyword{private}\<%
\\
\>[0][@{}l@{\AgdaIndent{0}}]%
\>[2]\AgdaKeyword{variable}\<%
\\
\>[2][@{}l@{\AgdaIndent{0}}]%
\>[4]\AgdaGeneralizable{ℓ}\AgdaSpace{}%
\AgdaGeneralizable{ℓ'}\AgdaSpace{}%
\AgdaSymbol{:}\AgdaSpace{}%
\AgdaPostulate{Level}\<%
\\
\\[\AgdaEmptyExtraSkip]%
\\[\AgdaEmptyExtraSkip]%
\>[0]\AgdaKeyword{infixl}\AgdaSpace{}%
\AgdaNumber{25}\AgdaSpace{}%
\AgdaInductiveConstructor{Λ}\<%
\\
\>[0]\AgdaKeyword{syntax}\AgdaSpace{}%
\AgdaInductiveConstructor{Λ}\AgdaSpace{}%
\AgdaBound{A}\AgdaSpace{}%
\AgdaSymbol{(λ}\AgdaSpace{}%
\AgdaBound{x}\AgdaSpace{}%
\AgdaSymbol{→}\AgdaSpace{}%
\AgdaBound{Δ}\AgdaSymbol{)}\AgdaSpace{}%
\AgdaSymbol{=}\AgdaSpace{}%
\AgdaInductiveConstructor{Λ⟨}\AgdaSpace{}%
\AgdaBound{x}\AgdaSpace{}%
\AgdaInductiveConstructor{∶}\AgdaSpace{}%
\AgdaBound{A}\AgdaSpace{}%
\AgdaInductiveConstructor{⟩}\AgdaSpace{}%
\AgdaBound{Δ}\<%
\\
\\[\AgdaEmptyExtraSkip]%
\\[\AgdaEmptyExtraSkip]%
\\[\AgdaEmptyExtraSkip]%
\>[0]\AgdaFunction{unitl}\AgdaSpace{}%
\AgdaSymbol{:}\AgdaSpace{}%
\AgdaSymbol{∀}\AgdaSpace{}%
\AgdaInductiveConstructor{(}\AgdaBound{Δ}\AgdaSpace{}%
\AgdaSymbol{:}\AgdaSpace{}%
\AgdaDatatype{sply}\AgdaSpace{}%
\AgdaGeneralizable{ℓ}\AgdaBound{)}\AgdaSpace{}%
\AgdaSymbol{→}\AgdaSpace{}%
\AgdaInductiveConstructor{◇}\AgdaSpace{}%
\AgdaOperator{\AgdaInductiveConstructor{⨾}}\AgdaSpace{}%
\AgdaBound{Δ}\AgdaSpace{}%
\AgdaOperator{\AgdaDatatype{⊩}}\AgdaSpace{}%
\AgdaBound{Δ}\<%
\\
\>[0]\AgdaFunction{unitl}\AgdaSpace{}%
\AgdaBound{Δ}\AgdaSpace{}%
\AgdaSymbol{=}\AgdaSpace{}%
\AgdaInductiveConstructor{unitr}\AgdaSpace{}%
\AgdaBound{Δ}\AgdaSpace{}%
\AgdaOperator{\AgdaInductiveConstructor{∘}}\AgdaSpace{}%
\AgdaInductiveConstructor{swap}\AgdaSpace{}%
\AgdaInductiveConstructor{◇}\AgdaSpace{}%
\AgdaBound{Δ}\<%
\\
\\[\AgdaEmptyExtraSkip]%
\>[0]\AgdaFunction{unitl'}\AgdaSpace{}%
\AgdaSymbol{:}\AgdaSpace{}%
\AgdaSymbol{∀}\AgdaSpace{}%
\AgdaSymbol{(}\AgdaBound{Δ}\AgdaSpace{}%
\AgdaSymbol{:}\AgdaSpace{}%
\AgdaDatatype{sply}\AgdaSpace{}%
\AgdaGeneralizable{ℓ}\AgdaSymbol{)}\AgdaSpace{}%
\AgdaSymbol{→}\AgdaSpace{}%
\AgdaBound{Δ}\AgdaSpace{}%
\AgdaOperator{\AgdaDatatype{⊩}}\AgdaSpace{}%
\AgdaInductiveConstructor{◇}\AgdaSpace{}%
\AgdaOperator{\AgdaInductiveConstructor{⨾}}\AgdaSpace{}%
\AgdaBound{Δ}\<%
\\
\>[0]\AgdaFunction{unitl'}\AgdaSpace{}%
\AgdaBound{Δ}\AgdaSpace{}%
\AgdaSymbol{=}\AgdaSpace{}%
\AgdaInductiveConstructor{swap}\AgdaSpace{}%
\AgdaBound{Δ}\AgdaSpace{}%
\AgdaInductiveConstructor{◇}\AgdaSpace{}%
\AgdaOperator{\AgdaInductiveConstructor{∘}}\AgdaSpace{}%
\AgdaInductiveConstructor{unitr'}\AgdaSpace{}%
\AgdaBound{Δ}\<%
\\
\\[\AgdaEmptyExtraSkip]%
\\[\AgdaEmptyExtraSkip]%
\\[\AgdaEmptyExtraSkip]%
\>[0]\AgdaFunction{assoc'}\AgdaSpace{}%
\AgdaSymbol{:}\AgdaSpace{}%
\AgdaSymbol{∀}\AgdaSpace{}%
\AgdaSymbol{(}\AgdaBound{Δ₀}\AgdaSpace{}%
\AgdaBound{Δ₁}\AgdaSpace{}%
\AgdaBound{Δ₂}\AgdaSpace{}%
\AgdaSymbol{:}\AgdaSpace{}%
\AgdaDatatype{sply}\AgdaSpace{}%
\AgdaGeneralizable{ℓ}\AgdaSymbol{)}\AgdaSpace{}%
\AgdaSymbol{→}\AgdaSpace{}%
\AgdaSymbol{(}\AgdaBound{Δ₀}\AgdaSpace{}%
\AgdaOperator{\AgdaInductiveConstructor{⨾}}\AgdaSpace{}%
\AgdaBound{Δ₁}\AgdaSymbol{)}\AgdaSpace{}%
\AgdaOperator{\AgdaInductiveConstructor{⨾}}\AgdaSpace{}%
\AgdaBound{Δ₂}\AgdaSpace{}%
\AgdaOperator{\AgdaDatatype{⊩}}\AgdaSpace{}%
\AgdaBound{Δ₀}\AgdaSpace{}%
\AgdaOperator{\AgdaInductiveConstructor{⨾}}\AgdaSpace{}%
\AgdaSymbol{(}\AgdaBound{Δ₁}\AgdaSpace{}%
\AgdaOperator{\AgdaInductiveConstructor{⨾}}\AgdaSpace{}%
\AgdaBound{Δ₂}\AgdaSymbol{)}\<%
\\
\>[0]\AgdaFunction{assoc'}\AgdaSpace{}%
\AgdaBound{x}\AgdaSpace{}%
\AgdaBound{y}\AgdaSpace{}%
\AgdaBound{z}\AgdaSpace{}%
\AgdaSymbol{=}\AgdaSpace{}%
\AgdaInductiveConstructor{swap}\AgdaSpace{}%
\AgdaSymbol{(}\AgdaBound{y}\AgdaSpace{}%
\AgdaOperator{\AgdaInductiveConstructor{⨾}}\AgdaSpace{}%
\AgdaBound{z}\AgdaSymbol{)}\AgdaSpace{}%
\AgdaBound{x}\<%
\\
\>[0][@{}l@{\AgdaIndent{0}}]%
\>[2]\AgdaOperator{\AgdaInductiveConstructor{∘}}\AgdaSpace{}%
\AgdaInductiveConstructor{assoc}\AgdaSpace{}%
\AgdaBound{y}\AgdaSpace{}%
\AgdaBound{z}\AgdaSpace{}%
\AgdaBound{x}\<%
\\
\>[2]\AgdaOperator{\AgdaInductiveConstructor{∘}}\AgdaSpace{}%
\AgdaInductiveConstructor{id}\AgdaSpace{}%
\AgdaBound{y}\AgdaSpace{}%
\AgdaOperator{\AgdaInductiveConstructor{⨾ᶠ}}\AgdaSpace{}%
\AgdaInductiveConstructor{swap}\AgdaSpace{}%
\AgdaBound{x}\AgdaSpace{}%
\AgdaBound{z}\<%
\\
\>[2]\AgdaOperator{\AgdaInductiveConstructor{∘}}\AgdaSpace{}%
\AgdaInductiveConstructor{swap}\AgdaSpace{}%
\AgdaSymbol{(}\AgdaBound{x}\AgdaSpace{}%
\AgdaOperator{\AgdaInductiveConstructor{⨾}}\AgdaSpace{}%
\AgdaBound{z}\AgdaSymbol{)}\AgdaSpace{}%
\AgdaBound{y}\<%
\\
\>[2]\AgdaOperator{\AgdaInductiveConstructor{∘}}\AgdaSpace{}%
\AgdaInductiveConstructor{swap}\AgdaSpace{}%
\AgdaBound{z}\AgdaSpace{}%
\AgdaBound{x}\AgdaSpace{}%
\AgdaOperator{\AgdaInductiveConstructor{⨾ᶠ}}\AgdaSpace{}%
\AgdaInductiveConstructor{id}\AgdaSpace{}%
\AgdaBound{y}\<%
\\
\>[2]\AgdaOperator{\AgdaInductiveConstructor{∘}}\AgdaSpace{}%
\AgdaInductiveConstructor{assoc}\AgdaSpace{}%
\AgdaBound{z}\AgdaSpace{}%
\AgdaBound{x}\AgdaSpace{}%
\AgdaBound{y}\<%
\\
\>[2]\AgdaOperator{\AgdaInductiveConstructor{∘}}\AgdaSpace{}%
\AgdaInductiveConstructor{swap}\AgdaSpace{}%
\AgdaSymbol{(}\AgdaBound{x}\AgdaSpace{}%
\AgdaOperator{\AgdaInductiveConstructor{⨾}}\AgdaSpace{}%
\AgdaBound{y}\AgdaSymbol{)}\AgdaSpace{}%
\AgdaBound{z}\<%
\\
\\[\AgdaEmptyExtraSkip]%
\\[\AgdaEmptyExtraSkip]%
\\[\AgdaEmptyExtraSkip]%
\>[0]\AgdaOperator{\AgdaFunction{\AgdaUnderscore{}\textasciicircum{}\AgdaUnderscore{}}}\AgdaSpace{}%
\AgdaSymbol{:}\AgdaSpace{}%
\AgdaDatatype{sply}\AgdaSpace{}%
\AgdaGeneralizable{ℓ}\AgdaSpace{}%
\AgdaSymbol{→}\AgdaSpace{}%
\AgdaDatatype{ℕ}\AgdaSpace{}%
\AgdaSymbol{→}\AgdaSpace{}%
\AgdaDatatype{sply}\AgdaSpace{}%
\AgdaGeneralizable{ℓ}\<%
\\
\>[0]\AgdaBound{Δ}\AgdaSpace{}%
\AgdaOperator{\AgdaFunction{\textasciicircum{}}}\AgdaSpace{}%
\AgdaInductiveConstructor{zero}\AgdaSpace{}%
\AgdaSymbol{=}\AgdaSpace{}%
\AgdaInductiveConstructor{◇}\<%
\\
\>[0]\AgdaBound{Δ}\AgdaSpace{}%
\AgdaOperator{\AgdaFunction{\textasciicircum{}}}\AgdaSpace{}%
\AgdaSymbol{(}\AgdaInductiveConstructor{suc}\AgdaSpace{}%
\AgdaBound{m}\AgdaSymbol{)}\AgdaSpace{}%
\AgdaSymbol{=}\AgdaSpace{}%
\AgdaBound{Δ}\AgdaSpace{}%
\AgdaOperator{\AgdaInductiveConstructor{⨾}}\AgdaSpace{}%
\AgdaSymbol{(}\AgdaBound{Δ}\AgdaSpace{}%
\AgdaOperator{\AgdaFunction{\textasciicircum{}}}\AgdaSpace{}%
\AgdaBound{m}\AgdaSymbol{)}\<%
\\
\\[\AgdaEmptyExtraSkip]%
\>[0]\AgdaKeyword{infixr}\AgdaSpace{}%
\AgdaNumber{50}\AgdaSpace{}%
\AgdaOperator{\AgdaFunction{\AgdaUnderscore{}\textasciicircum{}\AgdaUnderscore{}}}\<%
\\
\\[\AgdaEmptyExtraSkip]%
\>[0]\AgdaFunction{⊩\textasciicircum{}}\AgdaSpace{}%
\AgdaSymbol{:}\AgdaSpace{}%
\AgdaSymbol{∀\{}\AgdaBound{ℓ}\AgdaSymbol{\}\{}\AgdaBound{Δ₀}\AgdaSpace{}%
\AgdaBound{Δ₁}\AgdaSpace{}%
\AgdaSymbol{:}\AgdaSpace{}%
\AgdaDatatype{sply}\AgdaSpace{}%
\AgdaBound{ℓ}\AgdaSymbol{\}}\AgdaSpace{}%
\AgdaSymbol{→}\AgdaSpace{}%
\AgdaSymbol{(}\AgdaBound{m}\AgdaSpace{}%
\AgdaSymbol{:}\AgdaSpace{}%
\AgdaDatatype{ℕ}\AgdaSymbol{)}\AgdaSpace{}%
\AgdaSymbol{→}\AgdaSpace{}%
\AgdaBound{Δ₀}\AgdaSpace{}%
\AgdaOperator{\AgdaDatatype{⊩}}\AgdaSpace{}%
\AgdaBound{Δ₁}\AgdaSpace{}%
\AgdaSymbol{→}\AgdaSpace{}%
\AgdaSymbol{(}\AgdaBound{Δ₀}\AgdaSpace{}%
\AgdaOperator{\AgdaFunction{\textasciicircum{}}}\AgdaSpace{}%
\AgdaBound{m}\AgdaSymbol{)}\AgdaSpace{}%
\AgdaOperator{\AgdaDatatype{⊩}}\AgdaSpace{}%
\AgdaSymbol{(}\AgdaBound{Δ₁}\AgdaSpace{}%
\AgdaOperator{\AgdaFunction{\textasciicircum{}}}\AgdaSpace{}%
\AgdaBound{m}\AgdaSymbol{)}\<%
\\
\>[0]\AgdaFunction{⊩\textasciicircum{}}\AgdaSpace{}%
\AgdaInductiveConstructor{zero}\AgdaSpace{}%
\AgdaBound{δ}\AgdaSpace{}%
\AgdaSymbol{=}\AgdaSpace{}%
\AgdaInductiveConstructor{id}\AgdaSpace{}%
\AgdaInductiveConstructor{◇}\<%
\\
\>[0]\AgdaFunction{⊩\textasciicircum{}}\AgdaSpace{}%
\AgdaSymbol{(}\AgdaInductiveConstructor{suc}\AgdaSpace{}%
\AgdaBound{m}\AgdaSymbol{)}\AgdaSpace{}%
\AgdaBound{δ}\AgdaSpace{}%
\AgdaSymbol{=}\AgdaSpace{}%
\AgdaBound{δ}\AgdaSpace{}%
\AgdaOperator{\AgdaInductiveConstructor{⨾ᶠ}}\AgdaSpace{}%
\AgdaFunction{⊩\textasciicircum{}}\AgdaSpace{}%
\AgdaBound{m}\AgdaSpace{}%
\AgdaBound{δ}\<%
\\
\\[\AgdaEmptyExtraSkip]%
\>[0]\AgdaFunction{\textasciicircum{}asrec}\AgdaSpace{}%
\AgdaSymbol{:}\AgdaSpace{}%
\AgdaDatatype{sply}\AgdaSpace{}%
\AgdaGeneralizable{ℓ}\AgdaSpace{}%
\AgdaSymbol{→}\AgdaSpace{}%
\AgdaDatatype{ℕ}\AgdaSpace{}%
\AgdaSymbol{→}\AgdaSpace{}%
\AgdaDatatype{sply}\AgdaSpace{}%
\AgdaGeneralizable{ℓ}\<%
\\
\>[0]\AgdaFunction{\textasciicircum{}asrec}\AgdaSpace{}%
\AgdaSymbol{\{}\AgdaBound{ℓ}\AgdaSymbol{\}}\AgdaSpace{}%
\AgdaBound{Δ}\AgdaSpace{}%
\AgdaSymbol{=}\AgdaSpace{}%
\AgdaFunction{natElim}\AgdaSpace{}%
\AgdaSymbol{\{}\AgdaArgument{A}\AgdaSpace{}%
\AgdaSymbol{=}\AgdaSpace{}%
\AgdaSymbol{λ}\AgdaSpace{}%
\AgdaBound{\AgdaUnderscore{}}\AgdaSpace{}%
\AgdaSymbol{→}\AgdaSpace{}%
\AgdaDatatype{sply}\AgdaSpace{}%
\AgdaBound{ℓ}\AgdaSymbol{\}}\AgdaSpace{}%
\AgdaSymbol{(λ}\AgdaSpace{}%
\AgdaBound{\AgdaUnderscore{}}\AgdaSpace{}%
\AgdaSymbol{→}\AgdaSpace{}%
\AgdaOperator{\AgdaInductiveConstructor{\AgdaUnderscore{}⨾}}\AgdaSpace{}%
\AgdaBound{Δ}\AgdaSymbol{)}\AgdaSpace{}%
\AgdaInductiveConstructor{◇}\<%
\\
\\[\AgdaEmptyExtraSkip]%
\>[0]\AgdaFunction{⊩\textasciicircum{}asrec}\AgdaSpace{}%
\AgdaSymbol{:}\AgdaSpace{}%
\AgdaSymbol{∀\{}\AgdaBound{ℓ}\AgdaSymbol{\}\{}\AgdaBound{Δ₀}\AgdaSpace{}%
\AgdaBound{Δ₁}\AgdaSpace{}%
\AgdaSymbol{:}\AgdaSpace{}%
\AgdaDatatype{sply}\AgdaSpace{}%
\AgdaBound{ℓ}\AgdaSymbol{\}}\AgdaSpace{}%
\AgdaSymbol{→}\AgdaSpace{}%
\AgdaBound{Δ₀}\AgdaSpace{}%
\AgdaOperator{\AgdaDatatype{⊩}}\AgdaSpace{}%
\AgdaBound{Δ₁}\AgdaSpace{}%
\AgdaSymbol{→}\AgdaSpace{}%
\AgdaSymbol{(}\AgdaBound{m}\AgdaSpace{}%
\AgdaSymbol{:}\AgdaSpace{}%
\AgdaDatatype{ℕ}\AgdaSymbol{)}\AgdaSpace{}%
\AgdaSymbol{→}\AgdaSpace{}%
\AgdaSymbol{(}\AgdaFunction{\textasciicircum{}asrec}\AgdaSpace{}%
\AgdaBound{Δ₀}\AgdaSpace{}%
\AgdaBound{m}\AgdaSymbol{)}\AgdaSpace{}%
\AgdaOperator{\AgdaDatatype{⊩}}\AgdaSpace{}%
\AgdaSymbol{(}\AgdaFunction{\textasciicircum{}asrec}\AgdaSpace{}%
\AgdaBound{Δ₁}\AgdaSpace{}%
\AgdaBound{m}\AgdaSymbol{)}\<%
\\
\>[0]\AgdaFunction{⊩\textasciicircum{}asrec}\AgdaSpace{}%
\AgdaSymbol{\{}\AgdaBound{ℓ}\AgdaSymbol{\}}\AgdaSpace{}%
\AgdaSymbol{\{}\AgdaBound{Δ₀}\AgdaSymbol{\}}\AgdaSpace{}%
\AgdaSymbol{\{}\AgdaBound{Δ₁}\AgdaSymbol{\}}\AgdaSpace{}%
\AgdaBound{δ}\AgdaSpace{}%
\AgdaSymbol{=}\AgdaSpace{}%
\AgdaFunction{natElim}\AgdaSpace{}%
\AgdaSymbol{\{}\AgdaArgument{A}\AgdaSpace{}%
\AgdaSymbol{=}\AgdaSpace{}%
\AgdaSymbol{λ}\AgdaSpace{}%
\AgdaBound{m}\AgdaSpace{}%
\AgdaSymbol{→}\AgdaSpace{}%
\AgdaSymbol{(}\AgdaFunction{\textasciicircum{}asrec}\AgdaSpace{}%
\AgdaBound{Δ₀}\AgdaSpace{}%
\AgdaBound{m}\AgdaSymbol{)}\AgdaSpace{}%
\AgdaOperator{\AgdaDatatype{⊩}}\AgdaSpace{}%
\AgdaSymbol{(}\AgdaFunction{\textasciicircum{}asrec}\AgdaSpace{}%
\AgdaBound{Δ₁}\AgdaSpace{}%
\AgdaBound{m}\AgdaSymbol{)\}}\AgdaSpace{}%
\AgdaSymbol{(λ}\AgdaSpace{}%
\AgdaBound{\AgdaUnderscore{}}\AgdaSpace{}%
\AgdaSymbol{→}\AgdaSpace{}%
\AgdaOperator{\AgdaInductiveConstructor{\AgdaUnderscore{}⨾ᶠ}}\AgdaSpace{}%
\AgdaBound{δ}\AgdaSymbol{)}\AgdaSpace{}%
\AgdaSymbol{(}\AgdaInductiveConstructor{id}\AgdaSpace{}%
\AgdaInductiveConstructor{◇}\AgdaSymbol{)}\<%
\\
\\[\AgdaEmptyExtraSkip]%
\>[0]\AgdaOperator{\AgdaFunction{∣\AgdaUnderscore{}∣}}\AgdaSpace{}%
\AgdaSymbol{:}\AgdaSpace{}%
\AgdaDatatype{Bool}\AgdaSpace{}%
\AgdaSymbol{\{}\AgdaGeneralizable{ℓ}\AgdaSymbol{\}}\AgdaSpace{}%
\AgdaSymbol{→}\AgdaSpace{}%
\AgdaDatatype{ℕ}\<%
\\
\>[0]\AgdaOperator{\AgdaFunction{∣}}\AgdaSpace{}%
\AgdaInductiveConstructor{false}\AgdaSpace{}%
\AgdaOperator{\AgdaFunction{∣}}\AgdaSpace{}%
\AgdaSymbol{=}\AgdaSpace{}%
\AgdaNumber{0}\<%
\\
\>[0]\AgdaOperator{\AgdaFunction{∣}}\AgdaSpace{}%
\AgdaInductiveConstructor{true}\AgdaSpace{}%
\AgdaOperator{\AgdaFunction{∣}}\AgdaSpace{}%
\AgdaSymbol{=}\AgdaSpace{}%
\AgdaNumber{1}\<%
\\
\\[\AgdaEmptyExtraSkip]%
\\[\AgdaEmptyExtraSkip]%
\>[0]\AgdaKeyword{data}\AgdaSpace{}%
\AgdaDatatype{ltyp}\AgdaSpace{}%
\AgdaSymbol{(}\AgdaBound{ℓ}\AgdaSpace{}%
\AgdaSymbol{:}\AgdaSpace{}%
\AgdaPostulate{Level}\AgdaSymbol{)}\AgdaSpace{}%
\AgdaSymbol{:}\AgdaSpace{}%
\AgdaPrimitive{ityp}\AgdaSpace{}%
\AgdaSymbol{(}\AgdaPrimitive{ℓ-suc}\AgdaSpace{}%
\AgdaBound{ℓ}\AgdaSymbol{)}\<%
\\
\>[0]\<%
\end{code}
\newcommand{\AGDAltypsem}{%
\begin{code}%
\>[0]\AgdaOperator{\AgdaFunction{⟦\AgdaUnderscore{}⟧}}\AgdaSpace{}%
\AgdaSymbol{:}\AgdaSpace{}%
\AgdaDatatype{ltyp}\AgdaSpace{}%
\AgdaGeneralizable{ℓ}\AgdaSpace{}%
\AgdaSymbol{→}\AgdaSpace{}%
\AgdaFunction{Σ⟨}\AgdaSpace{}%
\AgdaBound{A}\AgdaSpace{}%
\AgdaFunction{∶}\AgdaSpace{}%
\AgdaPrimitive{ityp}\AgdaSpace{}%
\AgdaGeneralizable{ℓ}\AgdaSpace{}%
\AgdaFunction{⟩}\AgdaSpace{}%
\AgdaSymbol{(}\AgdaBound{A}\AgdaSpace{}%
\AgdaSymbol{→}\AgdaSpace{}%
\AgdaDatatype{sply}\AgdaSpace{}%
\AgdaGeneralizable{ℓ}\AgdaSymbol{)}\<%
\end{code}}
\begin{code}[hide]%
\>[0]\<%
\\
\>[0]\<%
\end{code}

\begin{code}[hide]%
\>[0]\<%
\\
\>[0]\AgdaKeyword{infix}\AgdaSpace{}%
\AgdaNumber{-4}\AgdaSpace{}%
\AgdaOperator{\AgdaFunction{\AgdaUnderscore{}．}}\<%
\\
\\[\AgdaEmptyExtraSkip]%
\>[0]\AgdaFunction{lin}\AgdaSpace{}%
\AgdaSymbol{:}\AgdaSpace{}%
\AgdaSymbol{(}\AgdaBound{A}\AgdaSpace{}%
\AgdaSymbol{:}\AgdaSpace{}%
\AgdaDatatype{ltyp}\AgdaSpace{}%
\AgdaGeneralizable{ℓ}\AgdaSymbol{)}\AgdaSpace{}%
\AgdaSymbol{→}\AgdaSpace{}%
\AgdaBound{A}\AgdaSpace{}%
\AgdaOperator{\AgdaFunction{．}}\AgdaSpace{}%
\AgdaSymbol{→}\AgdaSpace{}%
\AgdaDatatype{sply}\AgdaSpace{}%
\AgdaGeneralizable{ℓ}\<%
\\
\>[0]\AgdaFunction{lin}\AgdaSpace{}%
\AgdaBound{A}\AgdaSpace{}%
\AgdaBound{x}\AgdaSpace{}%
\AgdaSymbol{=}\AgdaSpace{}%
\AgdaOperator{\AgdaFunction{⟦}}\AgdaSpace{}%
\AgdaBound{A}\AgdaSpace{}%
\AgdaOperator{\AgdaFunction{⟧}}\AgdaSpace{}%
\AgdaSymbol{.}\AgdaField{snd}\AgdaSpace{}%
\AgdaBound{x}\<%
\\
\>[0]\AgdaKeyword{infix}\AgdaSpace{}%
\AgdaNumber{-10}\AgdaSpace{}%
\AgdaFunction{lin}\<%
\\
\>[0]\AgdaKeyword{syntax}\AgdaSpace{}%
\AgdaFunction{lin}\AgdaSpace{}%
\AgdaBound{A}\AgdaSpace{}%
\AgdaBound{x}\AgdaSpace{}%
\AgdaSymbol{=}\AgdaSpace{}%
\AgdaBound{x}\AgdaSpace{}%
\AgdaFunction{∶}\AgdaSpace{}%
\AgdaBound{A}\<%
\\
\\[\AgdaEmptyExtraSkip]%
\\[\AgdaEmptyExtraSkip]%
\>[0]\AgdaFunction{depltyp}\AgdaSpace{}%
\AgdaSymbol{:}\AgdaSpace{}%
\AgdaSymbol{\{}\AgdaBound{ℓ}\AgdaSpace{}%
\AgdaSymbol{:}\AgdaSpace{}%
\AgdaPostulate{Level}\AgdaSymbol{\}}\AgdaSpace{}%
\AgdaSymbol{→}\AgdaSpace{}%
\AgdaDatatype{ltyp}\AgdaSpace{}%
\AgdaBound{ℓ}\AgdaSpace{}%
\AgdaSymbol{→}\AgdaSpace{}%
\AgdaPrimitive{ityp}\AgdaSpace{}%
\AgdaSymbol{(}\AgdaPrimitive{ℓ-suc}\AgdaSpace{}%
\AgdaBound{ℓ}\AgdaSymbol{)}\<%
\\
\>[0]\AgdaFunction{depltyp}\AgdaSpace{}%
\AgdaSymbol{\{}\AgdaBound{ℓ}\AgdaSymbol{\}}\AgdaSpace{}%
\AgdaBound{A}\AgdaSpace{}%
\AgdaSymbol{=}\AgdaSpace{}%
\AgdaBound{A}\AgdaSpace{}%
\AgdaOperator{\AgdaFunction{．}}\AgdaSpace{}%
\AgdaSymbol{→}\AgdaSpace{}%
\AgdaDatatype{ltyp}\AgdaSpace{}%
\AgdaBound{ℓ}\<%
\\
\\[\AgdaEmptyExtraSkip]%
\>[0]\AgdaKeyword{syntax}\AgdaSpace{}%
\AgdaFunction{depltyp}\AgdaSpace{}%
\AgdaBound{A}\AgdaSpace{}%
\AgdaSymbol{=}\AgdaSpace{}%
\AgdaBound{A}\AgdaSpace{}%
\AgdaFunction{⇒}\AgdaSpace{}%
\AgdaFunction{ltyp}\<%
\\
\\[\AgdaEmptyExtraSkip]%
\>[0]\AgdaFunction{depfinltyp}\AgdaSpace{}%
\AgdaSymbol{:}\AgdaSpace{}%
\AgdaSymbol{\{}\AgdaBound{ℓ}\AgdaSpace{}%
\AgdaSymbol{:}\AgdaSpace{}%
\AgdaPostulate{Level}\AgdaSymbol{\}}\AgdaSpace{}%
\AgdaSymbol{→}\AgdaSpace{}%
\AgdaDatatype{ltyp}\AgdaSpace{}%
\AgdaBound{ℓ}\AgdaSpace{}%
\AgdaSymbol{→}\AgdaSpace{}%
\AgdaPrimitive{ityp}\AgdaSpace{}%
\AgdaSymbol{(}\AgdaPrimitive{ℓ-suc}\AgdaSpace{}%
\AgdaBound{ℓ}\AgdaSymbol{)}\<%
\\
\>[0]\AgdaFunction{depfinltyp}\AgdaSpace{}%
\AgdaSymbol{\{}\AgdaBound{ℓ}\AgdaSymbol{\}}\AgdaSpace{}%
\AgdaBound{A}\AgdaSpace{}%
\AgdaSymbol{=}\AgdaSpace{}%
\AgdaFunction{Σ⟨}\AgdaSpace{}%
\AgdaBound{B}\AgdaSpace{}%
\AgdaFunction{∶}\AgdaSpace{}%
\AgdaSymbol{(}\AgdaBound{A}\AgdaSpace{}%
\AgdaOperator{\AgdaFunction{．}}\AgdaSpace{}%
\AgdaSymbol{→}\AgdaSpace{}%
\AgdaDatatype{ltyp}\AgdaSpace{}%
\AgdaBound{ℓ}\AgdaSymbol{)}\AgdaSpace{}%
\AgdaFunction{⟩}\AgdaSpace{}%
\AgdaSymbol{((}\AgdaBound{x}\AgdaSpace{}%
\AgdaSymbol{:}\AgdaSpace{}%
\AgdaBound{A}\AgdaSpace{}%
\AgdaOperator{\AgdaFunction{．}}\AgdaSymbol{)}\AgdaSpace{}%
\AgdaSymbol{→}\AgdaSpace{}%
\AgdaDatatype{List}\AgdaSpace{}%
\AgdaSymbol{(}\AgdaBound{B}\AgdaSpace{}%
\AgdaBound{x}\AgdaSpace{}%
\AgdaOperator{\AgdaFunction{．}}\AgdaSymbol{))}\<%
\\
\\[\AgdaEmptyExtraSkip]%
\>[0]\AgdaKeyword{syntax}\AgdaSpace{}%
\AgdaFunction{depfinltyp}\AgdaSpace{}%
\AgdaBound{A}\AgdaSpace{}%
\AgdaSymbol{=}\AgdaSpace{}%
\AgdaBound{A}\AgdaSpace{}%
\AgdaFunction{⇒}\AgdaSpace{}%
\AgdaFunction{finltyp}\<%
\\
\\[\AgdaEmptyExtraSkip]%
\\[\AgdaEmptyExtraSkip]%
\>[0]\AgdaFunction{⨂}\AgdaSpace{}%
\AgdaSymbol{:}\AgdaSpace{}%
\AgdaSymbol{(}\AgdaBound{A}\AgdaSpace{}%
\AgdaSymbol{:}\AgdaSpace{}%
\AgdaDatatype{ltyp}\AgdaSpace{}%
\AgdaGeneralizable{ℓ}\AgdaSymbol{)}\AgdaSpace{}%
\AgdaSymbol{→}\AgdaSpace{}%
\AgdaSymbol{(}\AgdaDatatype{List}\AgdaSpace{}%
\AgdaSymbol{(}\AgdaBound{A}\AgdaSpace{}%
\AgdaOperator{\AgdaFunction{．}}\AgdaSymbol{))}\AgdaSpace{}%
\AgdaSymbol{→}\AgdaSpace{}%
\AgdaSymbol{(}\AgdaBound{Δ}\AgdaSpace{}%
\AgdaSymbol{:}\AgdaSpace{}%
\AgdaBound{A}\AgdaSpace{}%
\AgdaOperator{\AgdaFunction{．}}\AgdaSpace{}%
\AgdaSymbol{→}\AgdaSpace{}%
\AgdaDatatype{sply}\AgdaSpace{}%
\AgdaGeneralizable{ℓ}\AgdaSymbol{)}\AgdaSpace{}%
\AgdaSymbol{→}\AgdaSpace{}%
\AgdaDatatype{sply}\AgdaSpace{}%
\AgdaGeneralizable{ℓ}\<%
\\
\>[0]\AgdaFunction{⨂}\AgdaSpace{}%
\AgdaBound{A}\AgdaSpace{}%
\AgdaInductiveConstructor{[]}\AgdaSpace{}%
\AgdaBound{Δ}\AgdaSpace{}%
\AgdaSymbol{=}\AgdaSpace{}%
\AgdaInductiveConstructor{◇}\<%
\\
\>[0]\AgdaFunction{⨂}\AgdaSpace{}%
\AgdaBound{A}\AgdaSpace{}%
\AgdaSymbol{(}\AgdaBound{a}\AgdaSpace{}%
\AgdaOperator{\AgdaInductiveConstructor{∷}}\AgdaSpace{}%
\AgdaBound{as}\AgdaSymbol{)}\AgdaSpace{}%
\AgdaBound{Δ}\AgdaSpace{}%
\AgdaSymbol{=}\AgdaSpace{}%
\AgdaBound{Δ}\AgdaSpace{}%
\AgdaBound{a}\AgdaSpace{}%
\AgdaOperator{\AgdaInductiveConstructor{⨾}}\AgdaSpace{}%
\AgdaFunction{⨂}\AgdaSpace{}%
\AgdaBound{A}\AgdaSpace{}%
\AgdaBound{as}\AgdaSpace{}%
\AgdaBound{Δ}\<%
\\
\\[\AgdaEmptyExtraSkip]%
\>[0]\AgdaFunction{⨂ᶠ}\AgdaSpace{}%
\AgdaSymbol{:}\AgdaSpace{}%
\AgdaSymbol{\{}\AgdaBound{A}\AgdaSpace{}%
\AgdaSymbol{:}\AgdaSpace{}%
\AgdaDatatype{ltyp}\AgdaSpace{}%
\AgdaGeneralizable{ℓ}\AgdaSymbol{\}}\AgdaSpace{}%
\AgdaSymbol{\{}\AgdaBound{as}\AgdaSpace{}%
\AgdaSymbol{:}\AgdaSpace{}%
\AgdaDatatype{List}\AgdaSpace{}%
\AgdaSymbol{(}\AgdaBound{A}\AgdaSpace{}%
\AgdaOperator{\AgdaFunction{．}}\AgdaSymbol{)\}}\AgdaSpace{}%
\AgdaSymbol{\{}\AgdaBound{Δ₀}\AgdaSpace{}%
\AgdaBound{Δ₁}\AgdaSpace{}%
\AgdaSymbol{:}\AgdaSpace{}%
\AgdaBound{A}\AgdaSpace{}%
\AgdaOperator{\AgdaFunction{．}}\AgdaSpace{}%
\AgdaSymbol{→}\AgdaSpace{}%
\AgdaDatatype{sply}\AgdaSpace{}%
\AgdaGeneralizable{ℓ}\AgdaSymbol{\}}\<%
\\
\>[0][@{}l@{\AgdaIndent{0}}]%
\>[2]\AgdaSymbol{→}\AgdaSpace{}%
\AgdaSymbol{((}\AgdaBound{x}\AgdaSpace{}%
\AgdaSymbol{:}\AgdaSpace{}%
\AgdaBound{A}\AgdaSpace{}%
\AgdaOperator{\AgdaFunction{．}}\AgdaSymbol{)}\AgdaSpace{}%
\AgdaSymbol{→}\AgdaSpace{}%
\AgdaBound{Δ₀}\AgdaSpace{}%
\AgdaBound{x}\AgdaSpace{}%
\AgdaOperator{\AgdaDatatype{⊩}}\AgdaSpace{}%
\AgdaBound{Δ₁}\AgdaSpace{}%
\AgdaBound{x}\AgdaSymbol{)}\<%
\\
\>[2]\AgdaSymbol{→}\AgdaSpace{}%
\AgdaFunction{⨂}\AgdaSpace{}%
\AgdaBound{A}\AgdaSpace{}%
\AgdaBound{as}\AgdaSpace{}%
\AgdaBound{Δ₀}\AgdaSpace{}%
\AgdaOperator{\AgdaDatatype{⊩}}\AgdaSpace{}%
\AgdaFunction{⨂}\AgdaSpace{}%
\AgdaBound{A}\AgdaSpace{}%
\AgdaBound{as}\AgdaSpace{}%
\AgdaBound{Δ₁}\<%
\\
\>[0]\AgdaFunction{⨂ᶠ}\AgdaSpace{}%
\AgdaSymbol{\{}\AgdaArgument{as}\AgdaSpace{}%
\AgdaSymbol{=}\AgdaSpace{}%
\AgdaInductiveConstructor{[]}\AgdaSymbol{\}}\AgdaSpace{}%
\AgdaBound{f}\AgdaSpace{}%
\AgdaSymbol{=}\AgdaSpace{}%
\AgdaInductiveConstructor{id}\AgdaSpace{}%
\AgdaSymbol{\AgdaUnderscore{}}\<%
\\
\>[0]\AgdaFunction{⨂ᶠ}\AgdaSpace{}%
\AgdaSymbol{\{}\AgdaArgument{as}\AgdaSpace{}%
\AgdaSymbol{=}\AgdaSpace{}%
\AgdaBound{a}\AgdaSpace{}%
\AgdaOperator{\AgdaInductiveConstructor{∷}}\AgdaSpace{}%
\AgdaBound{as}\AgdaSymbol{\}}\AgdaSpace{}%
\AgdaBound{f}\AgdaSpace{}%
\AgdaSymbol{=}\AgdaSpace{}%
\AgdaBound{f}\AgdaSpace{}%
\AgdaBound{a}\AgdaSpace{}%
\AgdaOperator{\AgdaInductiveConstructor{⨾ᶠ}}\AgdaSpace{}%
\AgdaFunction{⨂ᶠ}\AgdaSpace{}%
\AgdaSymbol{\{}\AgdaArgument{as}\AgdaSpace{}%
\AgdaSymbol{=}\AgdaSpace{}%
\AgdaBound{as}\AgdaSymbol{\}}\AgdaSpace{}%
\AgdaBound{f}\<%
\\
\\[\AgdaEmptyExtraSkip]%
\>[0]\AgdaFunction{⨂lax}\AgdaSpace{}%
\AgdaSymbol{:}\AgdaSpace{}%
\AgdaSymbol{\{}\AgdaBound{A}\AgdaSpace{}%
\AgdaSymbol{:}\AgdaSpace{}%
\AgdaDatatype{ltyp}\AgdaSpace{}%
\AgdaGeneralizable{ℓ}\AgdaSymbol{\}}\AgdaSpace{}%
\AgdaSymbol{\{}\AgdaBound{as}\AgdaSpace{}%
\AgdaSymbol{:}\AgdaSpace{}%
\AgdaDatatype{List}\AgdaSpace{}%
\AgdaSymbol{(}\AgdaBound{A}\AgdaSpace{}%
\AgdaOperator{\AgdaFunction{．}}\AgdaSymbol{)\}}\AgdaSpace{}%
\AgdaSymbol{\{}\AgdaBound{Δ₀}\AgdaSpace{}%
\AgdaBound{Δ₁}\AgdaSpace{}%
\AgdaSymbol{:}\AgdaSpace{}%
\AgdaBound{A}\AgdaSpace{}%
\AgdaOperator{\AgdaFunction{．}}\AgdaSpace{}%
\AgdaSymbol{→}\AgdaSpace{}%
\AgdaDatatype{sply}\AgdaSpace{}%
\AgdaGeneralizable{ℓ}\AgdaSymbol{\}}\<%
\\
\>[0][@{}l@{\AgdaIndent{0}}]%
\>[2]\AgdaSymbol{→}\AgdaSpace{}%
\AgdaFunction{⨂}\AgdaSpace{}%
\AgdaBound{A}\AgdaSpace{}%
\AgdaBound{as}\AgdaSpace{}%
\AgdaBound{Δ₀}\AgdaSpace{}%
\AgdaOperator{\AgdaInductiveConstructor{⨾}}\AgdaSpace{}%
\AgdaFunction{⨂}\AgdaSpace{}%
\AgdaBound{A}\AgdaSpace{}%
\AgdaBound{as}\AgdaSpace{}%
\AgdaBound{Δ₁}\AgdaSpace{}%
\AgdaOperator{\AgdaDatatype{⊩}}\AgdaSpace{}%
\AgdaFunction{⨂}\AgdaSpace{}%
\AgdaBound{A}\AgdaSpace{}%
\AgdaBound{as}\AgdaSpace{}%
\AgdaSymbol{(λ}\AgdaSpace{}%
\AgdaBound{x}\AgdaSpace{}%
\AgdaSymbol{→}\AgdaSpace{}%
\AgdaBound{Δ₀}\AgdaSpace{}%
\AgdaBound{x}\AgdaSpace{}%
\AgdaOperator{\AgdaInductiveConstructor{⨾}}\AgdaSpace{}%
\AgdaBound{Δ₁}\AgdaSpace{}%
\AgdaBound{x}\AgdaSymbol{)}\<%
\\
\>[0]\AgdaFunction{⨂lax}\AgdaSpace{}%
\AgdaSymbol{\{}\AgdaArgument{as}\AgdaSpace{}%
\AgdaSymbol{=}\AgdaSpace{}%
\AgdaInductiveConstructor{[]}\AgdaSymbol{\}}\AgdaSpace{}%
\AgdaSymbol{=}\AgdaSpace{}%
\AgdaFunction{unitl}\AgdaSpace{}%
\AgdaSymbol{\AgdaUnderscore{}}\<%
\\
\>[0]\AgdaFunction{⨂lax}\AgdaSpace{}%
\AgdaSymbol{\{}\AgdaArgument{as}\AgdaSpace{}%
\AgdaSymbol{=}\AgdaSpace{}%
\AgdaBound{a}\AgdaSpace{}%
\AgdaOperator{\AgdaInductiveConstructor{∷}}\AgdaSpace{}%
\AgdaBound{as}\AgdaSymbol{\}}\AgdaSpace{}%
\AgdaSymbol{=}\AgdaSpace{}%
\AgdaInductiveConstructor{id}\AgdaSpace{}%
\AgdaSymbol{\AgdaUnderscore{}}\AgdaSpace{}%
\AgdaOperator{\AgdaInductiveConstructor{⨾ᶠ}}\AgdaSpace{}%
\AgdaFunction{⨂lax}\AgdaSpace{}%
\AgdaSymbol{\{}\AgdaArgument{as}\AgdaSpace{}%
\AgdaSymbol{=}\AgdaSpace{}%
\AgdaBound{as}\AgdaSymbol{\}}\AgdaSpace{}%
\AgdaOperator{\AgdaInductiveConstructor{∘}}\AgdaSpace{}%
\AgdaInductiveConstructor{assoc}\AgdaSpace{}%
\AgdaSymbol{\AgdaUnderscore{}}\AgdaSpace{}%
\AgdaSymbol{\AgdaUnderscore{}}\AgdaSpace{}%
\AgdaSymbol{\AgdaUnderscore{}}\AgdaSpace{}%
\AgdaOperator{\AgdaInductiveConstructor{∘}}\AgdaSpace{}%
\AgdaInductiveConstructor{id}\AgdaSpace{}%
\AgdaSymbol{\AgdaUnderscore{}}\AgdaSpace{}%
\AgdaOperator{\AgdaInductiveConstructor{⨾ᶠ}}\AgdaSpace{}%
\AgdaFunction{assoc'}\AgdaSpace{}%
\AgdaSymbol{\AgdaUnderscore{}}\AgdaSpace{}%
\AgdaSymbol{\AgdaUnderscore{}}\AgdaSpace{}%
\AgdaSymbol{\AgdaUnderscore{}}\AgdaSpace{}%
\AgdaOperator{\AgdaInductiveConstructor{∘}}\AgdaSpace{}%
\AgdaInductiveConstructor{id}\AgdaSpace{}%
\AgdaSymbol{\AgdaUnderscore{}}\AgdaSpace{}%
\AgdaOperator{\AgdaInductiveConstructor{⨾ᶠ}}\AgdaSpace{}%
\AgdaSymbol{(}\AgdaInductiveConstructor{swap}\AgdaSpace{}%
\AgdaSymbol{\AgdaUnderscore{}}\AgdaSpace{}%
\AgdaSymbol{\AgdaUnderscore{}}\AgdaSpace{}%
\AgdaOperator{\AgdaInductiveConstructor{⨾ᶠ}}\AgdaSpace{}%
\AgdaInductiveConstructor{id}\AgdaSpace{}%
\AgdaSymbol{\AgdaUnderscore{})}\AgdaSpace{}%
\AgdaOperator{\AgdaInductiveConstructor{∘}}\AgdaSpace{}%
\AgdaInductiveConstructor{id}\AgdaSpace{}%
\AgdaSymbol{\AgdaUnderscore{}}\AgdaSpace{}%
\AgdaOperator{\AgdaInductiveConstructor{⨾ᶠ}}\AgdaSpace{}%
\AgdaInductiveConstructor{assoc}\AgdaSpace{}%
\AgdaSymbol{\AgdaUnderscore{}}\AgdaSpace{}%
\AgdaSymbol{\AgdaUnderscore{}}\AgdaSpace{}%
\AgdaSymbol{\AgdaUnderscore{}}\AgdaSpace{}%
\AgdaOperator{\AgdaInductiveConstructor{∘}}\AgdaSpace{}%
\AgdaFunction{assoc'}\AgdaSpace{}%
\AgdaSymbol{\AgdaUnderscore{}}\AgdaSpace{}%
\AgdaSymbol{\AgdaUnderscore{}}\AgdaSpace{}%
\AgdaSymbol{\AgdaUnderscore{}}\<%
\\
\\[\AgdaEmptyExtraSkip]%
\\[\AgdaEmptyExtraSkip]%
\\[\AgdaEmptyExtraSkip]%
\>[0]\AgdaKeyword{data}\AgdaSpace{}%
\AgdaDatatype{ltyp}\AgdaSpace{}%
\AgdaBound{ℓ}\AgdaSpace{}%
\AgdaKeyword{where}\<%
\\
\>[0][@{}l@{\AgdaIndent{0}}]%
\>[2]\AgdaInductiveConstructor{Ground}\AgdaSpace{}%
\AgdaSymbol{:}\AgdaSpace{}%
\AgdaSymbol{(}\AgdaBound{A}\AgdaSpace{}%
\AgdaSymbol{:}\AgdaSpace{}%
\AgdaPrimitive{ityp}\AgdaSpace{}%
\AgdaBound{ℓ}\AgdaSymbol{)}\AgdaSpace{}%
\AgdaSymbol{→}\AgdaSpace{}%
\AgdaDatatype{ltyp}\AgdaSpace{}%
\AgdaBound{ℓ}\<%
\\
\>[2]\AgdaInductiveConstructor{⊥ᵒ}\AgdaSpace{}%
\AgdaSymbol{:}\AgdaSpace{}%
\AgdaDatatype{ltyp}\AgdaSpace{}%
\AgdaBound{ℓ}\<%
\\
\>[2]\AgdaInductiveConstructor{⊤ᵒ}\AgdaSpace{}%
\AgdaSymbol{:}\AgdaSpace{}%
\AgdaDatatype{ltyp}\AgdaSpace{}%
\AgdaBound{ℓ}\<%
\\
\>[2]\AgdaInductiveConstructor{Boolᵒ}\AgdaSpace{}%
\AgdaSymbol{:}\AgdaSpace{}%
\AgdaDatatype{ltyp}\AgdaSpace{}%
\AgdaBound{ℓ}\<%
\\
\>[2]\AgdaInductiveConstructor{Πᵒ}\AgdaSpace{}%
\AgdaSymbol{:}\AgdaSpace{}%
\AgdaSymbol{(}\AgdaBound{A}\AgdaSpace{}%
\AgdaSymbol{:}\AgdaSpace{}%
\AgdaDatatype{ltyp}\AgdaSpace{}%
\AgdaBound{ℓ}\AgdaSymbol{)}\AgdaSpace{}%
\AgdaSymbol{→}\AgdaSpace{}%
\AgdaDatatype{ℕ}\AgdaSpace{}%
\AgdaSymbol{→}\AgdaSpace{}%
\AgdaSymbol{(}\AgdaBound{A}\AgdaSpace{}%
\AgdaFunction{⇒}\AgdaSpace{}%
\AgdaFunction{ltyp}\AgdaSymbol{)}\AgdaSpace{}%
\AgdaSymbol{→}\AgdaSpace{}%
\AgdaDatatype{ltyp}\AgdaSpace{}%
\AgdaBound{ℓ}\<%
\\
\>[2]\AgdaInductiveConstructor{Π!}\AgdaSpace{}%
\AgdaSymbol{:}\AgdaSpace{}%
\AgdaSymbol{(}\AgdaBound{A}\AgdaSpace{}%
\AgdaSymbol{:}\AgdaSpace{}%
\AgdaDatatype{ltyp}\AgdaSpace{}%
\AgdaBound{ℓ}\AgdaSymbol{)}\AgdaSpace{}%
\AgdaSymbol{→}\AgdaSpace{}%
\AgdaSymbol{(}\AgdaBound{A}\AgdaSpace{}%
\AgdaFunction{⇒}\AgdaSpace{}%
\AgdaFunction{ltyp}\AgdaSymbol{)}\AgdaSpace{}%
\AgdaSymbol{→}\AgdaSpace{}%
\AgdaDatatype{ltyp}\AgdaSpace{}%
\AgdaBound{ℓ}\<%
\\
\>[2]\AgdaInductiveConstructor{Σᵒ}\AgdaSpace{}%
\AgdaSymbol{:}\AgdaSpace{}%
\AgdaSymbol{(}\AgdaBound{A}\AgdaSpace{}%
\AgdaSymbol{:}\AgdaSpace{}%
\AgdaDatatype{ltyp}\AgdaSpace{}%
\AgdaBound{ℓ}\AgdaSymbol{)}\AgdaSpace{}%
\AgdaSymbol{→}\AgdaSpace{}%
\AgdaDatatype{ℕ}\AgdaSpace{}%
\AgdaSymbol{→}\AgdaSpace{}%
\AgdaSymbol{(}\AgdaBound{A}\AgdaSpace{}%
\AgdaFunction{⇒}\AgdaSpace{}%
\AgdaFunction{ltyp}\AgdaSymbol{)}\AgdaSpace{}%
\AgdaSymbol{→}\AgdaSpace{}%
\AgdaDatatype{ltyp}\AgdaSpace{}%
\AgdaBound{ℓ}\<%
\\
\>[2]\AgdaInductiveConstructor{Wᵒ}\AgdaSpace{}%
\AgdaSymbol{:}\AgdaSpace{}%
\AgdaSymbol{(}\AgdaBound{A}\AgdaSpace{}%
\AgdaSymbol{:}\AgdaSpace{}%
\AgdaDatatype{ltyp}\AgdaSpace{}%
\AgdaBound{ℓ}\AgdaSymbol{)}\AgdaSpace{}%
\AgdaSymbol{→}\AgdaSpace{}%
\AgdaDatatype{ℕ}\AgdaSpace{}%
\AgdaSymbol{→}\AgdaSpace{}%
\AgdaSymbol{(}\AgdaBound{A}\AgdaSpace{}%
\AgdaFunction{⇒}\AgdaSpace{}%
\AgdaFunction{finltyp}\AgdaSymbol{)}\AgdaSpace{}%
\AgdaSymbol{→}\AgdaSpace{}%
\AgdaDatatype{ltyp}\AgdaSpace{}%
\AgdaBound{ℓ}\<%
\\
\\[\AgdaEmptyExtraSkip]%
\\[\AgdaEmptyExtraSkip]%
\\[\AgdaEmptyExtraSkip]%
\>[0]\AgdaOperator{\AgdaFunction{⟦}}\AgdaSpace{}%
\AgdaInductiveConstructor{Ground}\AgdaSpace{}%
\AgdaBound{A}\AgdaSpace{}%
\AgdaOperator{\AgdaFunction{⟧}}\AgdaSpace{}%
\AgdaSymbol{=}\AgdaSpace{}%
\AgdaBound{A}\AgdaSpace{}%
\AgdaOperator{\AgdaInductiveConstructor{,}}\AgdaSpace{}%
\AgdaInductiveConstructor{ι}\<%
\\
\>[0]\AgdaOperator{\AgdaFunction{⟦}}\AgdaSpace{}%
\AgdaInductiveConstructor{⊥ᵒ}\AgdaSpace{}%
\AgdaOperator{\AgdaFunction{⟧}}\AgdaSpace{}%
\AgdaSymbol{=}\AgdaSpace{}%
\AgdaDatatype{⊥}\AgdaSpace{}%
\AgdaOperator{\AgdaInductiveConstructor{,}}\AgdaSpace{}%
\AgdaSymbol{λ}\AgdaSpace{}%
\AgdaSymbol{()}\<%
\\
\>[0]\AgdaOperator{\AgdaFunction{⟦}}\AgdaSpace{}%
\AgdaInductiveConstructor{⊤ᵒ}\AgdaSpace{}%
\AgdaOperator{\AgdaFunction{⟧}}\AgdaSpace{}%
\AgdaSymbol{=}\AgdaSpace{}%
\AgdaFunction{⊤}\AgdaSpace{}%
\AgdaOperator{\AgdaInductiveConstructor{,}}\AgdaSpace{}%
\AgdaSymbol{(λ}\AgdaSpace{}%
\AgdaBound{\AgdaUnderscore{}}\AgdaSpace{}%
\AgdaSymbol{→}\AgdaSpace{}%
\AgdaInductiveConstructor{◇}\AgdaSymbol{)}\<%
\\
\>[0]\AgdaOperator{\AgdaFunction{⟦}}\AgdaSpace{}%
\AgdaInductiveConstructor{Boolᵒ}\AgdaSpace{}%
\AgdaOperator{\AgdaFunction{⟧}}\AgdaSpace{}%
\AgdaSymbol{=}\AgdaSpace{}%
\AgdaDatatype{Bool}\AgdaSpace{}%
\AgdaOperator{\AgdaInductiveConstructor{,}}\AgdaSpace{}%
\AgdaSymbol{(λ}\AgdaSpace{}%
\AgdaBound{\AgdaUnderscore{}}\AgdaSpace{}%
\AgdaSymbol{→}\AgdaSpace{}%
\AgdaInductiveConstructor{◇}\AgdaSymbol{)}\<%
\\
\>[0]\AgdaOperator{\AgdaFunction{⟦}}\AgdaSpace{}%
\AgdaInductiveConstructor{Πᵒ}\AgdaSpace{}%
\AgdaBound{A}\AgdaSpace{}%
\AgdaBound{m}\AgdaSpace{}%
\AgdaBound{B}\AgdaSpace{}%
\AgdaOperator{\AgdaFunction{⟧}}\AgdaSpace{}%
\AgdaSymbol{=}\AgdaSpace{}%
\AgdaSymbol{((}\AgdaBound{x}\AgdaSpace{}%
\AgdaSymbol{:}\AgdaSpace{}%
\AgdaBound{A}\AgdaSpace{}%
\AgdaOperator{\AgdaFunction{．}}\AgdaSymbol{)}\AgdaSpace{}%
\AgdaSymbol{→}\AgdaSpace{}%
\AgdaBound{B}\AgdaSpace{}%
\AgdaBound{x}\AgdaSpace{}%
\AgdaOperator{\AgdaFunction{．}}\AgdaSymbol{)}\AgdaSpace{}%
\AgdaOperator{\AgdaInductiveConstructor{,}}\AgdaSpace{}%
\AgdaSymbol{λ}\AgdaSpace{}%
\AgdaBound{f}\AgdaSpace{}%
\AgdaSymbol{→}\AgdaSpace{}%
\AgdaInductiveConstructor{Λ⟨}\AgdaSpace{}%
\AgdaBound{x}\AgdaSpace{}%
\AgdaInductiveConstructor{∶}\AgdaSpace{}%
\AgdaBound{A}\AgdaSpace{}%
\AgdaOperator{\AgdaFunction{．}}\AgdaSpace{}%
\AgdaInductiveConstructor{⟩}\AgdaSpace{}%
\AgdaOperator{\AgdaInductiveConstructor{[}}\AgdaSpace{}%
\AgdaSymbol{(}\AgdaBound{x}\AgdaSpace{}%
\AgdaFunction{∶}\AgdaSpace{}%
\AgdaBound{A}\AgdaSymbol{)}\AgdaSpace{}%
\AgdaOperator{\AgdaFunction{\textasciicircum{}}}\AgdaSpace{}%
\AgdaBound{m}\AgdaSpace{}%
\AgdaOperator{\AgdaInductiveConstructor{,}}\AgdaSpace{}%
\AgdaBound{f}\AgdaSpace{}%
\AgdaBound{x}\AgdaSpace{}%
\AgdaFunction{∶}\AgdaSpace{}%
\AgdaBound{B}\AgdaSpace{}%
\AgdaBound{x}\AgdaSpace{}%
\AgdaOperator{\AgdaInductiveConstructor{]}}\<%
\\
\>[0]\AgdaOperator{\AgdaFunction{⟦}}\AgdaSpace{}%
\AgdaInductiveConstructor{Π!}\AgdaSpace{}%
\AgdaBound{A}\AgdaSpace{}%
\AgdaBound{B}\AgdaSpace{}%
\AgdaOperator{\AgdaFunction{⟧}}\AgdaSpace{}%
\AgdaSymbol{=}\AgdaSpace{}%
\AgdaSymbol{((}\AgdaBound{x}\AgdaSpace{}%
\AgdaSymbol{:}\AgdaSpace{}%
\AgdaBound{A}\AgdaSpace{}%
\AgdaOperator{\AgdaFunction{．}}\AgdaSymbol{)}\AgdaSpace{}%
\AgdaSymbol{→}\AgdaSpace{}%
\AgdaBound{B}\AgdaSpace{}%
\AgdaBound{x}\AgdaSpace{}%
\AgdaOperator{\AgdaFunction{．}}\AgdaSymbol{)}\AgdaSpace{}%
\AgdaOperator{\AgdaInductiveConstructor{,}}\AgdaSpace{}%
\AgdaSymbol{λ}\AgdaSpace{}%
\AgdaBound{f}\AgdaSpace{}%
\AgdaSymbol{→}\AgdaSpace{}%
\AgdaInductiveConstructor{Λ⟨}\AgdaSpace{}%
\AgdaBound{x}\AgdaSpace{}%
\AgdaInductiveConstructor{∶}\AgdaSpace{}%
\AgdaBound{A}\AgdaSpace{}%
\AgdaOperator{\AgdaFunction{．}}\AgdaSpace{}%
\AgdaInductiveConstructor{⟩}\AgdaSpace{}%
\AgdaOperator{\AgdaInductiveConstructor{[}}\AgdaSpace{}%
\AgdaInductiveConstructor{!}\AgdaSpace{}%
\AgdaSymbol{(}\AgdaBound{x}\AgdaSpace{}%
\AgdaFunction{∶}\AgdaSpace{}%
\AgdaBound{A}\AgdaSymbol{)}\AgdaSpace{}%
\AgdaOperator{\AgdaInductiveConstructor{,}}\AgdaSpace{}%
\AgdaBound{f}\AgdaSpace{}%
\AgdaBound{x}\AgdaSpace{}%
\AgdaFunction{∶}\AgdaSpace{}%
\AgdaBound{B}\AgdaSpace{}%
\AgdaBound{x}\AgdaSpace{}%
\AgdaOperator{\AgdaInductiveConstructor{]}}\<%
\\
\>[0]\AgdaOperator{\AgdaFunction{⟦}}\AgdaSpace{}%
\AgdaInductiveConstructor{Σᵒ}\AgdaSpace{}%
\AgdaBound{A}\AgdaSpace{}%
\AgdaBound{m}\AgdaSpace{}%
\AgdaBound{B}\AgdaSpace{}%
\AgdaOperator{\AgdaFunction{⟧}}\AgdaSpace{}%
\AgdaSymbol{=}\AgdaSpace{}%
\AgdaRecord{Σ}\AgdaSpace{}%
\AgdaSymbol{(}\AgdaBound{A}\AgdaSpace{}%
\AgdaOperator{\AgdaFunction{．}}\AgdaSymbol{)}\AgdaSpace{}%
\AgdaSymbol{(λ}\AgdaSpace{}%
\AgdaBound{x}\AgdaSpace{}%
\AgdaSymbol{→}\AgdaSpace{}%
\AgdaBound{B}\AgdaSpace{}%
\AgdaBound{x}\AgdaSpace{}%
\AgdaOperator{\AgdaFunction{．}}\AgdaSymbol{)}\AgdaSpace{}%
\AgdaOperator{\AgdaInductiveConstructor{,}}\AgdaSpace{}%
\AgdaSymbol{λ}\AgdaSpace{}%
\AgdaSymbol{(}\AgdaBound{x}\AgdaSpace{}%
\AgdaOperator{\AgdaInductiveConstructor{,}}\AgdaSpace{}%
\AgdaBound{y}\AgdaSymbol{)}\AgdaSpace{}%
\AgdaSymbol{→}\AgdaSpace{}%
\AgdaSymbol{(}\AgdaBound{x}\AgdaSpace{}%
\AgdaFunction{∶}\AgdaSpace{}%
\AgdaBound{A}\AgdaSymbol{)}\AgdaSpace{}%
\AgdaOperator{\AgdaFunction{\textasciicircum{}}}\AgdaSpace{}%
\AgdaBound{m}\AgdaSpace{}%
\AgdaOperator{\AgdaInductiveConstructor{⨾}}\AgdaSpace{}%
\AgdaBound{y}\AgdaSpace{}%
\AgdaFunction{∶}\AgdaSpace{}%
\AgdaBound{B}\AgdaSpace{}%
\AgdaBound{x}\<%
\\
\>[0]\AgdaOperator{\AgdaFunction{⟦}}\AgdaSpace{}%
\AgdaInductiveConstructor{Wᵒ}\AgdaSpace{}%
\AgdaBound{A}\AgdaSpace{}%
\AgdaBound{m}\AgdaSpace{}%
\AgdaSymbol{(}\AgdaBound{B}\AgdaSpace{}%
\AgdaOperator{\AgdaInductiveConstructor{,}}\AgdaSpace{}%
\AgdaBound{Bfin}\AgdaSymbol{)}\AgdaSpace{}%
\AgdaOperator{\AgdaFunction{⟧}}\AgdaSpace{}%
\AgdaSymbol{=}\AgdaSpace{}%
\AgdaDatatype{W}\AgdaSpace{}%
\AgdaSymbol{(}\AgdaBound{A}\AgdaSpace{}%
\AgdaOperator{\AgdaFunction{．}}\AgdaSymbol{)}\AgdaSpace{}%
\AgdaSymbol{(λ}\AgdaSpace{}%
\AgdaBound{x}\AgdaSpace{}%
\AgdaSymbol{→}\AgdaSpace{}%
\AgdaBound{B}\AgdaSpace{}%
\AgdaBound{x}\AgdaSpace{}%
\AgdaOperator{\AgdaFunction{．}}\AgdaSymbol{)}\AgdaSpace{}%
\AgdaOperator{\AgdaInductiveConstructor{,}}\AgdaSpace{}%
\AgdaFunction{recW}\AgdaSpace{}%
\AgdaSymbol{(λ}\AgdaSpace{}%
\AgdaBound{x}\AgdaSpace{}%
\AgdaBound{Δ}\AgdaSpace{}%
\AgdaSymbol{→}\AgdaSpace{}%
\AgdaSymbol{(}\AgdaBound{x}\AgdaSpace{}%
\AgdaFunction{∶}\AgdaSpace{}%
\AgdaBound{A}\AgdaSymbol{)}\AgdaSpace{}%
\AgdaOperator{\AgdaFunction{\textasciicircum{}}}\AgdaSpace{}%
\AgdaBound{m}\AgdaSpace{}%
\AgdaOperator{\AgdaInductiveConstructor{⨾}}\AgdaSpace{}%
\AgdaFunction{⨂}\AgdaSpace{}%
\AgdaSymbol{(}\AgdaBound{B}\AgdaSpace{}%
\AgdaBound{x}\AgdaSymbol{)}\AgdaSpace{}%
\AgdaSymbol{(}\AgdaBound{Bfin}\AgdaSpace{}%
\AgdaBound{x}\AgdaSymbol{)}\AgdaSpace{}%
\AgdaBound{Δ}\AgdaSpace{}%
\AgdaSymbol{)}\<%
\\
\\[\AgdaEmptyExtraSkip]%
\\[\AgdaEmptyExtraSkip]%
\>[0]\AgdaKeyword{module}\AgdaSpace{}%
\AgdaModule{\AgdaUnderscore{}}\AgdaSpace{}%
\AgdaSymbol{\{}\AgdaBound{Δ}\AgdaSpace{}%
\AgdaSymbol{:}\AgdaSpace{}%
\AgdaDatatype{sply}\AgdaSpace{}%
\AgdaGeneralizable{ℓ}\AgdaSymbol{\}}\AgdaSpace{}%
\AgdaSymbol{\{}\AgdaBound{A}\AgdaSpace{}%
\AgdaSymbol{:}\AgdaSpace{}%
\AgdaDatatype{ltyp}\AgdaSpace{}%
\AgdaGeneralizable{ℓ}\AgdaSymbol{\}}\AgdaSpace{}%
\AgdaSymbol{\{}\AgdaBound{x}\AgdaSpace{}%
\AgdaSymbol{:}\AgdaSpace{}%
\AgdaBound{A}\AgdaSpace{}%
\AgdaOperator{\AgdaFunction{．}}\AgdaSymbol{\}}\AgdaSpace{}%
\AgdaKeyword{where}\<%
\end{code}
\newcommand{\AGDAeval}{%
\begin{code}%
\>[0][@{}l@{\AgdaIndent{1}}]%
\>[2]\AgdaFunction{eval}\AgdaSpace{}%
\AgdaSymbol{:}\AgdaSpace{}%
\AgdaBound{Δ}\AgdaSpace{}%
\AgdaOperator{\AgdaDatatype{⊩}}\AgdaSpace{}%
\AgdaBound{x}\AgdaSpace{}%
\AgdaFunction{∶}\AgdaSpace{}%
\AgdaBound{A}\AgdaSpace{}%
\AgdaSymbol{→}\AgdaSpace{}%
\AgdaBound{A}\AgdaSpace{}%
\AgdaOperator{\AgdaFunction{．}}\<%
\end{code}}
\begin{code}[hide]%
\>[2]\AgdaFunction{eval}\AgdaSpace{}%
\AgdaSymbol{\AgdaUnderscore{}}\AgdaSpace{}%
\AgdaSymbol{=}\AgdaSpace{}%
\AgdaBound{x}\<%
\\
\\[\AgdaEmptyExtraSkip]%
\>[0]\<%
\end{code}
\newcommand{\AGDAVar}{%
\begin{code}%
\>[0]\AgdaFunction{Varᵒ}\AgdaSpace{}%
\AgdaSymbol{:}\AgdaSpace{}%
\AgdaSymbol{(}\AgdaBound{A}\AgdaSpace{}%
\AgdaSymbol{:}\AgdaSpace{}%
\AgdaDatatype{ltyp}\AgdaSpace{}%
\AgdaGeneralizable{ℓ}\AgdaSymbol{)}\AgdaSpace{}%
\AgdaSymbol{→}\AgdaSpace{}%
\AgdaSymbol{(}\AgdaBound{x}\AgdaSpace{}%
\AgdaSymbol{:}\AgdaSpace{}%
\AgdaBound{A}\AgdaSpace{}%
\AgdaOperator{\AgdaFunction{．}}\AgdaSymbol{)}\AgdaSpace{}%
\AgdaSymbol{→}\AgdaSpace{}%
\AgdaBound{x}\AgdaSpace{}%
\AgdaFunction{∶}\AgdaSpace{}%
\AgdaBound{A}\AgdaSpace{}%
\AgdaOperator{\AgdaDatatype{⊩}}\AgdaSpace{}%
\AgdaBound{x}\AgdaSpace{}%
\AgdaFunction{∶}\AgdaSpace{}%
\AgdaBound{A}\<%
\\
\>[0]\AgdaFunction{Varᵒ}\AgdaSpace{}%
\AgdaBound{A}\AgdaSpace{}%
\AgdaBound{x}\AgdaSpace{}%
\AgdaSymbol{=}\AgdaSpace{}%
\AgdaInductiveConstructor{id}\AgdaSpace{}%
\AgdaSymbol{(}\AgdaBound{x}\AgdaSpace{}%
\AgdaFunction{∶}\AgdaSpace{}%
\AgdaBound{A}\AgdaSymbol{)}\<%
\end{code}}
\begin{code}[hide]%
\>[0]\<%
\\
\\[\AgdaEmptyExtraSkip]%
\\[\AgdaEmptyExtraSkip]%
\\[\AgdaEmptyExtraSkip]%
\>[0]\AgdaKeyword{infixr}\AgdaSpace{}%
\AgdaNumber{-9}\AgdaSpace{}%
\AgdaInductiveConstructor{Πᵒ}\<%
\\
\>[0]\AgdaKeyword{syntax}\AgdaSpace{}%
\AgdaInductiveConstructor{Πᵒ}\AgdaSpace{}%
\AgdaBound{A}\AgdaSpace{}%
\AgdaBound{m}\AgdaSpace{}%
\AgdaSymbol{(λ}\AgdaSpace{}%
\AgdaBound{x}\AgdaSpace{}%
\AgdaSymbol{→}\AgdaSpace{}%
\AgdaBound{B}\AgdaSymbol{)}\AgdaSpace{}%
\AgdaSymbol{=}\AgdaSpace{}%
\AgdaInductiveConstructor{⟨}\AgdaSpace{}%
\AgdaBound{x}\AgdaSpace{}%
\AgdaInductiveConstructor{∈}\AgdaSpace{}%
\AgdaBound{A}\AgdaSpace{}%
\AgdaInductiveConstructor{⟩\textasciicircum{}}\AgdaSpace{}%
\AgdaBound{m}\AgdaSpace{}%
\AgdaInductiveConstructor{⊸}\AgdaSpace{}%
\AgdaBound{B}\<%
\\
\\[\AgdaEmptyExtraSkip]%
\>[0]\AgdaFunction{Πᵒnd}\AgdaSpace{}%
\AgdaSymbol{:}\AgdaSpace{}%
\AgdaDatatype{ltyp}\AgdaSpace{}%
\AgdaGeneralizable{ℓ}\AgdaSpace{}%
\AgdaSymbol{→}\AgdaSpace{}%
\AgdaDatatype{ℕ}\AgdaSpace{}%
\AgdaSymbol{→}\AgdaSpace{}%
\AgdaDatatype{ltyp}\AgdaSpace{}%
\AgdaGeneralizable{ℓ}\AgdaSpace{}%
\AgdaSymbol{→}\AgdaSpace{}%
\AgdaDatatype{ltyp}\AgdaSpace{}%
\AgdaGeneralizable{ℓ}\<%
\\
\>[0]\AgdaFunction{Πᵒnd}\AgdaSpace{}%
\AgdaBound{A}\AgdaSpace{}%
\AgdaBound{m}\AgdaSpace{}%
\AgdaBound{B}\AgdaSpace{}%
\AgdaSymbol{=}\AgdaSpace{}%
\AgdaInductiveConstructor{Πᵒ}\AgdaSpace{}%
\AgdaBound{A}\AgdaSpace{}%
\AgdaBound{m}\AgdaSpace{}%
\AgdaSymbol{(λ}\AgdaSpace{}%
\AgdaBound{x}\AgdaSpace{}%
\AgdaSymbol{→}\AgdaSpace{}%
\AgdaBound{B}\AgdaSymbol{)}\<%
\\
\>[0]\AgdaKeyword{infixr}\AgdaSpace{}%
\AgdaNumber{-9}\AgdaSpace{}%
\AgdaFunction{Πᵒnd}\<%
\\
\>[0]\AgdaKeyword{syntax}\AgdaSpace{}%
\AgdaFunction{Πᵒnd}\AgdaSpace{}%
\AgdaBound{A}\AgdaSpace{}%
\AgdaBound{m}\AgdaSpace{}%
\AgdaBound{B}\AgdaSpace{}%
\AgdaSymbol{=}\AgdaSpace{}%
\AgdaFunction{⟨}\AgdaSpace{}%
\AgdaBound{A}\AgdaSpace{}%
\AgdaFunction{⟩\textasciicircum{}}\AgdaSpace{}%
\AgdaBound{m}\AgdaSpace{}%
\AgdaFunction{⊸}\AgdaSpace{}%
\AgdaBound{B}\<%
\\
\\[\AgdaEmptyExtraSkip]%
\>[0]\AgdaFunction{Πᵒnd1}\AgdaSpace{}%
\AgdaSymbol{:}\AgdaSpace{}%
\AgdaDatatype{ltyp}\AgdaSpace{}%
\AgdaGeneralizable{ℓ}\AgdaSpace{}%
\AgdaSymbol{→}\AgdaSpace{}%
\AgdaDatatype{ltyp}\AgdaSpace{}%
\AgdaGeneralizable{ℓ}\AgdaSpace{}%
\AgdaSymbol{→}\AgdaSpace{}%
\AgdaDatatype{ltyp}\AgdaSpace{}%
\AgdaGeneralizable{ℓ}\<%
\\
\>[0]\AgdaFunction{Πᵒnd1}\AgdaSpace{}%
\AgdaBound{A}\AgdaSpace{}%
\AgdaBound{B}\AgdaSpace{}%
\AgdaSymbol{=}\AgdaSpace{}%
\AgdaInductiveConstructor{Πᵒ}\AgdaSpace{}%
\AgdaBound{A}\AgdaSpace{}%
\AgdaNumber{1}\AgdaSpace{}%
\AgdaSymbol{(λ}\AgdaSpace{}%
\AgdaBound{x}\AgdaSpace{}%
\AgdaSymbol{→}\AgdaSpace{}%
\AgdaBound{B}\AgdaSymbol{)}\<%
\\
\>[0]\AgdaKeyword{infixr}\AgdaSpace{}%
\AgdaNumber{-9}\AgdaSpace{}%
\AgdaFunction{Πᵒnd1}\<%
\\
\>[0]\AgdaKeyword{syntax}\AgdaSpace{}%
\AgdaFunction{Πᵒnd1}\AgdaSpace{}%
\AgdaBound{A}\AgdaSpace{}%
\AgdaBound{B}\AgdaSpace{}%
\AgdaSymbol{=}\AgdaSpace{}%
\AgdaBound{A}\AgdaSpace{}%
\AgdaFunction{⊸}\AgdaSpace{}%
\AgdaBound{B}\<%
\\
\\[\AgdaEmptyExtraSkip]%
\\[\AgdaEmptyExtraSkip]%
\>[0]\AgdaKeyword{infixr}\AgdaSpace{}%
\AgdaNumber{-9}\AgdaSpace{}%
\AgdaInductiveConstructor{Π!}\<%
\\
\>[0]\AgdaKeyword{syntax}\AgdaSpace{}%
\AgdaInductiveConstructor{Π!}\AgdaSpace{}%
\AgdaBound{A}\AgdaSpace{}%
\AgdaSymbol{(λ}\AgdaSpace{}%
\AgdaBound{x}\AgdaSpace{}%
\AgdaSymbol{→}\AgdaSpace{}%
\AgdaBound{B}\AgdaSymbol{)}\AgdaSpace{}%
\AgdaSymbol{=}\AgdaSpace{}%
\AgdaInductiveConstructor{!⟨}\AgdaSpace{}%
\AgdaBound{x}\AgdaSpace{}%
\AgdaInductiveConstructor{∈}\AgdaSpace{}%
\AgdaBound{A}\AgdaSpace{}%
\AgdaInductiveConstructor{⟩}\AgdaSpace{}%
\AgdaInductiveConstructor{⊸}\AgdaSpace{}%
\AgdaBound{B}\<%
\\
\\[\AgdaEmptyExtraSkip]%
\>[0]\AgdaFunction{Π!nd}\AgdaSpace{}%
\AgdaSymbol{:}\AgdaSpace{}%
\AgdaDatatype{ltyp}\AgdaSpace{}%
\AgdaGeneralizable{ℓ}\AgdaSpace{}%
\AgdaSymbol{→}\AgdaSpace{}%
\AgdaDatatype{ltyp}\AgdaSpace{}%
\AgdaGeneralizable{ℓ}\AgdaSpace{}%
\AgdaSymbol{→}\AgdaSpace{}%
\AgdaDatatype{ltyp}\AgdaSpace{}%
\AgdaGeneralizable{ℓ}\<%
\\
\>[0]\AgdaFunction{Π!nd}\AgdaSpace{}%
\AgdaBound{A}\AgdaSpace{}%
\AgdaBound{B}\AgdaSpace{}%
\AgdaSymbol{=}\AgdaSpace{}%
\AgdaInductiveConstructor{Π!}\AgdaSpace{}%
\AgdaBound{A}\AgdaSpace{}%
\AgdaSymbol{(λ}\AgdaSpace{}%
\AgdaBound{\AgdaUnderscore{}}\AgdaSpace{}%
\AgdaSymbol{→}\AgdaSpace{}%
\AgdaBound{B}\AgdaSymbol{)}\<%
\\
\>[0]\AgdaKeyword{infixr}\AgdaSpace{}%
\AgdaNumber{-9}\AgdaSpace{}%
\AgdaFunction{Π!nd}\<%
\\
\>[0]\AgdaKeyword{syntax}\AgdaSpace{}%
\AgdaFunction{Π!nd}\AgdaSpace{}%
\AgdaBound{A}\AgdaSpace{}%
\AgdaBound{B}\AgdaSpace{}%
\AgdaSymbol{=}\AgdaSpace{}%
\AgdaFunction{!⟨}\AgdaSpace{}%
\AgdaBound{A}\AgdaSpace{}%
\AgdaFunction{⟩}\AgdaSpace{}%
\AgdaFunction{⊸}\AgdaSpace{}%
\AgdaBound{B}\<%
\\
\\[\AgdaEmptyExtraSkip]%
\\[\AgdaEmptyExtraSkip]%
\>[0]\AgdaKeyword{infixr}\AgdaSpace{}%
\AgdaNumber{-9}\AgdaSpace{}%
\AgdaInductiveConstructor{Σᵒ}\<%
\\
\>[0]\AgdaKeyword{syntax}\AgdaSpace{}%
\AgdaInductiveConstructor{Σᵒ}\AgdaSpace{}%
\AgdaBound{A}\AgdaSpace{}%
\AgdaBound{m}\AgdaSpace{}%
\AgdaSymbol{(λ}\AgdaSpace{}%
\AgdaBound{x}\AgdaSpace{}%
\AgdaSymbol{→}\AgdaSpace{}%
\AgdaBound{B}\AgdaSymbol{)}\AgdaSpace{}%
\AgdaSymbol{=}\AgdaSpace{}%
\AgdaInductiveConstructor{⟨}\AgdaSpace{}%
\AgdaBound{x}\AgdaSpace{}%
\AgdaInductiveConstructor{∈}\AgdaSpace{}%
\AgdaBound{A}\AgdaSpace{}%
\AgdaInductiveConstructor{⟩\textasciicircum{}}\AgdaSpace{}%
\AgdaBound{m}\AgdaSpace{}%
\AgdaInductiveConstructor{⊗}\AgdaSpace{}%
\AgdaBound{B}\<%
\\
\\[\AgdaEmptyExtraSkip]%
\\[\AgdaEmptyExtraSkip]%
\>[0]\AgdaFunction{Σᵒnd}\AgdaSpace{}%
\AgdaSymbol{:}\AgdaSpace{}%
\AgdaDatatype{ltyp}\AgdaSpace{}%
\AgdaGeneralizable{ℓ}\AgdaSpace{}%
\AgdaSymbol{→}\AgdaSpace{}%
\AgdaDatatype{ℕ}\AgdaSpace{}%
\AgdaSymbol{→}\AgdaSpace{}%
\AgdaDatatype{ltyp}\AgdaSpace{}%
\AgdaGeneralizable{ℓ}\AgdaSpace{}%
\AgdaSymbol{→}\AgdaSpace{}%
\AgdaDatatype{ltyp}\AgdaSpace{}%
\AgdaGeneralizable{ℓ}\<%
\\
\>[0]\AgdaFunction{Σᵒnd}\AgdaSpace{}%
\AgdaBound{A}\AgdaSpace{}%
\AgdaBound{m}\AgdaSpace{}%
\AgdaBound{B}\AgdaSpace{}%
\AgdaSymbol{=}\AgdaSpace{}%
\AgdaInductiveConstructor{Σᵒ}\AgdaSpace{}%
\AgdaBound{A}\AgdaSpace{}%
\AgdaBound{m}\AgdaSpace{}%
\AgdaSymbol{(λ}\AgdaSpace{}%
\AgdaBound{x}\AgdaSpace{}%
\AgdaSymbol{→}\AgdaSpace{}%
\AgdaBound{B}\AgdaSymbol{)}\<%
\\
\>[0]\AgdaKeyword{infixr}\AgdaSpace{}%
\AgdaNumber{-9}\AgdaSpace{}%
\AgdaFunction{Σᵒnd}\<%
\\
\>[0]\AgdaKeyword{syntax}\AgdaSpace{}%
\AgdaFunction{Σᵒnd}\AgdaSpace{}%
\AgdaBound{A}\AgdaSpace{}%
\AgdaBound{m}\AgdaSpace{}%
\AgdaBound{B}\AgdaSpace{}%
\AgdaSymbol{=}\AgdaSpace{}%
\AgdaFunction{⟨}\AgdaSpace{}%
\AgdaBound{A}\AgdaSpace{}%
\AgdaFunction{⟩\textasciicircum{}}\AgdaSpace{}%
\AgdaBound{m}\AgdaSpace{}%
\AgdaFunction{⊗}\AgdaSpace{}%
\AgdaBound{B}\<%
\\
\\[\AgdaEmptyExtraSkip]%
\>[0]\AgdaFunction{Σᵒnd1}\AgdaSpace{}%
\AgdaSymbol{:}\AgdaSpace{}%
\AgdaDatatype{ltyp}\AgdaSpace{}%
\AgdaGeneralizable{ℓ}\AgdaSpace{}%
\AgdaSymbol{→}\AgdaSpace{}%
\AgdaDatatype{ltyp}\AgdaSpace{}%
\AgdaGeneralizable{ℓ}\AgdaSpace{}%
\AgdaSymbol{→}\AgdaSpace{}%
\AgdaDatatype{ltyp}\AgdaSpace{}%
\AgdaGeneralizable{ℓ}\<%
\\
\>[0]\AgdaFunction{Σᵒnd1}\AgdaSpace{}%
\AgdaBound{A}\AgdaSpace{}%
\AgdaBound{B}\AgdaSpace{}%
\AgdaSymbol{=}\AgdaSpace{}%
\AgdaInductiveConstructor{Σᵒ}\AgdaSpace{}%
\AgdaBound{A}\AgdaSpace{}%
\AgdaNumber{1}\AgdaSpace{}%
\AgdaSymbol{(λ}\AgdaSpace{}%
\AgdaBound{x}\AgdaSpace{}%
\AgdaSymbol{→}\AgdaSpace{}%
\AgdaBound{B}\AgdaSymbol{)}\<%
\\
\>[0]\AgdaKeyword{infixr}\AgdaSpace{}%
\AgdaNumber{-9}\AgdaSpace{}%
\AgdaFunction{Σᵒnd1}\<%
\\
\>[0]\AgdaKeyword{syntax}\AgdaSpace{}%
\AgdaFunction{Σᵒnd1}\AgdaSpace{}%
\AgdaBound{A}\AgdaSpace{}%
\AgdaBound{B}\AgdaSpace{}%
\AgdaSymbol{=}\AgdaSpace{}%
\AgdaBound{A}\AgdaSpace{}%
\AgdaFunction{⊗}\AgdaSpace{}%
\AgdaBound{B}\<%
\\
\>[0]\<%
\end{code}

\begin{code}[hide]%
\>[0]\AgdaSymbol{\{-\#}\AgdaSpace{}%
\AgdaKeyword{OPTIONS}\AgdaSpace{}%
\AgdaPragma{--allow-unsolved-metas}\AgdaSpace{}%
\AgdaSymbol{\#-\}}\<%
\\
\>[0]\AgdaKeyword{module}\AgdaSpace{}%
\AgdaModule{Bang}\AgdaSpace{}%
\AgdaKeyword{where}\<%
\\
\\[\AgdaEmptyExtraSkip]%
\>[0]\AgdaKeyword{open}\AgdaSpace{}%
\AgdaKeyword{import}\AgdaSpace{}%
\AgdaModule{Base}\<%
\\
\\[\AgdaEmptyExtraSkip]%
\>[0]\AgdaKeyword{private}\<%
\\
\>[0][@{}l@{\AgdaIndent{0}}]%
\>[2]\AgdaKeyword{variable}\<%
\\
\>[2][@{}l@{\AgdaIndent{0}}]%
\>[4]\AgdaGeneralizable{ℓ}\AgdaSpace{}%
\AgdaGeneralizable{ℓ'}\AgdaSpace{}%
\AgdaSymbol{:}\AgdaSpace{}%
\AgdaPostulate{Level}\<%
\\
\\[\AgdaEmptyExtraSkip]%
\\[\AgdaEmptyExtraSkip]%
\\[\AgdaEmptyExtraSkip]%
\>[0]\AgdaKeyword{module}\AgdaSpace{}%
\AgdaModule{\AgdaUnderscore{}}\AgdaSpace{}%
\AgdaSymbol{\{}\AgdaBound{Δ}\AgdaSpace{}%
\AgdaSymbol{:}\AgdaSpace{}%
\AgdaDatatype{sply}\AgdaSpace{}%
\AgdaGeneralizable{ℓ}\AgdaSymbol{\}}\AgdaSpace{}%
\AgdaSymbol{\{}\AgdaBound{A}\AgdaSpace{}%
\AgdaSymbol{:}\AgdaSpace{}%
\AgdaDatatype{ltyp}\AgdaSpace{}%
\AgdaGeneralizable{ℓ}\AgdaSymbol{\}}\AgdaSpace{}%
\AgdaSymbol{\{}\AgdaBound{B}\AgdaSpace{}%
\AgdaSymbol{:}\AgdaSpace{}%
\AgdaBound{A}\AgdaSpace{}%
\AgdaFunction{⇒}\AgdaSpace{}%
\AgdaFunction{ltyp}\AgdaSymbol{\}}\AgdaSpace{}%
\AgdaSymbol{\{}\AgdaBound{b}\AgdaSpace{}%
\AgdaSymbol{:}\AgdaSpace{}%
\AgdaSymbol{\{}\AgdaBound{x}\AgdaSpace{}%
\AgdaSymbol{:}\AgdaSpace{}%
\AgdaBound{A}\AgdaSpace{}%
\AgdaOperator{\AgdaFunction{．}}\AgdaSymbol{\}}\AgdaSpace{}%
\AgdaSymbol{→}\AgdaSpace{}%
\AgdaBound{B}\AgdaSpace{}%
\AgdaBound{x}\AgdaSpace{}%
\AgdaOperator{\AgdaFunction{．}}\AgdaSymbol{\}}\AgdaSpace{}%
\AgdaKeyword{where}\<%
\\
\>[0][@{}l@{\AgdaIndent{0}}]%
\>[2]\AgdaFunction{⊸I}\AgdaSpace{}%
\AgdaSymbol{:}\AgdaSpace{}%
\AgdaSymbol{(\{}\AgdaBound{x}\AgdaSpace{}%
\AgdaSymbol{:}\AgdaSpace{}%
\AgdaBound{A}\AgdaSpace{}%
\AgdaOperator{\AgdaFunction{．}}\AgdaSymbol{\}}\AgdaSpace{}%
\AgdaSymbol{→}\AgdaSpace{}%
\AgdaBound{Δ}\AgdaSpace{}%
\AgdaOperator{\AgdaInductiveConstructor{⨾}}\AgdaSpace{}%
\AgdaInductiveConstructor{!}\AgdaSpace{}%
\AgdaSymbol{(}\AgdaBound{x}\AgdaSpace{}%
\AgdaFunction{∶}\AgdaSpace{}%
\AgdaBound{A}\AgdaSymbol{)}\AgdaSpace{}%
\AgdaOperator{\AgdaDatatype{⊩}}\AgdaSpace{}%
\AgdaBound{b}\AgdaSpace{}%
\AgdaFunction{∶}\AgdaSpace{}%
\AgdaBound{B}\AgdaSpace{}%
\AgdaBound{x}\AgdaSymbol{)}\<%
\\
\>[2][@{}l@{\AgdaIndent{0}}]%
\>[4]\AgdaSymbol{→}\AgdaSpace{}%
\AgdaBound{Δ}\AgdaSpace{}%
\AgdaOperator{\AgdaDatatype{⊩}}\AgdaSpace{}%
\AgdaSymbol{(λ}\AgdaSpace{}%
\AgdaBound{x}\AgdaSpace{}%
\AgdaSymbol{→}\AgdaSpace{}%
\AgdaBound{b}\AgdaSymbol{)}\AgdaSpace{}%
\AgdaFunction{∶}\AgdaSpace{}%
\AgdaInductiveConstructor{!⟨}\AgdaSpace{}%
\AgdaBound{x}\AgdaSpace{}%
\AgdaInductiveConstructor{∈}\AgdaSpace{}%
\AgdaBound{A}\AgdaSpace{}%
\AgdaInductiveConstructor{⟩}\AgdaSpace{}%
\AgdaInductiveConstructor{⊸}\AgdaSpace{}%
\AgdaBound{B}\AgdaSpace{}%
\AgdaBound{x}\<%
\\
\>[2]\AgdaFunction{⊸I}\AgdaSpace{}%
\AgdaBound{δ}\AgdaSpace{}%
\AgdaSymbol{=}\AgdaSpace{}%
\AgdaInductiveConstructor{bind}\AgdaSpace{}%
\AgdaSymbol{(λ}\AgdaSpace{}%
\AgdaBound{x}\AgdaSpace{}%
\AgdaSymbol{→}\AgdaSpace{}%
\AgdaInductiveConstructor{curry}\AgdaSpace{}%
\AgdaBound{δ}\AgdaSymbol{)}\<%
\\
\\[\AgdaEmptyExtraSkip]%
\\[\AgdaEmptyExtraSkip]%
\>[0]\AgdaKeyword{module}\AgdaSpace{}%
\AgdaModule{\AgdaUnderscore{}}\AgdaSpace{}%
\AgdaSymbol{\{}\AgdaBound{A}\AgdaSpace{}%
\AgdaSymbol{:}\AgdaSpace{}%
\AgdaDatatype{ltyp}\AgdaSpace{}%
\AgdaGeneralizable{ℓ}\AgdaSymbol{\}}\AgdaSpace{}%
\AgdaSymbol{\{}\AgdaBound{B}\AgdaSpace{}%
\AgdaSymbol{:}\AgdaSpace{}%
\AgdaBound{A}\AgdaSpace{}%
\AgdaFunction{⇒}\AgdaSpace{}%
\AgdaFunction{ltyp}\AgdaSymbol{\}}\AgdaSpace{}%
\AgdaSymbol{\{}\AgdaBound{Δ₀}\AgdaSpace{}%
\AgdaBound{Δ₁}\AgdaSpace{}%
\AgdaSymbol{:}\AgdaSpace{}%
\AgdaDatatype{sply}\AgdaSpace{}%
\AgdaGeneralizable{ℓ}\AgdaSymbol{\}}\AgdaSpace{}%
\AgdaSymbol{\{}\AgdaBound{m}\AgdaSpace{}%
\AgdaSymbol{:}\AgdaSpace{}%
\AgdaDatatype{ℕ}\AgdaSymbol{\}}\AgdaSpace{}%
\AgdaSymbol{\{}\AgdaBound{f}\AgdaSpace{}%
\AgdaSymbol{:}\AgdaSpace{}%
\AgdaSymbol{(}\AgdaBound{x}\AgdaSpace{}%
\AgdaSymbol{:}\AgdaSpace{}%
\AgdaBound{A}\AgdaSpace{}%
\AgdaOperator{\AgdaFunction{．}}\AgdaSymbol{)}\AgdaSpace{}%
\AgdaSymbol{→}\AgdaSpace{}%
\AgdaBound{B}\AgdaSpace{}%
\AgdaBound{x}\AgdaSpace{}%
\AgdaOperator{\AgdaFunction{．}}\AgdaSymbol{\}}\AgdaSpace{}%
\AgdaSymbol{\{}\AgdaBound{a}\AgdaSpace{}%
\AgdaSymbol{:}\AgdaSpace{}%
\AgdaBound{A}\AgdaSpace{}%
\AgdaOperator{\AgdaFunction{．}}\AgdaSymbol{\}}\AgdaSpace{}%
\AgdaKeyword{where}\<%
\\
\>[0][@{}l@{\AgdaIndent{0}}]%
\>[2]\AgdaFunction{⊸App}\AgdaSpace{}%
\AgdaSymbol{:}\AgdaSpace{}%
\AgdaComment{--\ \AgdaUnderscore{}﹫\AgdaUnderscore{}}\<%
\\
\>[2][@{}l@{\AgdaIndent{0}}]%
\>[4]\AgdaBound{Δ₀}\AgdaSpace{}%
\AgdaOperator{\AgdaDatatype{⊩}}\AgdaSpace{}%
\AgdaBound{f}\AgdaSpace{}%
\AgdaFunction{∶}\AgdaSpace{}%
\AgdaInductiveConstructor{!⟨}\AgdaSpace{}%
\AgdaBound{x}\AgdaSpace{}%
\AgdaInductiveConstructor{∈}\AgdaSpace{}%
\AgdaBound{A}\AgdaSpace{}%
\AgdaInductiveConstructor{⟩}\AgdaSpace{}%
\AgdaInductiveConstructor{⊸}\AgdaSpace{}%
\AgdaBound{B}\AgdaSpace{}%
\AgdaBound{x}\<%
\\
\>[4]\AgdaSymbol{→}\AgdaSpace{}%
\AgdaSymbol{(}\AgdaBound{Δ₁}\AgdaSpace{}%
\AgdaOperator{\AgdaDatatype{⊩}}\AgdaSpace{}%
\AgdaBound{a}\AgdaSpace{}%
\AgdaFunction{∶}\AgdaSpace{}%
\AgdaBound{A}\AgdaSymbol{)}\<%
\\
\>[4]\AgdaSymbol{→}\AgdaSpace{}%
\AgdaSymbol{(}\AgdaBound{Δ₀}\AgdaSpace{}%
\AgdaOperator{\AgdaInductiveConstructor{⨾}}\AgdaSpace{}%
\AgdaInductiveConstructor{!}\AgdaSpace{}%
\AgdaBound{Δ₁}\AgdaSpace{}%
\AgdaOperator{\AgdaDatatype{⊩}}\AgdaSpace{}%
\AgdaBound{f}\AgdaSpace{}%
\AgdaBound{a}\AgdaSpace{}%
\AgdaFunction{∶}\AgdaSpace{}%
\AgdaBound{B}\AgdaSpace{}%
\AgdaBound{a}\AgdaSymbol{)}\<%
\\
\>[2]\AgdaFunction{⊸App}\AgdaSpace{}%
\AgdaBound{δ₀}\AgdaSpace{}%
\AgdaBound{δ₁}\AgdaSpace{}%
\AgdaSymbol{=}\AgdaSpace{}%
\AgdaInductiveConstructor{uncurry}\AgdaSpace{}%
\AgdaSymbol{(}\AgdaInductiveConstructor{free}\AgdaSpace{}%
\AgdaBound{δ₀}\AgdaSpace{}%
\AgdaBound{a}\AgdaSymbol{)}\AgdaSpace{}%
\AgdaOperator{\AgdaInductiveConstructor{∘}}\AgdaSpace{}%
\AgdaInductiveConstructor{id}\AgdaSpace{}%
\AgdaBound{Δ₀}\AgdaSpace{}%
\AgdaOperator{\AgdaInductiveConstructor{⨾ᶠ}}\AgdaSpace{}%
\AgdaInductiveConstructor{!ᶠ}\AgdaSpace{}%
\AgdaBound{δ₁}\<%
\\
\\[\AgdaEmptyExtraSkip]%
\\[\AgdaEmptyExtraSkip]%
\>[0]\AgdaKeyword{open}\AgdaSpace{}%
\AgdaKeyword{import}\AgdaSpace{}%
\AgdaModule{Pair}\<%
\\
\\[\AgdaEmptyExtraSkip]%
\>[0]\AgdaKeyword{module}\AgdaSpace{}%
\AgdaModule{\AgdaUnderscore{}}\AgdaSpace{}%
\AgdaSymbol{\{}\AgdaBound{A}\AgdaSpace{}%
\AgdaBound{B}\AgdaSpace{}%
\AgdaSymbol{:}\AgdaSpace{}%
\AgdaDatatype{ltyp}\AgdaSpace{}%
\AgdaGeneralizable{ℓ}\AgdaSymbol{\}}\AgdaSpace{}%
\AgdaSymbol{\{}\AgdaBound{(}\AgdaBound{a}\AgdaSpace{}%
\AgdaOperator{\AgdaInductiveConstructor{,}}\AgdaSpace{}%
\AgdaBound{b}\AgdaBound{)}\AgdaSpace{}%
\AgdaSymbol{:}\AgdaSpace{}%
\AgdaSymbol{(}\AgdaBound{A}\AgdaSpace{}%
\AgdaFunction{⊗}\AgdaSpace{}%
\AgdaBound{B}\AgdaSymbol{)}\AgdaSpace{}%
\AgdaOperator{\AgdaFunction{．}}\AgdaSymbol{\}}\AgdaSpace{}%
\AgdaKeyword{where}\<%
\\
\>[0][@{}l@{\AgdaIndent{0}}]%
\>[2]\AgdaFunction{fst⊸}\AgdaSpace{}%
\AgdaSymbol{:}\<%
\end{code}
\newcommand{\AGDAbangfst}{%
\begin{code}%
\>[2][@{}l@{\AgdaIndent{1}}]%
\>[4]\AgdaInductiveConstructor{◇}\AgdaSpace{}%
\AgdaOperator{\AgdaDatatype{⊩}}\AgdaSpace{}%
\AgdaField{fst}\AgdaSpace{}%
\AgdaFunction{∶}\AgdaSpace{}%
\AgdaFunction{!⟨}\AgdaSpace{}%
\AgdaBound{A}\AgdaSpace{}%
\AgdaFunction{⊗}\AgdaSpace{}%
\AgdaBound{B}\AgdaSpace{}%
\AgdaFunction{⟩}\AgdaSpace{}%
\AgdaFunction{⊸}\AgdaSpace{}%
\AgdaBound{A}\<%
\end{code}}
\begin{code}[hide]%
\>[2]\AgdaFunction{fst⊸}\AgdaSpace{}%
\AgdaSymbol{=}\AgdaSpace{}%
\AgdaFunction{⊸I}\AgdaSpace{}%
\AgdaSymbol{\{}\AgdaArgument{A}\AgdaSpace{}%
\AgdaSymbol{=}\AgdaSpace{}%
\AgdaBound{A}\AgdaSpace{}%
\AgdaFunction{⊗}\AgdaSpace{}%
\AgdaBound{B}\AgdaSymbol{\}}\AgdaSpace{}%
\AgdaSymbol{\{λ}\AgdaSpace{}%
\AgdaBound{\AgdaUnderscore{}}\AgdaSpace{}%
\AgdaSymbol{→}\AgdaSpace{}%
\AgdaBound{A}\AgdaSymbol{\}}\AgdaSpace{}%
\AgdaSymbol{(λ}\AgdaSpace{}%
\AgdaSymbol{\{}\AgdaBound{(}\AgdaBound{x}\AgdaSpace{}%
\AgdaOperator{\AgdaInductiveConstructor{,}}\AgdaSpace{}%
\AgdaBound{y}\AgdaBound{)}\AgdaSymbol{\}}\AgdaSpace{}%
\AgdaSymbol{→}\AgdaSpace{}%
\AgdaInductiveConstructor{unitr}\AgdaSpace{}%
\AgdaSymbol{\AgdaUnderscore{}}\AgdaSpace{}%
\AgdaOperator{\AgdaInductiveConstructor{∘}}\AgdaSpace{}%
\AgdaInductiveConstructor{use}\AgdaSpace{}%
\AgdaOperator{\AgdaInductiveConstructor{⨾ᶠ}}\AgdaSpace{}%
\AgdaInductiveConstructor{erase}%
\>[73]\AgdaOperator{\AgdaInductiveConstructor{∘}}\AgdaSpace{}%
\AgdaInductiveConstructor{coh⨾'}\AgdaSpace{}%
\AgdaOperator{\AgdaInductiveConstructor{∘}}\AgdaSpace{}%
\AgdaSymbol{(}\AgdaInductiveConstructor{!ᶠ}\AgdaSpace{}%
\AgdaSymbol{(}\AgdaInductiveConstructor{unitr}\AgdaSpace{}%
\AgdaSymbol{\AgdaUnderscore{}}\AgdaSpace{}%
\AgdaOperator{\AgdaInductiveConstructor{⨾ᶠ}}\AgdaSpace{}%
\AgdaInductiveConstructor{id}\AgdaSpace{}%
\AgdaSymbol{\AgdaUnderscore{})}\AgdaSpace{}%
\AgdaOperator{\AgdaInductiveConstructor{∘}}\AgdaSpace{}%
\AgdaFunction{unitl}\AgdaSpace{}%
\AgdaSymbol{\AgdaUnderscore{}))}\<%
\\
\>[0]\<%
\end{code}

\begin{code}[hide]%
\>[0]\AgdaSymbol{\{-\#}\AgdaSpace{}%
\AgdaKeyword{OPTIONS}\AgdaSpace{}%
\AgdaPragma{--safe}\AgdaSpace{}%
\AgdaSymbol{\#-\}}\<%
\\
\\[\AgdaEmptyExtraSkip]%
\>[0]\AgdaKeyword{module}\AgdaSpace{}%
\AgdaModule{Pair}\AgdaSpace{}%
\AgdaKeyword{where}\<%
\\
\\[\AgdaEmptyExtraSkip]%
\>[0]\AgdaKeyword{open}\AgdaSpace{}%
\AgdaKeyword{import}\AgdaSpace{}%
\AgdaModule{Base}\<%
\\
\\[\AgdaEmptyExtraSkip]%
\>[0]\AgdaKeyword{private}\<%
\\
\>[0][@{}l@{\AgdaIndent{0}}]%
\>[2]\AgdaKeyword{variable}\<%
\\
\>[2][@{}l@{\AgdaIndent{0}}]%
\>[4]\AgdaGeneralizable{ℓ}\AgdaSpace{}%
\AgdaGeneralizable{ℓ'}\AgdaSpace{}%
\AgdaSymbol{:}\AgdaSpace{}%
\AgdaPostulate{Level}\<%
\\
\\[\AgdaEmptyExtraSkip]%
\\[\AgdaEmptyExtraSkip]%
\\[\AgdaEmptyExtraSkip]%
\>[0]\AgdaKeyword{module}\AgdaSpace{}%
\AgdaModule{\AgdaUnderscore{}}\AgdaSpace{}%
\AgdaSymbol{\{}\AgdaBound{A}\AgdaSpace{}%
\AgdaSymbol{:}\AgdaSpace{}%
\AgdaDatatype{ltyp}\AgdaSpace{}%
\AgdaGeneralizable{ℓ}\AgdaSymbol{\}}\AgdaSpace{}%
\AgdaSymbol{\{}\AgdaBound{B}\AgdaSpace{}%
\AgdaSymbol{:}\AgdaSpace{}%
\AgdaBound{A}\AgdaSpace{}%
\AgdaFunction{⇒}\AgdaSpace{}%
\AgdaFunction{ltyp}\AgdaSymbol{\}}\AgdaSpace{}%
\AgdaSymbol{\{}\AgdaBound{Δ₀}\AgdaSpace{}%
\AgdaBound{Δ₁}\AgdaSpace{}%
\AgdaSymbol{:}\AgdaSpace{}%
\AgdaDatatype{sply}\AgdaSpace{}%
\AgdaGeneralizable{ℓ}\AgdaSymbol{\}}\AgdaSpace{}%
\AgdaSymbol{\{}\AgdaBound{m}\AgdaSpace{}%
\AgdaSymbol{:}\AgdaSpace{}%
\AgdaDatatype{ℕ}\AgdaSymbol{\}}\AgdaSpace{}%
\AgdaSymbol{\{}\AgdaBound{a}\AgdaSpace{}%
\AgdaSymbol{:}\AgdaSpace{}%
\AgdaBound{A}\AgdaSpace{}%
\AgdaOperator{\AgdaFunction{．}}\AgdaSymbol{\}}\AgdaSpace{}%
\AgdaSymbol{\{}\AgdaBound{b}\AgdaSpace{}%
\AgdaSymbol{:}\AgdaSpace{}%
\AgdaBound{B}\AgdaSpace{}%
\AgdaBound{a}\AgdaSpace{}%
\AgdaOperator{\AgdaFunction{．}}\AgdaSymbol{\}}\AgdaSpace{}%
\AgdaKeyword{where}\<%
\\
\>[0][@{}l@{\AgdaIndent{0}}]%
\>[2]\AgdaFunction{⊗I}\AgdaSpace{}%
\AgdaSymbol{:}\AgdaSpace{}%
\AgdaSymbol{(}\AgdaBound{Δ₀}\AgdaSpace{}%
\AgdaOperator{\AgdaDatatype{⊩}}\AgdaSpace{}%
\AgdaBound{a}\AgdaSpace{}%
\AgdaFunction{∶}\AgdaSpace{}%
\AgdaBound{A}\AgdaSymbol{)}\<%
\\
\>[2][@{}l@{\AgdaIndent{0}}]%
\>[4]\AgdaSymbol{→}\AgdaSpace{}%
\AgdaSymbol{(}\AgdaBound{Δ₁}\AgdaSpace{}%
\AgdaOperator{\AgdaDatatype{⊩}}\AgdaSpace{}%
\AgdaBound{b}\AgdaSpace{}%
\AgdaFunction{∶}\AgdaSpace{}%
\AgdaBound{B}\AgdaSpace{}%
\AgdaBound{a}\AgdaSymbol{)}\<%
\\
\>[4]\AgdaSymbol{→}\AgdaSpace{}%
\AgdaBound{Δ₀}\AgdaSpace{}%
\AgdaOperator{\AgdaFunction{\textasciicircum{}}}\AgdaSpace{}%
\AgdaBound{m}\AgdaSpace{}%
\AgdaOperator{\AgdaInductiveConstructor{⨾}}\AgdaSpace{}%
\AgdaBound{Δ₁}\AgdaSpace{}%
\AgdaOperator{\AgdaDatatype{⊩}}\AgdaSpace{}%
\AgdaSymbol{(}\AgdaBound{a}\AgdaSpace{}%
\AgdaOperator{\AgdaInductiveConstructor{,}}\AgdaSpace{}%
\AgdaBound{b}\AgdaSymbol{)}\AgdaSpace{}%
\AgdaFunction{∶}\AgdaSpace{}%
\AgdaInductiveConstructor{⟨}\AgdaSpace{}%
\AgdaBound{x}\AgdaSpace{}%
\AgdaInductiveConstructor{∈}\AgdaSpace{}%
\AgdaBound{A}\AgdaSpace{}%
\AgdaInductiveConstructor{⟩\textasciicircum{}}\AgdaSpace{}%
\AgdaBound{m}\AgdaSpace{}%
\AgdaInductiveConstructor{⊗}\AgdaSpace{}%
\AgdaBound{B}\AgdaSpace{}%
\AgdaBound{x}\<%
\\
\>[2]\AgdaFunction{⊗I}\AgdaSpace{}%
\AgdaBound{δ₀}\AgdaSpace{}%
\AgdaBound{δ₁}\AgdaSpace{}%
\AgdaSymbol{=}\AgdaSpace{}%
\AgdaFunction{⊩\textasciicircum{}}\AgdaSpace{}%
\AgdaBound{m}\AgdaSpace{}%
\AgdaBound{δ₀}\AgdaSpace{}%
\AgdaOperator{\AgdaInductiveConstructor{⨾ᶠ}}\AgdaSpace{}%
\AgdaBound{δ₁}\<%
\\
\\[\AgdaEmptyExtraSkip]%
\>[0]\AgdaKeyword{module}\AgdaSpace{}%
\AgdaModule{\AgdaUnderscore{}}%
\>[10]\AgdaSymbol{\{}\AgdaBound{A}\AgdaSpace{}%
\AgdaSymbol{:}\AgdaSpace{}%
\AgdaDatatype{ltyp}\AgdaSpace{}%
\AgdaSymbol{(}\AgdaPrimitive{ℓ-suc}\AgdaSpace{}%
\AgdaGeneralizable{ℓ}\AgdaSymbol{)\}}\AgdaSpace{}%
\AgdaSymbol{\{}\AgdaBound{m}\AgdaSpace{}%
\AgdaSymbol{:}\AgdaSpace{}%
\AgdaDatatype{ℕ}\AgdaSymbol{\}}\AgdaSpace{}%
\AgdaSymbol{\{}\AgdaBound{B}\AgdaSpace{}%
\AgdaSymbol{:}\AgdaSpace{}%
\AgdaBound{A}\AgdaSpace{}%
\AgdaFunction{⇒}\AgdaSpace{}%
\AgdaFunction{ltyp}\AgdaSymbol{\}}\<%
\\
\>[0][@{}l@{\AgdaIndent{0}}]%
\>[2]\AgdaSymbol{\{}\AgdaBound{C}\AgdaSpace{}%
\AgdaSymbol{:}\AgdaSpace{}%
\AgdaInductiveConstructor{Σᵒ}\AgdaSpace{}%
\AgdaBound{A}\AgdaSpace{}%
\AgdaBound{m}\AgdaSpace{}%
\AgdaBound{B}\AgdaSpace{}%
\AgdaOperator{\AgdaFunction{．}}\AgdaSpace{}%
\AgdaSymbol{→}\AgdaSpace{}%
\AgdaDatatype{ltyp}\AgdaSpace{}%
\AgdaSymbol{(}\AgdaPrimitive{ℓ-suc}\AgdaSpace{}%
\AgdaGeneralizable{ℓ}\AgdaSymbol{)\}}\<%
\\
\>[2]\AgdaSymbol{\{}\AgdaBound{Δ₀}\AgdaSpace{}%
\AgdaBound{Δ₁}\AgdaSpace{}%
\AgdaSymbol{:}\AgdaSpace{}%
\AgdaDatatype{sply}\AgdaSpace{}%
\AgdaSymbol{(}\AgdaPrimitive{ℓ-suc}\AgdaSpace{}%
\AgdaGeneralizable{ℓ}\AgdaSymbol{)\}}\<%
\\
\>[2]\AgdaSymbol{\{}\AgdaBound{p}\AgdaSpace{}%
\AgdaSymbol{:}\AgdaSpace{}%
\AgdaInductiveConstructor{Σᵒ}\AgdaSpace{}%
\AgdaBound{A}\AgdaSpace{}%
\AgdaBound{m}\AgdaSpace{}%
\AgdaBound{B}\AgdaSpace{}%
\AgdaOperator{\AgdaFunction{．}}\AgdaSymbol{\}}\<%
\\
\>[2]\AgdaSymbol{\{}\AgdaBound{c}\AgdaSpace{}%
\AgdaSymbol{:}\AgdaSpace{}%
\AgdaSymbol{(}\AgdaBound{x}\AgdaSpace{}%
\AgdaSymbol{:}\AgdaSpace{}%
\AgdaBound{A}\AgdaSpace{}%
\AgdaOperator{\AgdaFunction{．}}\AgdaSymbol{)}\AgdaSpace{}%
\AgdaSymbol{(}\AgdaBound{y}\AgdaSpace{}%
\AgdaSymbol{:}\AgdaSpace{}%
\AgdaBound{B}\AgdaSpace{}%
\AgdaBound{x}\AgdaSpace{}%
\AgdaOperator{\AgdaFunction{．}}\AgdaSymbol{)}\AgdaSpace{}%
\AgdaSymbol{→}\AgdaSpace{}%
\AgdaBound{C}\AgdaSpace{}%
\AgdaSymbol{(}\AgdaBound{x}\AgdaSpace{}%
\AgdaOperator{\AgdaInductiveConstructor{,}}\AgdaSpace{}%
\AgdaBound{y}\AgdaSymbol{)}\AgdaSpace{}%
\AgdaOperator{\AgdaFunction{．}}\AgdaSymbol{\}}\<%
\\
\>[2]\AgdaKeyword{where}\<%
\\
\>[2]\AgdaFunction{⊗E}\AgdaSpace{}%
\AgdaSymbol{:}\<%
\\
\>[2][@{}l@{\AgdaIndent{0}}]%
\>[4]\AgdaSymbol{(}\AgdaBound{Δ₀}\AgdaSpace{}%
\AgdaOperator{\AgdaDatatype{⊩}}\AgdaSpace{}%
\AgdaBound{p}\AgdaSpace{}%
\AgdaFunction{∶}\AgdaSpace{}%
\AgdaInductiveConstructor{⟨}\AgdaSpace{}%
\AgdaBound{x}\AgdaSpace{}%
\AgdaInductiveConstructor{∈}\AgdaSpace{}%
\AgdaBound{A}\AgdaSpace{}%
\AgdaInductiveConstructor{⟩\textasciicircum{}}\AgdaSpace{}%
\AgdaBound{m}\AgdaSpace{}%
\AgdaInductiveConstructor{⊗}\AgdaSpace{}%
\AgdaBound{B}\AgdaSpace{}%
\AgdaBound{x}\AgdaSymbol{)}\<%
\\
\>[4]\AgdaSymbol{→}\AgdaSpace{}%
\AgdaSymbol{(}\AgdaBound{f}\AgdaSpace{}%
\AgdaSymbol{:}\AgdaSpace{}%
\AgdaSymbol{\{}\AgdaBound{x}\AgdaSpace{}%
\AgdaSymbol{:}\AgdaSpace{}%
\AgdaBound{A}\AgdaSpace{}%
\AgdaOperator{\AgdaFunction{．}}\AgdaSymbol{\}}\AgdaSpace{}%
\AgdaSymbol{\{}\AgdaBound{y}\AgdaSpace{}%
\AgdaSymbol{:}\AgdaSpace{}%
\AgdaBound{B}\AgdaSpace{}%
\AgdaBound{x}\AgdaSpace{}%
\AgdaOperator{\AgdaFunction{．}}\AgdaSymbol{\}}%
\>[34]\AgdaSymbol{→}\AgdaSpace{}%
\AgdaBound{Δ₁}\AgdaSpace{}%
\AgdaOperator{\AgdaInductiveConstructor{⨾}}\AgdaSpace{}%
\AgdaSymbol{(}\AgdaBound{x}\AgdaSpace{}%
\AgdaFunction{∶}\AgdaSpace{}%
\AgdaBound{A}\AgdaSymbol{)}\AgdaSpace{}%
\AgdaOperator{\AgdaFunction{\textasciicircum{}}}\AgdaSpace{}%
\AgdaBound{m}\AgdaSpace{}%
\AgdaOperator{\AgdaInductiveConstructor{⨾}}\AgdaSpace{}%
\AgdaBound{y}\AgdaSpace{}%
\AgdaFunction{∶}\AgdaSpace{}%
\AgdaBound{B}\AgdaSpace{}%
\AgdaBound{x}\AgdaSpace{}%
\AgdaOperator{\AgdaDatatype{⊩}}\AgdaSpace{}%
\AgdaBound{c}\AgdaSpace{}%
\AgdaBound{x}\AgdaSpace{}%
\AgdaBound{y}\AgdaSpace{}%
\AgdaFunction{∶}\AgdaSpace{}%
\AgdaBound{C}\AgdaSpace{}%
\AgdaSymbol{(}\AgdaBound{x}\AgdaSpace{}%
\AgdaOperator{\AgdaInductiveConstructor{,}}\AgdaSpace{}%
\AgdaBound{y}\AgdaSymbol{))}\<%
\\
\>[4]\AgdaSymbol{→}\AgdaSpace{}%
\AgdaBound{Δ₁}\AgdaSpace{}%
\AgdaOperator{\AgdaInductiveConstructor{⨾}}\AgdaSpace{}%
\AgdaBound{Δ₀}\AgdaSpace{}%
\AgdaOperator{\AgdaDatatype{⊩}}\AgdaSpace{}%
\AgdaFunction{elimΣ}\AgdaSpace{}%
\AgdaBound{c}\AgdaSpace{}%
\AgdaBound{p}\AgdaSpace{}%
\AgdaFunction{∶}\AgdaSpace{}%
\AgdaBound{C}\AgdaSpace{}%
\AgdaBound{p}\<%
\\
\>[2]\AgdaFunction{⊗E}\AgdaSpace{}%
\AgdaBound{δ₀}\AgdaSpace{}%
\AgdaBound{δ₁}\AgdaSpace{}%
\AgdaSymbol{=}\AgdaSpace{}%
\AgdaBound{δ₁}\AgdaSpace{}%
\AgdaSymbol{\{}\AgdaBound{p}\AgdaSpace{}%
\AgdaSymbol{.}\AgdaField{fst}\AgdaSymbol{\}}\AgdaSpace{}%
\AgdaSymbol{\{}\AgdaBound{p}\AgdaSpace{}%
\AgdaSymbol{.}\AgdaField{snd}\AgdaSymbol{\}}\AgdaSpace{}%
\AgdaOperator{\AgdaInductiveConstructor{∘}}\AgdaSpace{}%
\AgdaInductiveConstructor{id}\AgdaSpace{}%
\AgdaSymbol{\AgdaUnderscore{}}\AgdaSpace{}%
\AgdaOperator{\AgdaInductiveConstructor{⨾ᶠ}}\AgdaSpace{}%
\AgdaBound{δ₀}\<%
\\
\\[\AgdaEmptyExtraSkip]%
\>[0]\AgdaKeyword{module}\AgdaSpace{}%
\AgdaModule{\AgdaUnderscore{}}%
\>[10]\AgdaSymbol{\{}\AgdaBound{A}\AgdaSpace{}%
\AgdaSymbol{:}\AgdaSpace{}%
\AgdaDatatype{ltyp}\AgdaSpace{}%
\AgdaSymbol{(}\AgdaPrimitive{ℓ-suc}\AgdaSpace{}%
\AgdaGeneralizable{ℓ}\AgdaSymbol{)\}}\AgdaSpace{}%
\AgdaSymbol{\{}\AgdaBound{m}\AgdaSpace{}%
\AgdaSymbol{:}\AgdaSpace{}%
\AgdaDatatype{ℕ}\AgdaSymbol{\}}\AgdaSpace{}%
\AgdaSymbol{\{}\AgdaBound{B}\AgdaSpace{}%
\AgdaSymbol{:}\AgdaSpace{}%
\AgdaBound{A}\AgdaSpace{}%
\AgdaFunction{⇒}\AgdaSpace{}%
\AgdaFunction{ltyp}\AgdaSymbol{\}}\<%
\\
\>[0][@{}l@{\AgdaIndent{0}}]%
\>[2]\AgdaSymbol{\{}\AgdaBound{C}\AgdaSpace{}%
\AgdaSymbol{:}\AgdaSpace{}%
\AgdaInductiveConstructor{Σᵒ}\AgdaSpace{}%
\AgdaBound{A}\AgdaSpace{}%
\AgdaBound{m}\AgdaSpace{}%
\AgdaBound{B}\AgdaSpace{}%
\AgdaOperator{\AgdaFunction{．}}\AgdaSpace{}%
\AgdaSymbol{→}\AgdaSpace{}%
\AgdaDatatype{ltyp}\AgdaSpace{}%
\AgdaSymbol{(}\AgdaPrimitive{ℓ-suc}\AgdaSpace{}%
\AgdaGeneralizable{ℓ}\AgdaSymbol{)\}}\<%
\\
\>[2]\AgdaSymbol{\{}\AgdaBound{Δ}\AgdaSpace{}%
\AgdaSymbol{:}\AgdaSpace{}%
\AgdaDatatype{sply}\AgdaSpace{}%
\AgdaSymbol{(}\AgdaPrimitive{ℓ-suc}\AgdaSpace{}%
\AgdaGeneralizable{ℓ}\AgdaSymbol{)\}}\<%
\\
\>[2]\AgdaSymbol{\{}\AgdaBound{c}\AgdaSpace{}%
\AgdaSymbol{:}\AgdaSpace{}%
\AgdaSymbol{(}\AgdaBound{x}\AgdaSpace{}%
\AgdaSymbol{:}\AgdaSpace{}%
\AgdaBound{A}\AgdaSpace{}%
\AgdaOperator{\AgdaFunction{．}}\AgdaSymbol{)}\AgdaSpace{}%
\AgdaSymbol{(}\AgdaBound{y}\AgdaSpace{}%
\AgdaSymbol{:}\AgdaSpace{}%
\AgdaBound{B}\AgdaSpace{}%
\AgdaBound{x}\AgdaSpace{}%
\AgdaOperator{\AgdaFunction{．}}\AgdaSymbol{)}\AgdaSpace{}%
\AgdaSymbol{→}\AgdaSpace{}%
\AgdaBound{C}\AgdaSpace{}%
\AgdaSymbol{(}\AgdaBound{x}\AgdaSpace{}%
\AgdaOperator{\AgdaInductiveConstructor{,}}\AgdaSpace{}%
\AgdaBound{y}\AgdaSymbol{)}\AgdaSpace{}%
\AgdaOperator{\AgdaFunction{．}}\AgdaSymbol{\}}\<%
\\
\>[2]\AgdaKeyword{where}\<%
\\
\\[\AgdaEmptyExtraSkip]%
\>[2]\AgdaFunction{⊗β}\AgdaSpace{}%
\AgdaSymbol{:}\AgdaSpace{}%
\AgdaSymbol{(}\AgdaBound{f}\AgdaSpace{}%
\AgdaSymbol{:}\AgdaSpace{}%
\AgdaSymbol{\{}\AgdaBound{x}\AgdaSpace{}%
\AgdaSymbol{:}\AgdaSpace{}%
\AgdaBound{A}\AgdaSpace{}%
\AgdaOperator{\AgdaFunction{．}}\AgdaSymbol{\}}\AgdaSpace{}%
\AgdaSymbol{\{}\AgdaBound{y}\AgdaSpace{}%
\AgdaSymbol{:}\AgdaSpace{}%
\AgdaBound{B}\AgdaSpace{}%
\AgdaBound{x}\AgdaSpace{}%
\AgdaOperator{\AgdaFunction{．}}\AgdaSymbol{\}}%
\>[35]\AgdaSymbol{→}\AgdaSpace{}%
\AgdaBound{Δ}\AgdaSpace{}%
\AgdaOperator{\AgdaInductiveConstructor{⨾}}\AgdaSpace{}%
\AgdaSymbol{(}\AgdaBound{x}\AgdaSpace{}%
\AgdaFunction{∶}\AgdaSpace{}%
\AgdaBound{A}\AgdaSymbol{)}\AgdaSpace{}%
\AgdaOperator{\AgdaFunction{\textasciicircum{}}}\AgdaSpace{}%
\AgdaBound{m}\AgdaSpace{}%
\AgdaOperator{\AgdaInductiveConstructor{⨾}}\AgdaSpace{}%
\AgdaBound{y}\AgdaSpace{}%
\AgdaFunction{∶}\AgdaSpace{}%
\AgdaBound{B}\AgdaSpace{}%
\AgdaBound{x}\AgdaSpace{}%
\AgdaOperator{\AgdaDatatype{⊩}}\AgdaSpace{}%
\AgdaBound{c}\AgdaSpace{}%
\AgdaBound{x}\AgdaSpace{}%
\AgdaBound{y}\AgdaSpace{}%
\AgdaFunction{∶}\AgdaSpace{}%
\AgdaBound{C}\AgdaSpace{}%
\AgdaSymbol{(}\AgdaBound{x}\AgdaSpace{}%
\AgdaOperator{\AgdaInductiveConstructor{,}}\AgdaSpace{}%
\AgdaBound{y}\AgdaSymbol{))}\<%
\\
\>[2][@{}l@{\AgdaIndent{0}}]%
\>[4]\AgdaSymbol{→}\AgdaSpace{}%
\AgdaSymbol{\{}\AgdaBound{x}\AgdaSpace{}%
\AgdaSymbol{:}\AgdaSpace{}%
\AgdaBound{A}\AgdaSpace{}%
\AgdaOperator{\AgdaFunction{．}}\AgdaSymbol{\}}\AgdaSpace{}%
\AgdaSymbol{\{}\AgdaBound{y}\AgdaSpace{}%
\AgdaSymbol{:}\AgdaSpace{}%
\AgdaBound{B}\AgdaSpace{}%
\AgdaBound{x}\AgdaSpace{}%
\AgdaOperator{\AgdaFunction{．}}\AgdaSymbol{\}}\AgdaSpace{}%
\AgdaSymbol{→}\AgdaSpace{}%
\AgdaBound{Δ}\AgdaSpace{}%
\AgdaOperator{\AgdaInductiveConstructor{⨾}}\AgdaSpace{}%
\AgdaSymbol{(}\AgdaBound{x}\AgdaSpace{}%
\AgdaFunction{∶}\AgdaSpace{}%
\AgdaBound{A}\AgdaSymbol{)}\AgdaSpace{}%
\AgdaOperator{\AgdaFunction{\textasciicircum{}}}\AgdaSpace{}%
\AgdaBound{m}\AgdaSpace{}%
\AgdaOperator{\AgdaInductiveConstructor{⨾}}\AgdaSpace{}%
\AgdaBound{y}\AgdaSpace{}%
\AgdaFunction{∶}\AgdaSpace{}%
\AgdaBound{B}\AgdaSpace{}%
\AgdaBound{x}\AgdaSpace{}%
\AgdaOperator{\AgdaDatatype{⊩}}\AgdaSpace{}%
\AgdaFunction{elimΣ}\AgdaSpace{}%
\AgdaBound{c}\AgdaSpace{}%
\AgdaSymbol{(}\AgdaBound{x}\AgdaSpace{}%
\AgdaOperator{\AgdaInductiveConstructor{,}}\AgdaSpace{}%
\AgdaBound{y}\AgdaSymbol{)}\AgdaSpace{}%
\AgdaFunction{∶}\AgdaSpace{}%
\AgdaBound{C}\AgdaSpace{}%
\AgdaSymbol{(}\AgdaBound{x}\AgdaSpace{}%
\AgdaOperator{\AgdaInductiveConstructor{,}}\AgdaSpace{}%
\AgdaBound{y}\AgdaSymbol{)}\AgdaSpace{}%
\AgdaComment{--\ \ c\ x\ y}\<%
\\
\>[2]\AgdaFunction{⊗β}\AgdaSpace{}%
\AgdaBound{δ}\AgdaSpace{}%
\AgdaSymbol{\{}\AgdaBound{x}\AgdaSymbol{\}}\AgdaSpace{}%
\AgdaSymbol{\{}\AgdaBound{y}\AgdaSymbol{\}}\AgdaSpace{}%
\AgdaSymbol{=}\AgdaSpace{}%
\AgdaBound{δ}\<%
\\
\\[\AgdaEmptyExtraSkip]%
\\[\AgdaEmptyExtraSkip]%
\>[0]\AgdaKeyword{module}\AgdaSpace{}%
\AgdaModule{Σelimalreadyapplied}\AgdaSpace{}%
\AgdaSymbol{\{}\AgdaBound{A}\AgdaSpace{}%
\AgdaBound{B}\AgdaSpace{}%
\AgdaSymbol{:}\AgdaSpace{}%
\AgdaDatatype{ltyp}\AgdaSpace{}%
\AgdaSymbol{(}\AgdaPrimitive{ℓ-suc}\AgdaSpace{}%
\AgdaGeneralizable{ℓ}\AgdaSymbol{)\}}\AgdaSpace{}%
\AgdaSymbol{\{}\AgdaBound{(}\AgdaBound{x}\AgdaSpace{}%
\AgdaOperator{\AgdaInductiveConstructor{,}}\AgdaSpace{}%
\AgdaBound{y}\AgdaBound{)}\AgdaSpace{}%
\AgdaSymbol{:}\AgdaSpace{}%
\AgdaSymbol{(}\AgdaBound{A}\AgdaSpace{}%
\AgdaFunction{⊗}\AgdaSpace{}%
\AgdaBound{B}\AgdaSymbol{)}\AgdaSpace{}%
\AgdaOperator{\AgdaFunction{．}}\AgdaSymbol{\}}\AgdaSpace{}%
\AgdaKeyword{where}\<%
\\
\>[0][@{}l@{\AgdaIndent{0}}]%
\>[2]\AgdaFunction{swapᵒ}\AgdaSpace{}%
\AgdaSymbol{:}\<%
\end{code}

\begin{code}[hide]%
\>[2]\AgdaFunction{swapᵒ}\AgdaSpace{}%
\AgdaSymbol{=}\AgdaSpace{}%
\AgdaFunction{⊗E}\AgdaSpace{}%
\AgdaSymbol{\{}\AgdaArgument{A}\AgdaSpace{}%
\AgdaSymbol{=}\AgdaSpace{}%
\AgdaBound{A}\AgdaSymbol{\}}\AgdaSpace{}%
\AgdaSymbol{\{}\AgdaArgument{m}\AgdaSpace{}%
\AgdaSymbol{=}\AgdaSpace{}%
\AgdaNumber{1}\AgdaSymbol{\}}\AgdaSpace{}%
\AgdaSymbol{\{λ}\AgdaSpace{}%
\AgdaBound{\AgdaUnderscore{}}\AgdaSpace{}%
\AgdaSymbol{→}\AgdaSpace{}%
\AgdaBound{B}\AgdaSymbol{\}}\AgdaSpace{}%
\AgdaSymbol{\{}\AgdaArgument{C}\AgdaSpace{}%
\AgdaSymbol{=}\AgdaSpace{}%
\AgdaSymbol{λ}\AgdaSpace{}%
\AgdaBound{\AgdaUnderscore{}}\AgdaSpace{}%
\AgdaSymbol{→}\AgdaSpace{}%
\AgdaBound{B}\AgdaSpace{}%
\AgdaFunction{⊗}\AgdaSpace{}%
\AgdaBound{A}\AgdaSymbol{\}}\AgdaSpace{}%
\AgdaSymbol{\{}\AgdaArgument{Δ₀}\AgdaSpace{}%
\AgdaSymbol{=}\AgdaSpace{}%
\AgdaSymbol{(}\AgdaBound{x}\AgdaSpace{}%
\AgdaOperator{\AgdaInductiveConstructor{,}}\AgdaSpace{}%
\AgdaBound{y}\AgdaSymbol{)}\AgdaSpace{}%
\AgdaFunction{∶}\AgdaSpace{}%
\AgdaSymbol{(}\AgdaBound{A}\AgdaSpace{}%
\AgdaFunction{⊗}\AgdaSpace{}%
\AgdaBound{B}\AgdaSymbol{)\}}\AgdaSpace{}%
\AgdaSymbol{\{}\AgdaInductiveConstructor{◇}\AgdaSymbol{\}}\AgdaSpace{}%
\AgdaSymbol{(}\AgdaInductiveConstructor{id}\AgdaSpace{}%
\AgdaSymbol{\AgdaUnderscore{})}\AgdaSpace{}%
\AgdaSymbol{(λ}\AgdaSpace{}%
\AgdaSymbol{\{}\AgdaBound{x}\AgdaSymbol{\}}\AgdaSpace{}%
\AgdaSymbol{\{}\AgdaBound{y}\AgdaSymbol{\}}\AgdaSpace{}%
\AgdaSymbol{→}\AgdaSpace{}%
\AgdaFunction{⊗I}\AgdaSpace{}%
\AgdaSymbol{\{}\AgdaArgument{A}\AgdaSpace{}%
\AgdaSymbol{=}\AgdaSpace{}%
\AgdaBound{B}\AgdaSymbol{\}}\AgdaSpace{}%
\AgdaSymbol{\{λ}\AgdaSpace{}%
\AgdaBound{\AgdaUnderscore{}}\AgdaSpace{}%
\AgdaSymbol{→}\AgdaSpace{}%
\AgdaBound{A}\AgdaSymbol{\}}\AgdaSpace{}%
\AgdaSymbol{(}\AgdaFunction{Varᵒ}\AgdaSpace{}%
\AgdaBound{B}\AgdaSpace{}%
\AgdaBound{y}\AgdaSymbol{)}\AgdaSpace{}%
\AgdaSymbol{(}\AgdaFunction{Varᵒ}\AgdaSpace{}%
\AgdaBound{A}\AgdaSpace{}%
\AgdaBound{x}\AgdaSymbol{)}\<%
\\
\>[2][@{}l@{\AgdaIndent{0}}]%
\>[4]\AgdaOperator{\AgdaInductiveConstructor{∘}}\AgdaSpace{}%
\AgdaInductiveConstructor{swap}\AgdaSpace{}%
\AgdaSymbol{\AgdaUnderscore{}}\AgdaSpace{}%
\AgdaSymbol{\AgdaUnderscore{}}\AgdaSpace{}%
\AgdaOperator{\AgdaInductiveConstructor{∘}}\AgdaSpace{}%
\AgdaSymbol{(}\AgdaInductiveConstructor{id}\AgdaSpace{}%
\AgdaSymbol{\AgdaUnderscore{}}\AgdaSpace{}%
\AgdaOperator{\AgdaInductiveConstructor{⨾ᶠ}}\AgdaSpace{}%
\AgdaInductiveConstructor{unitr'}\AgdaSpace{}%
\AgdaSymbol{\AgdaUnderscore{}}\AgdaSpace{}%
\AgdaOperator{\AgdaInductiveConstructor{∘}}\AgdaSpace{}%
\AgdaSymbol{(}\AgdaInductiveConstructor{unitr}\AgdaSpace{}%
\AgdaSymbol{\AgdaUnderscore{}}\AgdaSpace{}%
\AgdaOperator{\AgdaInductiveConstructor{⨾ᶠ}}\AgdaSpace{}%
\AgdaInductiveConstructor{id}\AgdaSpace{}%
\AgdaSymbol{\AgdaUnderscore{}}\AgdaSpace{}%
\AgdaOperator{\AgdaInductiveConstructor{∘}}\AgdaSpace{}%
\AgdaFunction{unitl}\AgdaSpace{}%
\AgdaSymbol{\AgdaUnderscore{})))}\AgdaSpace{}%
\AgdaOperator{\AgdaInductiveConstructor{∘}}%
\>[70]\AgdaFunction{unitl'}\AgdaSpace{}%
\AgdaSymbol{\AgdaUnderscore{}}\<%
\\
\\[\AgdaEmptyExtraSkip]%
\>[0]\AgdaKeyword{module}\AgdaSpace{}%
\AgdaModule{\AgdaUnderscore{}}\AgdaSpace{}%
\AgdaSymbol{\{}\AgdaBound{A}\AgdaSpace{}%
\AgdaBound{B}\AgdaSpace{}%
\AgdaSymbol{:}\AgdaSpace{}%
\AgdaDatatype{ltyp}\AgdaSpace{}%
\AgdaSymbol{(}\AgdaPrimitive{ℓ-suc}\AgdaSpace{}%
\AgdaGeneralizable{ℓ}\AgdaSymbol{)\}}\AgdaSpace{}%
\AgdaSymbol{\{}\AgdaBound{z}\AgdaSpace{}%
\AgdaSymbol{:}\AgdaSpace{}%
\AgdaSymbol{(}\AgdaBound{A}\AgdaSpace{}%
\AgdaFunction{⊗}\AgdaSpace{}%
\AgdaBound{B}\AgdaSymbol{)}\AgdaSpace{}%
\AgdaOperator{\AgdaFunction{．}}\AgdaSymbol{\}}\AgdaSpace{}%
\AgdaKeyword{where}\<%
\\
\>[0][@{}l@{\AgdaIndent{0}}]%
\>[2]\AgdaFunction{sndᵒ}\AgdaSpace{}%
\AgdaSymbol{:}\AgdaSpace{}%
\AgdaBound{z}\AgdaSpace{}%
\AgdaFunction{∶}\AgdaSpace{}%
\AgdaSymbol{(}\AgdaFunction{⟨}\AgdaSpace{}%
\AgdaBound{A}\AgdaSpace{}%
\AgdaFunction{⟩\textasciicircum{}}\AgdaSpace{}%
\AgdaNumber{0}\AgdaSpace{}%
\AgdaFunction{⊗}\AgdaSpace{}%
\AgdaBound{B}\AgdaSymbol{)}\AgdaSpace{}%
\AgdaOperator{\AgdaDatatype{⊩}}\AgdaSpace{}%
\AgdaField{snd}\AgdaSpace{}%
\AgdaBound{z}\AgdaSpace{}%
\AgdaFunction{∶}\AgdaSpace{}%
\AgdaBound{B}\<%
\\
\>[2]\AgdaFunction{sndᵒ}\AgdaSpace{}%
\AgdaSymbol{=}\AgdaSpace{}%
\AgdaFunction{⊗E}\AgdaSpace{}%
\AgdaSymbol{\{}\AgdaArgument{A}\AgdaSpace{}%
\AgdaSymbol{=}\AgdaSpace{}%
\AgdaBound{A}\AgdaSymbol{\}}\AgdaSpace{}%
\AgdaSymbol{\{}\AgdaNumber{0}\AgdaSymbol{\}}\AgdaSpace{}%
\AgdaSymbol{\{λ}\AgdaSpace{}%
\AgdaBound{\AgdaUnderscore{}}\AgdaSpace{}%
\AgdaSymbol{→}\AgdaSpace{}%
\AgdaBound{B}\AgdaSymbol{\}}\AgdaSpace{}%
\AgdaSymbol{\{}\AgdaArgument{C}\AgdaSpace{}%
\AgdaSymbol{=}\AgdaSpace{}%
\AgdaSymbol{λ}\AgdaSpace{}%
\AgdaBound{\AgdaUnderscore{}}\AgdaSpace{}%
\AgdaSymbol{→}\AgdaSpace{}%
\AgdaBound{B}\AgdaSymbol{\}}\AgdaSpace{}%
\AgdaSymbol{\{}\AgdaBound{z}\AgdaSpace{}%
\AgdaFunction{∶}\AgdaSpace{}%
\AgdaSymbol{(}\AgdaFunction{⟨}\AgdaSpace{}%
\AgdaBound{A}\AgdaSpace{}%
\AgdaFunction{⟩\textasciicircum{}}\AgdaSpace{}%
\AgdaNumber{0}\AgdaSpace{}%
\AgdaFunction{⊗}\AgdaSpace{}%
\AgdaBound{B}\AgdaSymbol{)\}}\AgdaSpace{}%
\AgdaSymbol{\{}\AgdaInductiveConstructor{◇}\AgdaSymbol{\}}\AgdaSpace{}%
\AgdaSymbol{\{}\AgdaBound{z}\AgdaSymbol{\}}\AgdaSpace{}%
\AgdaSymbol{\{λ}\AgdaSpace{}%
\AgdaBound{\AgdaUnderscore{}}\AgdaSpace{}%
\AgdaBound{y}\AgdaSpace{}%
\AgdaSymbol{→}\AgdaSpace{}%
\AgdaBound{y}\AgdaSymbol{\}}\<%
\\
\>[2][@{}l@{\AgdaIndent{0}}]%
\>[4]\AgdaSymbol{(}\AgdaFunction{Varᵒ}\AgdaSpace{}%
\AgdaSymbol{(}\AgdaFunction{⟨}\AgdaSpace{}%
\AgdaBound{A}\AgdaSpace{}%
\AgdaFunction{⟩\textasciicircum{}}\AgdaSpace{}%
\AgdaNumber{0}\AgdaSpace{}%
\AgdaFunction{⊗}\AgdaSpace{}%
\AgdaBound{B}\AgdaSymbol{)}\AgdaSpace{}%
\AgdaBound{z}\AgdaSymbol{)}\AgdaSpace{}%
\AgdaSymbol{(λ}\AgdaSpace{}%
\AgdaSymbol{\{}\AgdaBound{\AgdaUnderscore{}}\AgdaSymbol{\}}\AgdaSpace{}%
\AgdaSymbol{\{}\AgdaBound{y}\AgdaSymbol{\}}\AgdaSpace{}%
\AgdaSymbol{→}\AgdaSpace{}%
\AgdaFunction{Varᵒ}\AgdaSpace{}%
\AgdaBound{B}\AgdaSpace{}%
\AgdaBound{y}\AgdaSpace{}%
\AgdaOperator{\AgdaInductiveConstructor{∘}}\AgdaSpace{}%
\AgdaSymbol{(}\AgdaFunction{unitl}\AgdaSpace{}%
\AgdaSymbol{\AgdaUnderscore{}}\AgdaSpace{}%
\AgdaOperator{\AgdaInductiveConstructor{∘}}\AgdaSpace{}%
\AgdaFunction{unitl}\AgdaSpace{}%
\AgdaSymbol{\AgdaUnderscore{}))}\AgdaSpace{}%
\AgdaOperator{\AgdaInductiveConstructor{∘}}\AgdaSpace{}%
\AgdaInductiveConstructor{id}\AgdaSpace{}%
\AgdaInductiveConstructor{◇}\AgdaSpace{}%
\AgdaOperator{\AgdaInductiveConstructor{⨾ᶠ}}\AgdaSpace{}%
\AgdaFunction{unitl'}\AgdaSpace{}%
\AgdaSymbol{\AgdaUnderscore{}}\<%
\\
\>[0]\<%
\end{code}

\begin{code}[hide]%
\>[0]\AgdaSymbol{\{-\#}\AgdaSpace{}%
\AgdaKeyword{OPTIONS}\AgdaSpace{}%
\AgdaPragma{--safe}\AgdaSpace{}%
\AgdaSymbol{\#-\}}\<%
\\
\>[0]\AgdaKeyword{module}\AgdaSpace{}%
\AgdaModule{BTree}\AgdaSpace{}%
\AgdaKeyword{where}\<%
\\
\\[\AgdaEmptyExtraSkip]%
\>[0]\AgdaKeyword{open}\AgdaSpace{}%
\AgdaKeyword{import}\AgdaSpace{}%
\AgdaModule{Base}\<%
\\
\>[0]\AgdaKeyword{open}\AgdaSpace{}%
\AgdaKeyword{import}\AgdaSpace{}%
\AgdaModule{StructuralLemmas}\<%
\\
\>[0]\AgdaKeyword{open}\AgdaSpace{}%
\AgdaKeyword{import}\AgdaSpace{}%
\AgdaModule{Sum}\<%
\\
\>[0]\AgdaKeyword{open}\AgdaSpace{}%
\AgdaKeyword{import}\AgdaSpace{}%
\AgdaModule{W}\<%
\\
\\[\AgdaEmptyExtraSkip]%
\>[0]\AgdaKeyword{private}\<%
\\
\>[0][@{}l@{\AgdaIndent{0}}]%
\>[2]\AgdaKeyword{variable}\<%
\\
\>[2][@{}l@{\AgdaIndent{0}}]%
\>[4]\AgdaGeneralizable{ℓ}\AgdaSpace{}%
\AgdaSymbol{:}\AgdaSpace{}%
\AgdaPostulate{Level}\<%
\\
\\[\AgdaEmptyExtraSkip]%
\>[0]\AgdaFunction{treeswᵒ}\AgdaSpace{}%
\AgdaSymbol{:}\AgdaSpace{}%
\AgdaSymbol{\{}\AgdaBound{A}\AgdaSpace{}%
\AgdaSymbol{:}\AgdaSpace{}%
\AgdaDatatype{ltyp}\AgdaSpace{}%
\AgdaGeneralizable{ℓ}\AgdaSymbol{\}}\AgdaSpace{}%
\AgdaSymbol{→}\AgdaSpace{}%
\AgdaSymbol{(}\AgdaInductiveConstructor{⊤ᵒ}\AgdaSpace{}%
\AgdaOperator{\AgdaFunction{⊕}}\AgdaSpace{}%
\AgdaBound{A}\AgdaSymbol{)}\AgdaSpace{}%
\AgdaFunction{⇒}\AgdaSpace{}%
\AgdaFunction{finltyp}\<%
\\
\>[0]\AgdaFunction{treeswᵒ}\AgdaSpace{}%
\AgdaSymbol{.}\AgdaField{fst}\AgdaSpace{}%
\AgdaSymbol{(}\AgdaInductiveConstructor{inl}\AgdaSpace{}%
\AgdaSymbol{\AgdaUnderscore{})}\AgdaSpace{}%
\AgdaSymbol{=}\AgdaSpace{}%
\AgdaInductiveConstructor{Boolᵒ}\<%
\\
\>[0]\AgdaFunction{treeswᵒ}\AgdaSpace{}%
\AgdaSymbol{.}\AgdaField{fst}\AgdaSpace{}%
\AgdaSymbol{(}\AgdaInductiveConstructor{inr}\AgdaSpace{}%
\AgdaBound{a}\AgdaSymbol{)}\AgdaSpace{}%
\AgdaSymbol{=}\AgdaSpace{}%
\AgdaInductiveConstructor{⊥ᵒ}\<%
\\
\>[0]\AgdaFunction{treeswᵒ}\AgdaSpace{}%
\AgdaSymbol{.}\AgdaField{snd}\AgdaSpace{}%
\AgdaSymbol{(}\AgdaInductiveConstructor{inl}\AgdaSpace{}%
\AgdaSymbol{\AgdaUnderscore{})}\AgdaSpace{}%
\AgdaSymbol{=}\AgdaSpace{}%
\AgdaInductiveConstructor{true}\AgdaSpace{}%
\AgdaOperator{\AgdaInductiveConstructor{∷}}\AgdaSpace{}%
\AgdaInductiveConstructor{false}\AgdaSpace{}%
\AgdaOperator{\AgdaInductiveConstructor{∷}}\AgdaSpace{}%
\AgdaInductiveConstructor{[]}\<%
\\
\>[0]\AgdaFunction{treeswᵒ}\AgdaSpace{}%
\AgdaSymbol{.}\AgdaField{snd}\AgdaSpace{}%
\AgdaSymbol{(}\AgdaInductiveConstructor{inr}\AgdaSpace{}%
\AgdaBound{a}\AgdaSymbol{)}\AgdaSpace{}%
\AgdaSymbol{=}\AgdaSpace{}%
\AgdaInductiveConstructor{[]}\<%
\\
\\[\AgdaEmptyExtraSkip]%
\\[\AgdaEmptyExtraSkip]%
\>[0]\AgdaFunction{BTree}\AgdaSpace{}%
\AgdaSymbol{:}\AgdaSpace{}%
\AgdaSymbol{(}\AgdaBound{A}\AgdaSpace{}%
\AgdaSymbol{:}\AgdaSpace{}%
\AgdaDatatype{ltyp}\AgdaSpace{}%
\AgdaGeneralizable{ℓ}\AgdaSymbol{)}\AgdaSpace{}%
\AgdaSymbol{→}\AgdaSpace{}%
\AgdaDatatype{ltyp}\AgdaSpace{}%
\AgdaGeneralizable{ℓ}\<%
\\
\>[0]\AgdaFunction{BTree}\AgdaSpace{}%
\AgdaBound{A}\AgdaSpace{}%
\AgdaSymbol{=}\AgdaSpace{}%
\AgdaInductiveConstructor{Wᵒ}\AgdaSpace{}%
\AgdaSymbol{(}\AgdaInductiveConstructor{⊤ᵒ}\AgdaSpace{}%
\AgdaOperator{\AgdaFunction{⊕}}\AgdaSpace{}%
\AgdaBound{A}\AgdaSymbol{)}\AgdaSpace{}%
\AgdaNumber{1}\AgdaSpace{}%
\AgdaFunction{treeswᵒ}\<%
\\
\\[\AgdaEmptyExtraSkip]%
\>[0]\AgdaFunction{leafs}\AgdaSpace{}%
\AgdaSymbol{:}\AgdaSpace{}%
\AgdaSymbol{\{}\AgdaBound{A}\AgdaSpace{}%
\AgdaSymbol{:}\AgdaSpace{}%
\AgdaDatatype{ltyp}\AgdaSpace{}%
\AgdaGeneralizable{ℓ}\AgdaSymbol{\}}\AgdaSpace{}%
\AgdaSymbol{→}\AgdaSpace{}%
\AgdaFunction{BTree}\AgdaSpace{}%
\AgdaBound{A}\AgdaSpace{}%
\AgdaOperator{\AgdaFunction{．}}\AgdaSpace{}%
\AgdaSymbol{→}\AgdaSpace{}%
\AgdaDatatype{ℕ}\<%
\\
\>[0]\AgdaFunction{leafs}\AgdaSpace{}%
\AgdaSymbol{(}\AgdaInductiveConstructor{sup}\AgdaSpace{}%
\AgdaSymbol{(}\AgdaInductiveConstructor{inl}\AgdaSpace{}%
\AgdaSymbol{\AgdaUnderscore{})}\AgdaSpace{}%
\AgdaBound{f}\AgdaSymbol{)}\AgdaSpace{}%
\AgdaSymbol{=}\AgdaSpace{}%
\AgdaFunction{leafs}\AgdaSpace{}%
\AgdaSymbol{(}\AgdaBound{f}\AgdaSpace{}%
\AgdaInductiveConstructor{true}\AgdaSymbol{)}\AgdaSpace{}%
\AgdaOperator{\AgdaPrimitive{+}}\AgdaSpace{}%
\AgdaFunction{leafs}\AgdaSpace{}%
\AgdaSymbol{(}\AgdaBound{f}\AgdaSpace{}%
\AgdaInductiveConstructor{false}\AgdaSymbol{)}\<%
\\
\>[0]\AgdaFunction{leafs}\AgdaSpace{}%
\AgdaSymbol{(}\AgdaInductiveConstructor{sup}\AgdaSpace{}%
\AgdaSymbol{(}\AgdaInductiveConstructor{inr}\AgdaSpace{}%
\AgdaBound{a}\AgdaSymbol{)}\AgdaSpace{}%
\AgdaBound{x}\AgdaSymbol{)}\AgdaSpace{}%
\AgdaSymbol{=}\AgdaSpace{}%
\AgdaNumber{1}\<%
\\
\\[\AgdaEmptyExtraSkip]%
\\[\AgdaEmptyExtraSkip]%
\\[\AgdaEmptyExtraSkip]%
\>[0]\AgdaKeyword{module}\AgdaSpace{}%
\AgdaModule{\AgdaUnderscore{}}\<%
\\
\>[0][@{}l@{\AgdaIndent{0}}]%
\>[2]\AgdaSymbol{\{}\AgdaBound{A}\AgdaSpace{}%
\AgdaSymbol{:}\AgdaSpace{}%
\AgdaDatatype{ltyp}\AgdaSpace{}%
\AgdaGeneralizable{ℓ}\AgdaSymbol{\}}\<%
\\
\>[2]\AgdaSymbol{\{}\AgdaBound{C}\AgdaSpace{}%
\AgdaSymbol{:}\AgdaSpace{}%
\AgdaDatatype{ltyp}\AgdaSpace{}%
\AgdaGeneralizable{ℓ}\AgdaSymbol{\}}\<%
\\
\>[2]\AgdaSymbol{\{}\AgdaBound{Δ₀}\AgdaSpace{}%
\AgdaBound{Δ₁}\AgdaSpace{}%
\AgdaSymbol{:}\AgdaSpace{}%
\AgdaDatatype{sply}\AgdaSpace{}%
\AgdaGeneralizable{ℓ}\AgdaSymbol{\}}\<%
\\
\>[2]\AgdaSymbol{\{}\AgdaBound{l}\AgdaSpace{}%
\AgdaSymbol{:}\AgdaSpace{}%
\AgdaBound{A}\AgdaSpace{}%
\AgdaOperator{\AgdaFunction{．}}\AgdaSpace{}%
\AgdaSymbol{→}\AgdaSpace{}%
\AgdaBound{C}\AgdaSpace{}%
\AgdaOperator{\AgdaFunction{．}}\AgdaSymbol{\}}\<%
\\
\>[2]\AgdaSymbol{\{}\AgdaBound{n}\AgdaSpace{}%
\AgdaSymbol{:}\AgdaSpace{}%
\AgdaBound{C}\AgdaSpace{}%
\AgdaOperator{\AgdaFunction{．}}\AgdaSpace{}%
\AgdaSymbol{→}\AgdaSpace{}%
\AgdaBound{C}\AgdaSpace{}%
\AgdaOperator{\AgdaFunction{．}}\AgdaSpace{}%
\AgdaSymbol{→}\AgdaSpace{}%
\AgdaBound{C}\AgdaSpace{}%
\AgdaOperator{\AgdaFunction{．}}\AgdaSymbol{\}}\<%
\\
\>[2]\AgdaSymbol{(}\AgdaBound{δ₀}\AgdaSpace{}%
\AgdaSymbol{:}\AgdaSpace{}%
\AgdaSymbol{\{}\AgdaBound{x}\AgdaSpace{}%
\AgdaSymbol{:}\AgdaSpace{}%
\AgdaBound{A}\AgdaSpace{}%
\AgdaOperator{\AgdaFunction{．}}\AgdaSymbol{\}}\AgdaSpace{}%
\AgdaSymbol{→}\AgdaSpace{}%
\AgdaBound{Δ₁}\AgdaSpace{}%
\AgdaOperator{\AgdaInductiveConstructor{⨾}}\AgdaSpace{}%
\AgdaBound{x}\AgdaSpace{}%
\AgdaFunction{∶}\AgdaSpace{}%
\AgdaBound{A}\AgdaSpace{}%
\AgdaOperator{\AgdaDatatype{⊩}}\AgdaSpace{}%
\AgdaBound{l}\AgdaSpace{}%
\AgdaBound{x}\AgdaSpace{}%
\AgdaFunction{∶}\AgdaSpace{}%
\AgdaBound{C}\AgdaSymbol{)}\<%
\\
\>[2]\AgdaSymbol{(}\AgdaBound{δ₁}\AgdaSpace{}%
\AgdaSymbol{:}\AgdaSpace{}%
\AgdaSymbol{\{}\AgdaBound{p}\AgdaSpace{}%
\AgdaBound{q}\AgdaSpace{}%
\AgdaSymbol{:}\AgdaSpace{}%
\AgdaBound{C}\AgdaSpace{}%
\AgdaOperator{\AgdaFunction{．}}\AgdaSymbol{\}}\AgdaSpace{}%
\AgdaSymbol{→}\AgdaSpace{}%
\AgdaBound{p}\AgdaSpace{}%
\AgdaFunction{∶}\AgdaSpace{}%
\AgdaBound{C}\AgdaSpace{}%
\AgdaOperator{\AgdaInductiveConstructor{⨾}}\AgdaSpace{}%
\AgdaBound{q}\AgdaSpace{}%
\AgdaFunction{∶}\AgdaSpace{}%
\AgdaBound{C}\AgdaSpace{}%
\AgdaOperator{\AgdaDatatype{⊩}}\AgdaSpace{}%
\AgdaBound{n}\AgdaSpace{}%
\AgdaBound{p}\AgdaSpace{}%
\AgdaBound{q}\AgdaSpace{}%
\AgdaFunction{∶}\AgdaSpace{}%
\AgdaBound{C}\AgdaSpace{}%
\AgdaSymbol{)}\<%
\\
\>[2]\AgdaKeyword{where}\<%
\\
\\[\AgdaEmptyExtraSkip]%
\>[2]\AgdaFunction{c}\AgdaSpace{}%
\AgdaSymbol{:}\AgdaSpace{}%
\AgdaSymbol{(}\AgdaBound{a}\AgdaSpace{}%
\AgdaSymbol{:}\AgdaSpace{}%
\AgdaSymbol{(}\AgdaInductiveConstructor{⊤ᵒ}\AgdaSpace{}%
\AgdaOperator{\AgdaFunction{⊕}}\AgdaSpace{}%
\AgdaBound{A}\AgdaSymbol{)}\AgdaSpace{}%
\AgdaOperator{\AgdaFunction{．}}\AgdaSymbol{)}\AgdaSpace{}%
\AgdaSymbol{→}\AgdaSpace{}%
\AgdaSymbol{(}\AgdaFunction{treeswᵒ}\AgdaSpace{}%
\AgdaSymbol{.}\AgdaField{fst}\AgdaSpace{}%
\AgdaBound{a}\AgdaSpace{}%
\AgdaOperator{\AgdaFunction{．}}\AgdaSpace{}%
\AgdaSymbol{→}\AgdaSpace{}%
\AgdaBound{C}\AgdaSpace{}%
\AgdaOperator{\AgdaFunction{．}}\AgdaSymbol{)}\AgdaSpace{}%
\AgdaSymbol{→}\AgdaSpace{}%
\AgdaBound{C}\AgdaSpace{}%
\AgdaOperator{\AgdaFunction{．}}\<%
\\
\>[2]\AgdaFunction{c}\AgdaSpace{}%
\AgdaSymbol{(}\AgdaInductiveConstructor{inl}\AgdaSpace{}%
\AgdaSymbol{\AgdaUnderscore{})}\AgdaSpace{}%
\AgdaBound{f}\AgdaSpace{}%
\AgdaSymbol{=}\AgdaSpace{}%
\AgdaBound{n}\AgdaSpace{}%
\AgdaSymbol{(}\AgdaBound{f}\AgdaSpace{}%
\AgdaInductiveConstructor{true}\AgdaSymbol{)}\AgdaSpace{}%
\AgdaSymbol{(}\AgdaBound{f}\AgdaSpace{}%
\AgdaInductiveConstructor{false}\AgdaSymbol{)}\<%
\\
\>[2]\AgdaFunction{c}\AgdaSpace{}%
\AgdaSymbol{(}\AgdaInductiveConstructor{inr}\AgdaSpace{}%
\AgdaBound{x}\AgdaSymbol{)}\AgdaSpace{}%
\AgdaBound{f}\AgdaSpace{}%
\AgdaSymbol{=}\AgdaSpace{}%
\AgdaBound{l}\AgdaSpace{}%
\AgdaBound{x}\<%
\\
\\[\AgdaEmptyExtraSkip]%
\>[2]\AgdaFunction{Δc}\AgdaSpace{}%
\AgdaSymbol{:}\AgdaSpace{}%
\AgdaSymbol{(}\AgdaInductiveConstructor{⊤ᵒ}\AgdaSpace{}%
\AgdaOperator{\AgdaFunction{⊕}}\AgdaSpace{}%
\AgdaBound{A}\AgdaSymbol{)}\AgdaSpace{}%
\AgdaOperator{\AgdaFunction{．}}\AgdaSpace{}%
\AgdaSymbol{→}\AgdaSpace{}%
\AgdaDatatype{sply}\AgdaSpace{}%
\AgdaBound{ℓ}\<%
\\
\>[2]\AgdaFunction{Δc}\AgdaSpace{}%
\AgdaSymbol{(}\AgdaInductiveConstructor{inl}\AgdaSpace{}%
\AgdaSymbol{\AgdaUnderscore{})}\AgdaSpace{}%
\AgdaSymbol{=}\AgdaSpace{}%
\AgdaInductiveConstructor{◇}\<%
\\
\>[2]\AgdaFunction{Δc}\AgdaSpace{}%
\AgdaSymbol{(}\AgdaInductiveConstructor{inr}\AgdaSpace{}%
\AgdaBound{x}\AgdaSymbol{)}\AgdaSpace{}%
\AgdaSymbol{=}\AgdaSpace{}%
\AgdaBound{Δ₁}\<%
\\
\\[\AgdaEmptyExtraSkip]%
\\[\AgdaEmptyExtraSkip]%
\>[2]\AgdaFunction{δm}\AgdaSpace{}%
\AgdaSymbol{:}%
\>[8]\AgdaSymbol{\{}\AgdaBound{x}\AgdaSpace{}%
\AgdaSymbol{:}\AgdaSpace{}%
\AgdaSymbol{(}\AgdaInductiveConstructor{⊤ᵒ}\AgdaSpace{}%
\AgdaOperator{\AgdaFunction{⊕}}\AgdaSpace{}%
\AgdaBound{A}\AgdaSymbol{)}\AgdaSpace{}%
\AgdaOperator{\AgdaFunction{．}}\AgdaSymbol{\}}\AgdaSpace{}%
\AgdaSymbol{\{}\AgdaBound{g}\AgdaSpace{}%
\AgdaSymbol{:}\AgdaSpace{}%
\AgdaFunction{treeswᵒ}\AgdaSpace{}%
\AgdaSymbol{.}\AgdaField{fst}\AgdaSpace{}%
\AgdaBound{x}\AgdaSpace{}%
\AgdaOperator{\AgdaFunction{．}}\AgdaSpace{}%
\AgdaSymbol{→}\AgdaSpace{}%
\AgdaBound{C}\AgdaSpace{}%
\AgdaOperator{\AgdaFunction{．}}\AgdaSymbol{\}}\<%
\\
\>[2][@{}l@{\AgdaIndent{0}}]%
\>[4]\AgdaSymbol{→}\AgdaSpace{}%
\AgdaFunction{lin}\AgdaSpace{}%
\AgdaSymbol{(}\AgdaInductiveConstructor{⊤ᵒ}\AgdaSpace{}%
\AgdaOperator{\AgdaFunction{⊕}}\AgdaSpace{}%
\AgdaBound{A}\AgdaSymbol{)}\AgdaSpace{}%
\AgdaBound{x}\AgdaSpace{}%
\AgdaOperator{\AgdaFunction{\textasciicircum{}}}\AgdaSpace{}%
\AgdaNumber{1}\AgdaSpace{}%
\AgdaOperator{\AgdaInductiveConstructor{⨾}}\<%
\\
\>[4]\AgdaFunction{Δc}\AgdaSpace{}%
\AgdaBound{x}\AgdaSpace{}%
\AgdaOperator{\AgdaInductiveConstructor{⨾}}\<%
\\
\>[4]\AgdaFunction{⨂}\AgdaSpace{}%
\AgdaSymbol{(}\AgdaFunction{treeswᵒ}\AgdaSpace{}%
\AgdaSymbol{.}\AgdaField{fst}\AgdaSpace{}%
\AgdaBound{x}\AgdaSymbol{)}\AgdaSpace{}%
\AgdaSymbol{(}\AgdaFunction{treeswᵒ}\AgdaSpace{}%
\AgdaSymbol{.}\AgdaField{snd}\AgdaSpace{}%
\AgdaBound{x}\AgdaSymbol{)}\AgdaSpace{}%
\AgdaSymbol{(λ}\AgdaSpace{}%
\AgdaBound{y}\AgdaSpace{}%
\AgdaSymbol{→}\AgdaSpace{}%
\AgdaOperator{\AgdaFunction{⟦}}\AgdaSpace{}%
\AgdaBound{C}\AgdaSpace{}%
\AgdaOperator{\AgdaFunction{⟧}}\AgdaSpace{}%
\AgdaSymbol{.}\AgdaField{snd}\AgdaSpace{}%
\AgdaSymbol{(}\AgdaBound{g}\AgdaSpace{}%
\AgdaBound{y}\AgdaSymbol{))}\<%
\\
\>[4]\AgdaOperator{\AgdaDatatype{⊩}}\AgdaSpace{}%
\AgdaFunction{c}\AgdaSpace{}%
\AgdaBound{x}\AgdaSpace{}%
\AgdaBound{g}\AgdaSpace{}%
\AgdaFunction{∶}\AgdaSpace{}%
\AgdaBound{C}\<%
\\
\>[2]\AgdaFunction{δm}\AgdaSpace{}%
\AgdaSymbol{\{}\AgdaArgument{x}\AgdaSpace{}%
\AgdaSymbol{=}\AgdaSpace{}%
\AgdaInductiveConstructor{inl}\AgdaSpace{}%
\AgdaSymbol{\AgdaUnderscore{}\}}\AgdaSpace{}%
\AgdaSymbol{\{}\AgdaBound{g}\AgdaSymbol{\}}\AgdaSpace{}%
\AgdaSymbol{=}\AgdaSpace{}%
\AgdaBound{δ₁}\AgdaSpace{}%
\AgdaSymbol{\{}\AgdaBound{g}\AgdaSpace{}%
\AgdaInductiveConstructor{true}\AgdaSymbol{\}}\AgdaSpace{}%
\AgdaSymbol{\{}\AgdaBound{g}\AgdaSpace{}%
\AgdaInductiveConstructor{false}\AgdaSymbol{\}}\AgdaSpace{}%
\AgdaOperator{\AgdaInductiveConstructor{∘}}\AgdaSpace{}%
\AgdaSymbol{(}\AgdaInductiveConstructor{id}\AgdaSpace{}%
\AgdaSymbol{\AgdaUnderscore{}}\AgdaSpace{}%
\AgdaOperator{\AgdaInductiveConstructor{⨾ᶠ}}\AgdaSpace{}%
\AgdaInductiveConstructor{unitr}\AgdaSpace{}%
\AgdaSymbol{\AgdaUnderscore{}}\AgdaSpace{}%
\AgdaOperator{\AgdaInductiveConstructor{∘}}\AgdaSpace{}%
\AgdaFunction{unitl}\AgdaSpace{}%
\AgdaSymbol{\AgdaUnderscore{}}\AgdaSpace{}%
\AgdaOperator{\AgdaInductiveConstructor{∘}}\AgdaSpace{}%
\AgdaSymbol{(}\AgdaFunction{unitl}\AgdaSpace{}%
\AgdaSymbol{\AgdaUnderscore{}}\AgdaSpace{}%
\AgdaOperator{\AgdaInductiveConstructor{∘}}\AgdaSpace{}%
\AgdaInductiveConstructor{unitr}\AgdaSpace{}%
\AgdaSymbol{\AgdaUnderscore{})}\AgdaSpace{}%
\AgdaOperator{\AgdaInductiveConstructor{⨾ᶠ}}\AgdaSpace{}%
\AgdaInductiveConstructor{id}\AgdaSpace{}%
\AgdaSymbol{\AgdaUnderscore{}}\AgdaSpace{}%
\AgdaOperator{\AgdaInductiveConstructor{∘}}\AgdaSpace{}%
\AgdaInductiveConstructor{id}\AgdaSpace{}%
\AgdaSymbol{\AgdaUnderscore{}}\AgdaSpace{}%
\AgdaOperator{\AgdaInductiveConstructor{⨾ᶠ}}\AgdaSpace{}%
\AgdaFunction{unitl}\AgdaSpace{}%
\AgdaSymbol{\AgdaUnderscore{})}\<%
\\
\>[2]\AgdaFunction{δm}\AgdaSpace{}%
\AgdaSymbol{\{}\AgdaArgument{x}\AgdaSpace{}%
\AgdaSymbol{=}\AgdaSpace{}%
\AgdaInductiveConstructor{inr}\AgdaSpace{}%
\AgdaBound{x}\AgdaSymbol{\}}\AgdaSpace{}%
\AgdaSymbol{\{}\AgdaBound{g}\AgdaSymbol{\}}\AgdaSpace{}%
\AgdaSymbol{=}\AgdaSpace{}%
\AgdaBound{δ₀}\AgdaSpace{}%
\AgdaOperator{\AgdaInductiveConstructor{∘}}\AgdaSpace{}%
\AgdaSymbol{(}\AgdaInductiveConstructor{swap}\AgdaSpace{}%
\AgdaSymbol{\AgdaUnderscore{}}\AgdaSpace{}%
\AgdaSymbol{\AgdaUnderscore{}}\AgdaSpace{}%
\AgdaOperator{\AgdaInductiveConstructor{∘}}\AgdaSpace{}%
\AgdaFunction{unitl}\AgdaSpace{}%
\AgdaSymbol{\AgdaUnderscore{}}\AgdaSpace{}%
\AgdaOperator{\AgdaInductiveConstructor{⨾ᶠ}}\AgdaSpace{}%
\AgdaInductiveConstructor{id}\AgdaSpace{}%
\AgdaSymbol{\AgdaUnderscore{}}\AgdaSpace{}%
\AgdaOperator{\AgdaInductiveConstructor{∘}}\AgdaSpace{}%
\AgdaInductiveConstructor{unitr}\AgdaSpace{}%
\AgdaSymbol{\AgdaUnderscore{}}\AgdaSpace{}%
\AgdaOperator{\AgdaInductiveConstructor{⨾ᶠ}}\AgdaSpace{}%
\AgdaInductiveConstructor{id}\AgdaSpace{}%
\AgdaSymbol{\AgdaUnderscore{}}\AgdaSpace{}%
\AgdaOperator{\AgdaInductiveConstructor{∘}}\AgdaSpace{}%
\AgdaInductiveConstructor{id}\AgdaSpace{}%
\AgdaSymbol{\AgdaUnderscore{}}\AgdaSpace{}%
\AgdaOperator{\AgdaInductiveConstructor{⨾ᶠ}}\AgdaSpace{}%
\AgdaInductiveConstructor{unitr}\AgdaSpace{}%
\AgdaSymbol{\AgdaUnderscore{})}\<%
\\
\\[\AgdaEmptyExtraSkip]%
\>[2]\AgdaFunction{lemma}\AgdaSpace{}%
\AgdaSymbol{:}\AgdaSpace{}%
\AgdaSymbol{(}\AgdaBound{a}\AgdaSpace{}%
\AgdaSymbol{:}\AgdaSpace{}%
\AgdaSymbol{(}\AgdaInductiveConstructor{⊤ᵒ}\AgdaSpace{}%
\AgdaOperator{\AgdaFunction{⊕}}\AgdaSpace{}%
\AgdaBound{A}\AgdaSymbol{)}\AgdaSpace{}%
\AgdaOperator{\AgdaFunction{．}}\AgdaSymbol{)}\<%
\\
\>[2][@{}l@{\AgdaIndent{0}}]%
\>[4]\AgdaSymbol{(}\AgdaBound{f}\AgdaSpace{}%
\AgdaSymbol{:}\AgdaSpace{}%
\AgdaFunction{treeswᵒ}\AgdaSpace{}%
\AgdaSymbol{.}\AgdaField{fst}\AgdaSpace{}%
\AgdaBound{a}\AgdaSpace{}%
\AgdaOperator{\AgdaFunction{．}}\AgdaSpace{}%
\AgdaSymbol{→}\AgdaSpace{}%
\AgdaDatatype{W}\AgdaSpace{}%
\AgdaSymbol{((}\AgdaInductiveConstructor{⊤ᵒ}\AgdaSpace{}%
\AgdaOperator{\AgdaFunction{⊕}}\AgdaSpace{}%
\AgdaBound{A}\AgdaSymbol{)}\AgdaSpace{}%
\AgdaOperator{\AgdaFunction{．}}\AgdaSymbol{)}\AgdaSpace{}%
\AgdaSymbol{(λ}\AgdaSpace{}%
\AgdaBound{x}\AgdaSpace{}%
\AgdaSymbol{→}\AgdaSpace{}%
\AgdaFunction{treeswᵒ}\AgdaSpace{}%
\AgdaSymbol{.}\AgdaField{fst}\AgdaSpace{}%
\AgdaBound{x}\AgdaSpace{}%
\AgdaOperator{\AgdaFunction{．}}\AgdaSymbol{))}\AgdaSpace{}%
\AgdaSymbol{→}\<%
\\
\>[4]\AgdaSymbol{((}\AgdaBound{b}\AgdaSpace{}%
\AgdaSymbol{:}\AgdaSpace{}%
\AgdaFunction{treeswᵒ}\AgdaSpace{}%
\AgdaSymbol{.}\AgdaField{fst}\AgdaSpace{}%
\AgdaBound{a}\AgdaSpace{}%
\AgdaOperator{\AgdaFunction{．}}\AgdaSymbol{)}\AgdaSpace{}%
\AgdaSymbol{→}\<%
\\
\>[4]\AgdaBound{Δ₁}\AgdaSpace{}%
\AgdaOperator{\AgdaFunction{\textasciicircum{}}}\AgdaSpace{}%
\AgdaFunction{leafs}\AgdaSpace{}%
\AgdaSymbol{(}\AgdaBound{f}\AgdaSpace{}%
\AgdaBound{b}\AgdaSymbol{)}\AgdaSpace{}%
\AgdaOperator{\AgdaDatatype{⊩}}\<%
\\
\>[4]\AgdaFunction{recW}\AgdaSpace{}%
\AgdaSymbol{(λ}\AgdaSpace{}%
\AgdaBound{a₁}\AgdaSpace{}%
\AgdaBound{Δ'}\AgdaSpace{}%
\AgdaSymbol{→}\AgdaSpace{}%
\AgdaFunction{Δc}\AgdaSpace{}%
\AgdaBound{a₁}\AgdaSpace{}%
\AgdaOperator{\AgdaInductiveConstructor{⨾}}\AgdaSpace{}%
\AgdaFunction{⨂}\AgdaSpace{}%
\AgdaSymbol{(}\AgdaFunction{treeswᵒ}\AgdaSpace{}%
\AgdaSymbol{.}\AgdaField{fst}\AgdaSpace{}%
\AgdaBound{a₁}\AgdaSymbol{)}\AgdaSpace{}%
\AgdaSymbol{(}\AgdaFunction{treeswᵒ}\AgdaSpace{}%
\AgdaSymbol{.}\AgdaField{snd}\AgdaSpace{}%
\AgdaBound{a₁}\AgdaSymbol{)}\AgdaSpace{}%
\AgdaBound{Δ'}\AgdaSymbol{)}\<%
\\
\>[4]\AgdaSymbol{(}\AgdaBound{f}\AgdaSpace{}%
\AgdaBound{b}\AgdaSymbol{))}\AgdaSpace{}%
\AgdaSymbol{→}\<%
\\
\>[4]\AgdaBound{Δ₁}\AgdaSpace{}%
\AgdaOperator{\AgdaFunction{\textasciicircum{}}}\AgdaSpace{}%
\AgdaFunction{leafs}\AgdaSpace{}%
\AgdaSymbol{(}\AgdaInductiveConstructor{sup}\AgdaSpace{}%
\AgdaBound{a}\AgdaSpace{}%
\AgdaBound{f}\AgdaSymbol{)}\AgdaSpace{}%
\AgdaOperator{\AgdaDatatype{⊩}}\<%
\\
\>[4]\AgdaFunction{recW}\AgdaSpace{}%
\AgdaSymbol{(λ}\AgdaSpace{}%
\AgdaBound{a₁}\AgdaSpace{}%
\AgdaBound{Δ'}\AgdaSpace{}%
\AgdaSymbol{→}\AgdaSpace{}%
\AgdaFunction{Δc}\AgdaSpace{}%
\AgdaBound{a₁}\AgdaSpace{}%
\AgdaOperator{\AgdaInductiveConstructor{⨾}}\AgdaSpace{}%
\AgdaFunction{⨂}\AgdaSpace{}%
\AgdaSymbol{(}\AgdaFunction{treeswᵒ}\AgdaSpace{}%
\AgdaSymbol{.}\AgdaField{fst}\AgdaSpace{}%
\AgdaBound{a₁}\AgdaSymbol{)}\AgdaSpace{}%
\AgdaSymbol{(}\AgdaFunction{treeswᵒ}\AgdaSpace{}%
\AgdaSymbol{.}\AgdaField{snd}\AgdaSpace{}%
\AgdaBound{a₁}\AgdaSymbol{)}\AgdaSpace{}%
\AgdaBound{Δ'}\AgdaSymbol{)}\<%
\\
\>[4]\AgdaSymbol{(}\AgdaInductiveConstructor{sup}\AgdaSpace{}%
\AgdaBound{a}\AgdaSpace{}%
\AgdaBound{f}\AgdaSymbol{)}\<%
\\
\>[2]\AgdaFunction{lemma}\AgdaSpace{}%
\AgdaSymbol{(}\AgdaInductiveConstructor{inl}\AgdaSpace{}%
\AgdaSymbol{\AgdaUnderscore{})}\AgdaSpace{}%
\AgdaBound{f}\AgdaSpace{}%
\AgdaBound{g}%
\>[21]\AgdaSymbol{=}\AgdaSpace{}%
\AgdaFunction{unitl'}\AgdaSpace{}%
\AgdaSymbol{\AgdaUnderscore{}}\AgdaSpace{}%
\AgdaOperator{\AgdaInductiveConstructor{∘}}\AgdaSpace{}%
\AgdaSymbol{(}\AgdaInductiveConstructor{id}\AgdaSpace{}%
\AgdaSymbol{\AgdaUnderscore{}}\AgdaSpace{}%
\AgdaOperator{\AgdaInductiveConstructor{⨾ᶠ}}\AgdaSpace{}%
\AgdaInductiveConstructor{unitr'}\AgdaSpace{}%
\AgdaSymbol{\AgdaUnderscore{}}\AgdaSpace{}%
\AgdaOperator{\AgdaInductiveConstructor{∘}}\AgdaSpace{}%
\AgdaSymbol{(}\AgdaSpace{}%
\AgdaBound{g}\AgdaSpace{}%
\AgdaInductiveConstructor{true}\AgdaSpace{}%
\AgdaOperator{\AgdaInductiveConstructor{⨾ᶠ}}\AgdaSpace{}%
\AgdaBound{g}\AgdaSpace{}%
\AgdaInductiveConstructor{false}\AgdaSpace{}%
\AgdaOperator{\AgdaInductiveConstructor{∘}}\AgdaSpace{}%
\AgdaFunction{splyadd}\AgdaSymbol{))}\<%
\\
\>[2]\AgdaFunction{lemma}\AgdaSpace{}%
\AgdaSymbol{(}\AgdaInductiveConstructor{inr}\AgdaSpace{}%
\AgdaBound{x}\AgdaSymbol{)}\AgdaSpace{}%
\AgdaBound{f}\AgdaSpace{}%
\AgdaBound{g}\AgdaSpace{}%
\AgdaSymbol{=}\AgdaSpace{}%
\AgdaInductiveConstructor{id}\AgdaSpace{}%
\AgdaSymbol{\AgdaUnderscore{}}\<%
\\
\\[\AgdaEmptyExtraSkip]%
\>[2]\AgdaFunction{recBTreeᵒ}\AgdaSpace{}%
\AgdaSymbol{:}\AgdaSpace{}%
\AgdaSymbol{\{}\AgdaBound{t}\AgdaSpace{}%
\AgdaSymbol{:}\AgdaSpace{}%
\AgdaFunction{BTree}\AgdaSpace{}%
\AgdaBound{A}\AgdaSpace{}%
\AgdaOperator{\AgdaFunction{．}}\AgdaSymbol{\}}\AgdaSpace{}%
\AgdaSymbol{→}\<%
\\
\>[2][@{}l@{\AgdaIndent{0}}]%
\>[4]\AgdaSymbol{(}\AgdaBound{Δ₀}\AgdaSpace{}%
\AgdaOperator{\AgdaDatatype{⊩}}\AgdaSpace{}%
\AgdaBound{t}\AgdaSpace{}%
\AgdaFunction{∶}\AgdaSpace{}%
\AgdaFunction{BTree}\AgdaSpace{}%
\AgdaBound{A}\AgdaSymbol{)}\<%
\\
\>[4]\AgdaSymbol{→}\AgdaSpace{}%
\AgdaBound{Δ₀}\AgdaSpace{}%
\AgdaOperator{\AgdaInductiveConstructor{⨾}}\AgdaSpace{}%
\AgdaBound{Δ₁}\AgdaSpace{}%
\AgdaOperator{\AgdaFunction{\textasciicircum{}}}\AgdaSpace{}%
\AgdaFunction{leafs}\AgdaSpace{}%
\AgdaBound{t}\AgdaSpace{}%
\AgdaOperator{\AgdaDatatype{⊩}}\AgdaSpace{}%
\AgdaFunction{recW}\AgdaSpace{}%
\AgdaFunction{c}\AgdaSpace{}%
\AgdaBound{t}\AgdaSpace{}%
\AgdaFunction{∶}\AgdaSpace{}%
\AgdaBound{C}\<%
\\
\>[2]\AgdaFunction{recBTreeᵒ}\AgdaSpace{}%
\AgdaSymbol{\{}\AgdaBound{w}\AgdaSymbol{\}}\AgdaSpace{}%
\AgdaBound{δ}\AgdaSpace{}%
\AgdaSymbol{=}\AgdaSpace{}%
\AgdaFunction{recWᵒ}\AgdaSpace{}%
\AgdaSymbol{\{}\AgdaArgument{A}\AgdaSpace{}%
\AgdaSymbol{=}\AgdaSpace{}%
\AgdaInductiveConstructor{⊤ᵒ}\AgdaSpace{}%
\AgdaOperator{\AgdaFunction{⊕}}\AgdaSpace{}%
\AgdaBound{A}\AgdaSymbol{\}}\AgdaSpace{}%
\AgdaSymbol{\{}\AgdaNumber{1}\AgdaSymbol{\}}\AgdaSpace{}%
\AgdaSymbol{\{}\AgdaFunction{treeswᵒ}\AgdaSymbol{\}}\AgdaSpace{}%
\AgdaSymbol{\{}\AgdaBound{C}\AgdaSymbol{\}}\AgdaSpace{}%
\AgdaSymbol{\{}\AgdaFunction{c}\AgdaSymbol{\}}\AgdaSpace{}%
\AgdaSymbol{\{}\AgdaArgument{p}\AgdaSpace{}%
\AgdaSymbol{=}\AgdaSpace{}%
\AgdaBound{w}\AgdaSymbol{\}}\AgdaSpace{}%
\AgdaFunction{δm}\AgdaSpace{}%
\AgdaBound{δ}\AgdaSpace{}%
\AgdaOperator{\AgdaInductiveConstructor{∘}}\AgdaSpace{}%
\AgdaInductiveConstructor{id}\AgdaSpace{}%
\AgdaSymbol{\AgdaUnderscore{}}\AgdaSpace{}%
\AgdaOperator{\AgdaInductiveConstructor{⨾ᶠ}}\AgdaSpace{}%
\AgdaFunction{elimW}\AgdaSpace{}%
\AgdaFunction{lemma}\AgdaSpace{}%
\AgdaBound{w}\<%
\\
\\[\AgdaEmptyExtraSkip]%
\\[\AgdaEmptyExtraSkip]%
\>[2]\AgdaComment{--\ for\ type-setting:}\<%
\\
\>[2]\AgdaFunction{recBTree}\AgdaSpace{}%
\AgdaSymbol{:}\AgdaSpace{}%
\AgdaSymbol{(}\AgdaBound{C}\AgdaSpace{}%
\AgdaOperator{\AgdaFunction{．}}\AgdaSpace{}%
\AgdaSymbol{→}\AgdaSpace{}%
\AgdaBound{C}\AgdaSpace{}%
\AgdaOperator{\AgdaFunction{．}}\AgdaSpace{}%
\AgdaSymbol{→}\AgdaSpace{}%
\AgdaBound{C}\AgdaSpace{}%
\AgdaOperator{\AgdaFunction{．}}\AgdaSymbol{)}\AgdaSpace{}%
\AgdaSymbol{→}\AgdaSpace{}%
\AgdaSymbol{(}\AgdaBound{A}\AgdaSpace{}%
\AgdaOperator{\AgdaFunction{．}}\AgdaSpace{}%
\AgdaSymbol{→}\AgdaSpace{}%
\AgdaBound{C}\AgdaSpace{}%
\AgdaOperator{\AgdaFunction{．}}\AgdaSymbol{)}\AgdaSpace{}%
\AgdaSymbol{→}\AgdaSpace{}%
\AgdaFunction{BTree}\AgdaSpace{}%
\AgdaBound{A}\AgdaSpace{}%
\AgdaOperator{\AgdaFunction{．}}\AgdaSpace{}%
\AgdaSymbol{→}\AgdaSpace{}%
\AgdaBound{C}\AgdaSpace{}%
\AgdaOperator{\AgdaFunction{．}}\<%
\\
\>[2]\AgdaFunction{recBTree}\AgdaSpace{}%
\AgdaBound{fork}\AgdaSpace{}%
\AgdaBound{leaf}\AgdaSpace{}%
\AgdaSymbol{=}\AgdaSpace{}%
\AgdaFunction{recW}\AgdaSpace{}%
\AgdaFunction{c}\<%
\\
\\[\AgdaEmptyExtraSkip]%
\>[2]\AgdaFunction{recBTreeᵒ'}\AgdaSpace{}%
\AgdaSymbol{:}\AgdaSpace{}%
\AgdaSymbol{\{}\AgdaBound{t}\AgdaSpace{}%
\AgdaSymbol{:}\AgdaSpace{}%
\AgdaFunction{BTree}\AgdaSpace{}%
\AgdaBound{A}\AgdaSpace{}%
\AgdaOperator{\AgdaFunction{．}}\AgdaSpace{}%
\AgdaSymbol{\}}\AgdaSpace{}%
\AgdaSymbol{→}\<%
\end{code}
\newcommand{\AGDArecBTree}{%
\begin{code}%
\>[2][@{}l@{\AgdaIndent{1}}]%
\>[6]\AgdaSymbol{(}\AgdaBound{Δ₀}\AgdaSpace{}%
\AgdaOperator{\AgdaDatatype{⊩}}\AgdaSpace{}%
\AgdaBound{t}\AgdaSpace{}%
\AgdaFunction{∶}\AgdaSpace{}%
\AgdaFunction{BTree}\AgdaSpace{}%
\AgdaBound{A}\AgdaSymbol{)}\<%
\\
\>[2][@{}l@{\AgdaIndent{1}}]%
\>[4]\AgdaSymbol{→}\AgdaSpace{}%
\AgdaSymbol{(\{}\AgdaBound{y}\AgdaSpace{}%
\AgdaBound{z}\AgdaSpace{}%
\AgdaSymbol{:}\AgdaSpace{}%
\AgdaBound{C}\AgdaSpace{}%
\AgdaOperator{\AgdaFunction{．}}\AgdaSymbol{\}}\AgdaSpace{}%
\AgdaSymbol{→}\AgdaSpace{}%
\AgdaBound{y}\AgdaSpace{}%
\AgdaFunction{∶}\AgdaSpace{}%
\AgdaBound{C}\AgdaSpace{}%
\AgdaOperator{\AgdaInductiveConstructor{⨾}}\AgdaSpace{}%
\AgdaBound{z}\AgdaSpace{}%
\AgdaFunction{∶}\AgdaSpace{}%
\AgdaBound{C}\AgdaSpace{}%
\AgdaOperator{\AgdaDatatype{⊩}}\AgdaSpace{}%
\AgdaBound{n}\AgdaSpace{}%
\AgdaBound{y}\AgdaSpace{}%
\AgdaBound{z}\AgdaSpace{}%
\AgdaFunction{∶}\AgdaSpace{}%
\AgdaBound{C}\AgdaSpace{}%
\AgdaSymbol{)}\<%
\\
\>[4]\AgdaSymbol{→}\AgdaSpace{}%
\AgdaSymbol{(\{}\AgdaBound{x}\AgdaSpace{}%
\AgdaSymbol{:}\AgdaSpace{}%
\AgdaBound{A}\AgdaSpace{}%
\AgdaOperator{\AgdaFunction{．}}\AgdaSymbol{\}}\AgdaSpace{}%
\AgdaSymbol{→}\AgdaSpace{}%
\AgdaBound{Δ₁}\AgdaSpace{}%
\AgdaOperator{\AgdaInductiveConstructor{⨾}}\AgdaSpace{}%
\AgdaBound{x}\AgdaSpace{}%
\AgdaFunction{∶}\AgdaSpace{}%
\AgdaBound{A}\AgdaSpace{}%
\AgdaOperator{\AgdaDatatype{⊩}}\AgdaSpace{}%
\AgdaBound{l}\AgdaSpace{}%
\AgdaBound{x}\AgdaSpace{}%
\AgdaFunction{∶}\AgdaSpace{}%
\AgdaBound{C}\AgdaSymbol{)}\<%
\\
\>[4]\AgdaSymbol{→}\AgdaSpace{}%
\AgdaBound{Δ₀}\AgdaSpace{}%
\AgdaOperator{\AgdaInductiveConstructor{⨾}}\AgdaSpace{}%
\AgdaBound{Δ₁}\AgdaSpace{}%
\AgdaOperator{\AgdaFunction{\textasciicircum{}}}\AgdaSpace{}%
\AgdaFunction{leafs}\AgdaSpace{}%
\AgdaBound{t}\AgdaSpace{}%
\AgdaOperator{\AgdaDatatype{⊩}}\AgdaSpace{}%
\AgdaFunction{recBTree}\AgdaSpace{}%
\AgdaBound{n}\AgdaSpace{}%
\AgdaBound{l}\AgdaSpace{}%
\AgdaBound{t}\AgdaSpace{}%
\AgdaFunction{∶}\AgdaSpace{}%
\AgdaBound{C}\<%
\end{code}}
\begin{code}[hide]%
\>[2]\AgdaFunction{recBTreeᵒ'}\AgdaSpace{}%
\AgdaSymbol{\{}\AgdaBound{t}\AgdaSymbol{\}}\AgdaSpace{}%
\AgdaBound{x}\AgdaSpace{}%
\AgdaBound{x₁}\AgdaSpace{}%
\AgdaBound{x₂}\AgdaSpace{}%
\AgdaSymbol{=}\AgdaSpace{}%
\AgdaFunction{recBTreeᵒ}\AgdaSpace{}%
\AgdaSymbol{\{}\AgdaBound{t}\AgdaSymbol{\}}\AgdaSpace{}%
\AgdaBound{x}\<%
\\
\\[\AgdaEmptyExtraSkip]%
\>[0]\<%
\end{code}

\newcommand{\fcmp}{\mathrel{,}}
\newcommand{\mdlgwhtdiamond}{\diamond}
\newcommand{\mathratio}{\mathrel{:}}

\begin{document}

\title{Dependent Multiplicities in Dependent Linear Type
  Theory}

\author{Maximilian Doré}
\email{maximilian.dore@cs.ox.ac.uk}
\orcid{0000-0003-4012-6401}

\begin{abstract}
We present a novel dependent linear type theory in which the multiplicity of
some variable---i.e., the number of times the variable can be used in a
program---can depend on other variables. This allows us to give precise
resource annotations to programs involving branching and recursion that cannot
be adequately typed in other theories. Our type system is obtained by
embedding linear logic into dependent type theory and specifying how the
embedded logic interacts with the host theory. We can then use the natural
numbers of the dependent type theory to derive a quantitative typing system
with dependent multiplicities. Our theory supports W-types, thereby giving a
principled resource-aware treatment of a large class of inductive types. We
characterise the semantics as Categories with Families indexed in symmetric
monoidal categories, thereby generalising Quantitative Categories with
Families. Existing dependently typed languages can easily be extended with our
system, which we demonstrate with an implementation in Agda.
\end{abstract}



\received{20 February 2007}
\received[revised]{12 March 2009}
\received[accepted]{5 June 2009}

\maketitle

\section{Introduction}
\label{sec:intro}

Girard's linear logic \citep{girard87_linear_logic} has seen exciting and
diverse applications in computer science, ranging from concurrency
\citep{caires10_session_types_intuit_linear_propos} over quantum programming
languages \citep{abramsky06_categ_quant_logic} to complexity theory
\citep{baillot10, dal11_linear_depen_types_relat_compl}. When we view
propositions as types, linear logic gives rise to a programming language whose
type system ensures that any program uses exactly the resources it received. A
natural extension of linear logic is to not only allow single usage of a
variable, but to instead equip each variable with a \emph{multiplicity} which
specifies how often that variable is to be used. For example, a copying function
will have multiplicity 2 for the input variable in such quantitative type
systems. Linear Haskell~\citep{bernardy17_linear_haskel}, Quantitative Type
Theory~(QTT)~ \citep{mcbride16_i_got_plent_onutt,
  atkey18_syntax_seman_quant_type_theor} and
Graded~Modal~Type~Theory~\citep{orchard19_quant_progr_reason_with_graded_modal_types,abel23_graded_modal_depen_type_theor}
have applied this idea fruitfully to give practical programming languages in
which the type of a program specifies how all inputs are being used, giving
guidance to both the programmer who wants to be confident in the correctness of
a program and to the compiler which attempts to find efficient ways to execute
it. However, the typing disciplines of those systems quickly becomes restrictive
when we use data types that do not represent resources but instead implement
program logic and organise the actual resources (which might, e.g., be values on
the heap, file handlers, messages passed along some channel, etc.
\citep{brady21_idris,reinking21_perceus}). Consider the booleans, which has the
following elimination principle in such a linear or quantitative system.%
\begin{mathpar}
  \inferrule*{
    \isof.b:\Bool \and
    \isof<\Delta>.x:A \and
    \isof<\Delta>.y:A
  }{ \isof<\Delta>.\elimBool{b}{x}{y}:{A} }
\end{mathpar}

Some boolean $b$, which does not require any resources, is used to decide which
of the two elements of type $A$ we return. Both elements of type $A$ have to be
constructed from the same resources, which is a steep ask: different
if-then-else branches in a program will generally do different things, and hence
use different resources.


The limitations of current resource-sensitive type theories get more present if
we introduce more features to our programming language. In a higher-order
function, how some parameter is used might depend on the values of other
parameters, making it impossible to precisely type many functions. Similarly,
inductive types introduce a dynamic usage of resource that cannot be captured.
Consider an inductive type of binary trees, $\BTree{A}$, which stores values of
type $A$ in the leaves. Suppose we want to map a function $f$ onto such trees,
i.e., a program which would have the following type in
Agda~\citep{norell07_towar}.
\begin{mathpar}
  \Fun[t]{\BTree{A}}{ \Fun[f]{ \Fun{A}{B} }{ \BTree{B}}}
\end{mathpar}

An implementation of this type should take a binary tree $t$ and invoke $f$ for
any leaf of the $t$. However, no existing linear or quantitative system allows
us to express this use of $f$. Again, the issue is that the usage of some
resource \emph{depends} on some value that is only known at runtime.

Did somebody say a type depends on some value? Enter Martin-Löf's dependent type
theory~(DTT)~\citep{martin-loef84_intuit}. DTT equips intuitionistic logic with
predicates, which allows for expressing detailed specifications in the type of a
program~\citep{martin-loef82_const_mathem_comput_progr}. It seems natural to
expect that a combination of DTT with linear logic should be able to deal with
the phenomena described above, and the promise of combining both systems has
long excited
researchers~\citep{cervesato02_linear_logic_framew,dal11_linear_depen_types_relat_compl,
  vakar15_categ_seman_linear_logic_framew,krishnaswami15_integ_linear_depen_types}.
However, work on dependent linear type theory was long plagued by an apparent
conundrum: in DTT, a type can depend on a term, does this constitute a use of
the term? The answer most researchers gave was an unemphatic ``yes'' and
disallowed dependency on linear variables, until
\citet{mcbride16_i_got_plent_onutt} argued that the answer should be ``no''
since types are just there for contemplation, and not for computation. In his
QTT, further worked out by \citet{atkey18_syntax_seman_quant_type_theor},
variables are equipped with multiplicities drawn from some resource algebra and
the structural rules of DTT are restricted to take these multiplicities into
account. While QTT provides a full combination of linearity and dependency, it
still has a crucial shortcoming: multiplicities are static elements of the
resource algebra, thereby making it impossible for some multiplicity to depend
on some other variables or to capture the runtime dynamics of a recursive
program.

In this paper we show how to take the approach of
\citeauthor{mcbride16_i_got_plent_onutt} further: we should contemplate terms
not only in types, but also in multiplicities. By combining DTT with linear
logic in a quite natural fashion, we obtain a highly expressive type system in
which we can capture all of the above described phenomena.




\paragraph*{Approach}
Our theory comes with two entailment relations $\vdash$ and $\Vdash$, where the
former incarnates a standard intuitionistic DTT, called the \emph{host theory},
while the latter is a resource-aware calculus (our setup can be thought of akin
to the 0- and 1-fragments of QTT). We cannot hypothesise over new variables in
$\Vdash$, but only repeat values that were constructeded in the $\vdash$
fragment, consequently, we call the hypotheses of $\Vdash$ \emph{supplies} (and
not ``contexts''). These supplies are subject to the standard structural rules
of linear logic. Our setup can be considered a deep embedding of linear logic
into DTT. In particular, supplies form a type in the host theory that we can
eliminate into, which allows us compute for a given natural number
$\isof[\Gamma].m:\Nat$ the $m$-fold copy of a supply $\Delta$, denoted
$\expsply{\Delta}{m}$. This operation allows us to equip variables with
multiplicities, giving rise to a quantitative type system. Since $m$ can depend
on variables in $\Gamma$, our notion of multiplicity is dynamic, allowing us to
express that a multiplicity \emph{depends} on some other variables.

We present linear versions of the usual type formers of DTT, such as functions
$\Natfun[x]{A}[m]{\tyapp{B}{x}}$ whose inputs can be used $m$ times (we will
omit $\expsply{}{m}$ in case $m$ is 1), and booleans $\Boolo$, whose eliminator
allows for different resources to be used in each branch (we write
$\booltonat{-}$ for a function which embeds $\Boolo$ into $\Nat$).%
\begin{mathpar}
  \inferrule*[]{
    \isof.b:\Boolo \and
    \isof<\Delta_0>.x:A \and
    \isof<\Delta_1>.y:A
  }{ \isof<\expsply{\Delta_0}{\booltonat{b}} \splysep \expsply{\Delta_1}{\booltonat{\neg b}}>.\elimBool{b}{x}{y}:{A} }
\end{mathpar}

Moreover, we can give precise types to many higher-order functions, in
particular those involving inductive types. We can implement a map function for
binary trees whose type precisely captures how often the mapped function is
used, where $\mathsf{leafs}~t$ computes the number of leaves in $t$.%
\begin{mathpar}
  \Natfun[t]{\BTree{A}}{ \Natfun[f]{ \Natfun{A}{B}
    }[(\mathsf{leafs}~t)]{\BTree{B}}}
\end{mathpar}

Semantics for our language can be characterised as a rather straightforward
combination of models of DTT, namely Categories with
Families~\citep{dybjer96_inter}, with models of linear logic, namely symmetric
monoidal categories; we call the resulting structure \emph{linear Categories
  with Families}. Our notion of model of dependent linear type theory has been
foreshadowed by~\citet{vakar15_categ_seman_linear_logic_framew}, the main
contribution of our semantics lies not in the novelty of the given structure,
but rather that it is enough to interpret our highly expressive syntax. In
particular, we show how our semantics justifies intricate introduction and
elimination rules for linear W-types, thereby giving a principled answer to how
inductive types can be treated in a linear system. Our notion generalises
certain Quantitative Categories with Families, which have been proposed as the
semantics for QTT \citep{atkey18_syntax_seman_quant_type_theor}, further
substantiating our claim that the presented theory is a more expressive version
of QTT. In particular, we can mimic the realisability model given by
\citet{atkey18_syntax_seman_quant_type_theor}, which justifies erasing the
$\vdash$ fragment and only keeping terms derived in $\Vdash$ for computation.

Since we do not change the structural rules of our host theory, we can
relatively easily add our system to existing dependently typed languages. We
demonstrate this with an implementation in Agda, using the semantic
interpretation of the linear judgments and type formers as a guide for a highly
rigorous (if not particularly practical) type-checker.

\paragraph{Summary}

\begin{itemize}
\item We present syntax for a novel \emph{dependent linear type theory}
  (\autoref{sec:syntax}) in which variables have \emph{dependent
    multiplicities}, allowing us to precisely type programs that cannot be typed
  in other linear or quantitative systems. Our language has function types, pair
  types and standard algebraic data types, more specifically W-types whose
  positions are finite. We give a syntactic characterisation of a class of
  \emph{clearly linear} programs which can all be precisely typed and are hence
  valid programs in our linear typing discipline (\autoref{lemma:completeness}).
\item Models of our syntax can be characterised as a rather direct combination
  of standards model of DTT and linear logic that we call \emph{linear
    Categories with Families} (\autoref{sec:semantics}); we will show how to
  interpret all type formers (Lemmas~\ref{lemma:pair}---\ref{lemma:fun}); evince
  a realisability model which demonstrates that the underlying DTT can be erased
  for computation (\autoref{exp:real}); and show how our model generalises
  Quantitative Categories with
  Families~\citep{atkey18_syntax_seman_quant_type_theor}
  (\autoref{lemma:genqtt}).
\item We show in \autoref{sec:bang} that the full power of intuitionistic logic
  can be recovered in $\Vdash$ in standard fashion with a linear exponential
  comonad ``!'' to annotate variables which can be used in unrestricted fashion
  (\autoref{lemma:bang}).
\item Since our theory is a combination of a standard DTT with an embedded
  linear logic, we can add our type system to existing languages, which we
  demonstrate with an implementation of our theory in Agda
  (\autoref{sec:implementation}).
\end{itemize}

We will close with a discussion of related (\autoref{sec:related}) and future
work (\autoref{sec:conclusions}). The accompanying artefact\footnote{\url{https://github.com/anonforlics26/dltt}} contains the
implementation of our type system in Agda with all examples, and provides a
formalisation of Lemmas~\ref{lemma:pair}---\ref{lemma:fun}.

\section{Syntax for Dependent Linear Type Theory}
\label{sec:syntax}

Our type system is based on intuitionistic type theory in the style of
\citet{martin-loef84_intuit}, which we call the \emph{host theory} or
\emph{intuitionistic} theory (the linear logic we will embed is also
non-classical and hence ``intuitionistic'', but we will use this adjective only
for the host theory). We have the standard judgements $\Gamma \isctxt$ for
$\Gamma$ being a \emph{context}; $\isof[\Gamma]{A \isityp}$ for $A$ being an
\emph{intuitionistic type} in context $\Gamma$; and $\isof[\Gamma].a:A$ for $a$
being an \emph{intuitionistic term} of type $A$ in context $\Gamma$. We assume
that we have a cumulative hierarchy of universes, for which also simply write
$\ityp$, omitting universe levels (which can be consistently assigned, as
demonstrated in the artefact).
Our theory has explicit simultaneous substitutions, where we write
$\isof[\Gamma'].\gamma:\Gamma$ for a substitution $\gamma$ which turns types and
terms in context $\Gamma$ into ones living in $\Gamma'$. The substitution
calculus is standard and we refer to \citet{angiuli25_princ_depen_type_theor}
for a modern presentation. We write $\isof[\Gamma]{a \equiv a'}$ if two terms
$\isof.{a,a'}:{A}$ are definitionally equal and $\Eq{A}{a}{a'}$ for a
propositional identity type which we assume to be intensional. Such an equality
is present in, e.g., the type theories underlying Agda~\citep{norell07_towar}
and Roqc~\citep{barras97_coq_proof_assis_refer_manual}, but also in homotopy
type theory~\citep{hott} and cubical type
theory~\citep{bezem14_model_type_theor_cubic_sets}.

We distinguish a collection of ground types, denoted $A \isgtyp$, any one of
which is also an intuitionistic type. 
Our theory comes with standard data types and type formers where we write
$\isof.\tt:\Unit$ for the single inhabitant of the unit type; $\Empty$ for the
type with no inhabitants; $\isof.{\true , \false}:\Bool$ for the booleans;
$\isof.{\pair{x}{y}}:{\Pair[x]{A}{\tyapp{B}{x}}}$ for the dependent pair type,
which we define positively with pair constructor $\pair{-}{-}$ and an eliminator
$\Pairelim[x][y]{-}{-}$; and $\isof.{\fun{x}{b}}:{\Fun[x]{A}{\tyapp{B}{x}}}$ for
the dependent function type. Moreover, our host theory comes with W-types
\citep{martin-loef84_intuit}, written $\W[x]{A}{\tyapp{B}{x}}$, where we call
$A$ the \emph{constructors} and $\tyapp{B}{x}$ the \emph{positions}. When
eliminating a given $\isof.p:{\W[x]{A}{\tyapp{B}{x}}}$ into some dependent
motive $C$ using $\elimW{afg.r}{p}$, $r$ can use the given constructor $a$,
subtree $\isof.{f}:{\Fun{\tyapp{B}{a}}{\W[x]{A}{\tyapp{B}{x}}}}$ and recursively
computed value $\isof.{g}:{\Fun[y]{\tyapp{B}{a}}{\tyapp{C}{\app{f}{y}}}}$. In
case we are eliminating into non-dependent types, we do not need the $f$ and
will simply write $\elimW{ag.r}{p}$.

We write $\zero$ and $\succ{}$ for the constructors of the natural numbers
$\Nat$, defined as the standard W-type, and $\elimNat{s}{z}{m}$ for the
eliminator of $\isof.m:{\Nat}$ with inductive case $s$ and base case $z$.
The map $\isof.{\booltonat{-}}:{\Fun{\Bool}{\Nat}}$ sends $\false$ to $0$ and
$\true$ to $1$.

\subsection{Linear Judgements and Structural Rules}
\label{ssec:syntaxstructure}

We now embed linear logic into our intuitionistic host theory by adding the
following judgements.
\begin{itemize}
\item $\isof[\Gamma]{\Delta \issply}$ asserts that $\Delta$ is a \emph{supply} in
  context $\Gamma$.
\item $\isof[\Gamma]{A \isltyp}$ asserts that $A$ is a \emph{linear type} in
  context $\Gamma$.
\item $\isof<\Delta>.a:A$ asserts that $a$ is a \emph{linear term} of linear
  type $A$ derivable from $\Delta$, where $\isof[\Gamma]{A \isltyp}$ and
  $\isof[\Gamma]{\Delta \issply}$.
\end{itemize}

As a first intuition, supplies can be thought of as ``linear contexts'', and we
will see that the standard structural rules of linear logic apply to supplies.
However, supplies will not introduce new variables, but rather arbitrary terms.
They are hence used to specify which resources are to be used in a construction,
but they themselves live in some intuitionistic context. Similarly, our linear
types live in some intuitionistic context. We embed linear supplies and types
into our host theory by stipulating that these form in fact \emph{intuitionistic
  types}. This will allow us to compute with resources annotations in our host
theory.
\begin{mathpar}
  \RULEsplyityp{}

  \RULEltypityp{}
\end{mathpar}

Note that as intuitionistic types, $\sply$ and $\ltyp$ are quite unusual since
we will not stipulate any elimination rules. Instead, we find it helpful to
think of the intuitionistic host theory as giving the universe of discourse for
our linear calculus.

Our distinction between intuitionistic types and linear types should not be
taken akin to the dichotomoy introduced by, e.g.,
\citet{cervesato02_linear_logic_framew},
\citet{vakar15_categ_seman_linear_logic_framew} or
\citet{krishnaswami15_integ_linear_depen_types}; rather, linear types should be
thought of as substructural, more fine-grained versions of the intuitionistic
types. Throughout this section will define by structural induction an operation
$\ltoi{(-)}$ on the linear type formers which turns any linear type into an
intuitionistic type, giving rise to the following admissible rule.%
\begin{mathpar}
  \inferrule*[]{
    \isof[\Gamma]{A \isltyp}
  }{
    \isof[\Gamma]{\ltoi{A} \isityp}
  }
\end{mathpar}

Similarly, we will also be able to regard any linear term as an intuitionistic
term, but we will do this translation implicitly. Whenever a term appears in a
type, it is an (implicitly translated) intuitionistic term, e.g., we might
consider $\tyapp{B}{a}$ where $a$ might have originally been a linear term. Akin
to the approach of \citet{mcbride16_i_got_plent_onutt} and
\citet{atkey18_syntax_seman_quant_type_theor}, we use the host theory to model
dependencies between linear types---i.e., a linear type $B$ depending on some
other linear type $A$ is represented by the judgement
$\isof[\isof.x:{\ltoi{A}}]{\tyapp{B}{x} \isltyp}$.

The first linear types that we introduce are the ground types. The values of
ground types are those that we actually care about as resources, while all the
type formers introduced in \autoref{ssec:syntaxtypes} only allow us to rearrange
and combine such values.
\begin{mathpar}
  \RULEltyp{}
\end{mathpar}

For ground types $A$ we define $\ltoi{A} := A$.

\paragraph*{Structural rules}

The rules governing supplies are standard for a linear calculus. The following
introduce the empty supply and the join of two supplies living in the same
intuitionistic context.
\begin{mathpar}
  \RULEemptysply{}

  \RULEjoinsply{}
\end{mathpar}

Additionally, we can create supplies which hypothesise over a variable of a
linear type, which lives in a context extended with a variable of the underlying
intuitionistic type. Using this supply we can then stipulate the \emph{variable}
rule of linear logic, sometimes also called the identity or axiom rule.
\begin{mathpar}
  \RULEvarsply{}

  \RULElvar{}
\end{mathpar}

Our host theory therefore takes care of introducing new variables, which our
supplies may incorporate.

Lastly, we introduce the central structural rule for linear logic which allows
us to reorder the assumptions of a supply.%
\begin{mathpar}
  \RULEexchange{}
\end{mathpar}

We furthermore stipulate structural rules which allow us to remove $\di$ from
non-empty contexts and treat ``$\splysep$'' as an associative operation: we have
rules that make $\isof< \Delta \splysep \di >{\mathcal{J}}$ and $\isof< \Delta
>{\mathcal{J}}$ interderivable; as well as $\isof< (\Delta_0 \splysep \Delta_1)
\splysep \Delta_2 >{\mathcal{J}}$ and $\isof< \Delta_0 \splysep (\Delta_1
\splysep \Delta_2) >{\mathcal{J}}$.


\paragraph*{Substitutions}

Our host theory has taken care of most of the non-linear structural part of our
theory, such as modelling type dependencies, and we will also just reuse the
intuitionistic substitutions for the $\Vdash$ fragment of our theory. In
particular, we can substitute a linear term for a variable using its
intuitionistic translation.

The central substitution rule for $\Vdash$ establishes that when applying a
substitution, we have to apply it to the supply, linear type and linear term at
the same time.
\begin{mathpar}
  \RULEltmsubst{}
\end{mathpar}

We have omitted rules which state that $\subst{\Delta}{\gamma}$ and
$\subst{A}{\gamma}$ also live in $\Gamma'$, and that substitutions factor
through supplies, e.g., $\subst{(\Delta_0 \splysep \Delta_1)}{\gamma} \equiv
\subst{\Delta_0}{\gamma} \splysep \subst{\Delta_1}{\gamma}$, until they are at
the level of variables where substitution is defined as usual for intuitionistic
variables.

Our substitution rule allows us to put arbitrary terms into supplies, e.g., we
can derive for any term $\isof.a:A$ that $\isof<\isof.a:A>a:A$ by substituting
$a$ for $x$ in the variable judgement $\isof<\isof.x:A>.x:A$. Supplies are hence
not ``contexts'', since they can contain arbitrary terms. Note that
substitutions might in general duplicate or drop resources---but since we apply
a substitution to both the assumption and conclusion of a linear judgement, we
maintain that $\Vdash$ captures linear derivability.
In particular, we can apply the weakening substitution of the host theory to a
linear judgement, which means $\isof.x:A$ is derivable from supply $\isof.x:A$
in context $\isof.x:{\ltoi{A}} \ctxtsep
\isof.y:{\ltoi{\tyapp{B}{x}}}$---crucially, we have only weakened the context in
which our linear derivation lives, but not the resources given by our supply.
Also note that we can reorder terms which ever way we like in a supply, e.g.,
$\isof.y:{\tyapp{B}{x}} \splysep \isof.x:A$ is a supply living in the
aforementioned context. Since our host theory takes care of variable
dependencies, we can unperturbedly apply the structural rules of linear logic.

We stipulate the usual type conversion rule for linear types, i.e., if two
linear types $A$ and $A'$ are equal and $\isof<\Delta>.a:A$, then
$\isof<\Delta>.a:A'$. We also add a rule which allows us to use a \emph{proof}
that two supplies are equal to coerce a linear term judgment.
\begin{mathpar}
  \RULEsplyconv
\end{mathpar}


\paragraph*{Computing with supplies}

So far, we have introduced an intutionistic type theory with an embedded linear
type system without mentioning multiplicities in the spirit of
QTT. 
It turns out that we do not have to stipulate any more rules to obtain a
quantitative calculus---since our supplies are embedded as intuitionistic types,
we can eliminate into supplies, and thereby compute with these. It is not only
convenient that our structural rules do not have to take into account
multiplicities, we will see that this way of deriving multiplicities is
significantly more expressive since it allows us to specify dynamic resource
annotations.

A very useful type to compute supplies with are the natural numbers, and these
will serve as our main notion of multiplicity.

\begin{definition} \label{def:expsply}
  Let $\expsply{\Delta}{m} := \elimNat{ (- \splysep \Delta) }{\di}{m}$ for
  $\isof[\Gamma]{\Delta \issply}$ and $\isof[\Gamma].m:\Nat$.
\end{definition}

Note that the multiplicity $m$ can mention other variables in the context. The
exact number of $\Delta$'s generated by $\expsply{\Delta}{m}$ can therefore
differ at runtime, depending on how the variables in the context are
initialised.

\paragraph*{Relation with quantitative type theory}

We often talk about our language as an extension to
QTT~\citep{mcbride16_i_got_plent_onutt, atkey18_syntax_seman_quant_type_theor},
for readers familiar with QTT it might be helpful to make the relation between
both theories explicit (we will study the relation between our theory and QTT
semantically in \autoref{ssec:semanticsresults}).

The term judgement of QTT is $\Gamma \vdash a \overset{\sigma}{:} A$, where
$\sigma$ is either 0 or 1---in our setting, the 0-fragment corresponds to
$\vdash$, while the 1-fragment corresponds to $\Vdash$.

In QTT, a context $\Gamma$ is of the form $x_1 \overset{\rho_1}{:} A_1 , \ldots
, x_n \overset{\rho_n}{:} A_n$, which corresponds to the supply
$\isof.{x_1}:{A_1}[\rho_1] \splysep \ldots \splysep \isof.{x_n}:{A_n}[\rho_n] $,
living in context $\isof.{x_1}:{\ltoi{A_1}} \ctxtsep \ldots \ctxtsep
\isof.{x_n}:{\ltoi{A_n}}$. The 0-ing operation of QTT hence just amounts to
considering the context underlying a supply.

QTT is parametrised over the resource algebra that is used for the
multiplicities, in case of the natural numbers we can recover the structural
rules of QTT in our theory, where we only make use of closed, i.e., static
natural numbers $m$ in $\expsply{\Delta}{m}$.
Context addition of QTT is simply taking the join of two supplies, while context
scaling is derived in our setting in \autoref{def:expsply}. The equations for
these operations (which hold definitionally in QTT) can be shown propositionally
by induction on the natural numbers.


The variable rule of QTT is reflected in our setting by weakening
$\lino{\text{Var}}$ appropriately often.

\subsection{Type Formers with Dynamic Multiplicities}
\label{ssec:syntaxtypes}

We will now introduce linear type formers, which are akin to the usual
intuitionistic type formers, but annotated with resource specifications, making
use of \autoref{def:expsply} rules to hypothesise over, e.g., $m$ copies of some
supply for some open term of the natural numbers. We have already stipulated
with $\iota \gtyp$ that any ground type is a linear type, and in fact, only
values of ground types are considered resources in our system. All type formers
introduced in this section will just serve to program with and organise
resources, but do not constitute resources themselves.

\paragraph*{Function types}

We will denote the linear analogue of the dependent function type as usual with
a ``$\multimap$''. The rules of this type are similar to those of
QTT~\citep{atkey18_syntax_seman_quant_type_theor}, with the crucial difference
that multiplicities are now terms of the natural numbers built from the same
context as the linear types, instead of static elements of some resource
algebra.
\begin{mathpar}
\RULEnatfunform{}

\RULEnatfunintro{}

\hspace*{-1.8em}\RULEnatfunapp{}

\RULEnatfuncomp{}
\end{mathpar}

In the introduction rule, the supply $\Delta \splysep \isof.x:A[m]$ lives in
context $\Gamma \ctxtsep \isof.x:{\ltoi{A}}$, while in the conclusion $\Delta$
lives just in $\Gamma$. When introducing a function, we hence abstract over
the variable in both the intuitionistic context and the linear supply.

Note that the ``$\expsply{}{m}$'' in the linear function type is part of the
syntax, while exponentiation in the supplies is just using \autoref{def:expsply}
(we will see in the semantics in \autoref{ssec:semanticstypes} that both notions
of exponentiation coincide however).

We will not spell out how substitutions act on our linear type and term formers
as this works analogously to their intuitionistic counterparts, only taking into
account also the multiplicity parameters. For example, substitutions pass
through the function type in the expected way.
\begin{mathpar}
  \subst{(\Natfun[x]{A}[m]{\tyapp{B}{x}})}{\gamma} \equiv
  \Natfun[x]{\subst{A}{\gamma}}[\subst{m}{\gamma}]{\subst{\tyapp{B}{x}}{\gamma}}
\end{mathpar}

The intuitionistic type underlying a linear function type is given as
$\ltoi{(\Natfun[x]{A}[m]{\tyapp{B}{x}})} :=
\Fun[x]{\ltoi{A}}{\ltoi{\tyapp{B}{x}}}$. We will also implicitly translate
functions $\natfun{x}[m]{b}$ to their intuitionistic counterparts $\fun{x}{b}$.
We will omit any multiplicities in types if they are 1.

\paragraph*{Pair types}

In a similar vein, our linear pairs equip the first argument with a
multiplicity, which again is a potentially open term of the natural numbers.

\begin{mathpar}
  \RULEnatpairform{}

  \RULEnatpairintro{}
\end{mathpar}

Define $\ltoi{(\Natpair[x]{A}[m]{\tyapp{B}{x}})} :=
\Pair[x]{\ltoi{A}}{\ltoi{\tyapp{B}{x}}}$, and let us implicitly translate a
linear term $\natpair{x}[m]y$ to an intuitionistic term $\pair{x}{y}$ where
necessary. Eliminating into some linear type ${C}$, which depends on the pair
type and hence lives in $\Gamma \ctxtsep
\isof.{z}:{\Pair[x]{\ltoi{A}}{\ltoi{\tyapp{B}{x}}}}$,
works similarly to QTT, with both $\Delta_0$ and $\Delta_1$ living in $\Gamma$.

\begin{mathpar}
  \RULEnatpairelim{}
\end{mathpar}

Going from the last premise to the conclusion, we abstract over the variables
$x$ and $y$ in both the context and supply. The pair type is subject to the
usual substitution \citep{angiuli25_princ_depen_type_theor} and computation
rules \citep{barber96_dual}.


  Recall that our host theory has standard dependent pairs, so we can define
  projections for intuitionistic pairs $p$.
\begin{mathpar}
  \mathsf{fst}~p := \Pairelim[x][y]{p}{x} \qquad
  \mathsf{snd}~p := \Pairelim[x][y]{p}{y}
\end{mathpar}

We hence have a substitution from context $\isof.x:{\ltoi{A}}$ 
to context $\isof.z:{\Pair[x]{\ltoi{A}}{\ltoi{\tyapp{B}{x}}}}$ which replaces a
variable $x$ with $\mathsf{fst}~z$. Applying this to the variable judgement
$\isof<\isof.x:A>.x:A$ allows us to derive the linear judgement
$\isof<\isof.{\mathsf{fst}~z}:A>.{\mathsf{fst}~z}:A$. Note that while this
linear judgement lives in context
$\isof.z:{\Pair[x]{\ltoi{A}}{\ltoi{\tyapp{B}{x}}}}$, all it shows is that the
first projection of $z$ is enough to derive itself. Prima facie, it might seem
that we endanger our linear typing discipline by applying intuitionistic
substitutions to our linear terms, but this turns out to be fine since we also
subject our supplies to these substitutions, thereby maintaining a calculus
which neither drops nor duplicates values.

\paragraph*{Empty, unit and booleans}

With our theory we want to capture \emph{value linearity}, which means that we
do not consider elements of data types such as the unit or booleans as
resources. It would also be sensible to consider a unit type as
resource-relevant akin to the diamond type of
\citet{hofmann99_linear_types_non_size_increas} and
\citet{atkey24_polyn_time_depen_types} to reason about the runtime of programs;
or to consider booleans a resource to capture reversible computations
\citep{abramsky05_struc_approac_to_rever_comput}. While these are exciting lines
of work that we hope to explore in the future, we will in the following focus on
value linearity as in QTT~\citep{mcbride16_i_got_plent_onutt,
  atkey18_syntax_seman_quant_type_theor}, ensuring that any program
written in $\Vdash$ uses exactly the provided values of ground types.

The unit type works exactly like in QTT. Constructing the unique element of the
unit type does not require any resources, and we need to explicitly eliminate
elements of the unit type, which will require the joint resources of the element
we are eliminating and the motive $C$ which depends on the unit type.
\begin{mathpar}
  \RULEunitointro{}

  \RULEunitoelim{}
\end{mathpar}

The computation and substitution rules follow again in standard fashion and we
define $\ltoi{(\Unito)} := \Unit$.

In order to devise interesting inductive types as W-types we will also need a
linear version of the empty type, which can be introduced in any context, does
not have any constructors, and gives a linear version of the explosion principle
for any linear type $C$ depending on $\Emptyo$.
\begin{mathpar}
  \RULEemptyoelim{}
\end{mathpar}

We define $\ltoi{(\Emptyo)} := \Empty$.

Our dynamic notion of multiplicity becomes very useful for booleans. While our
introduction rules are similar to those of QTT---we assume that we do not
require any resources to introduce a boolean---our elimination rule is
considerably more general since the resources used can differ between the
branches.
\begin{mathpar}
  \RULEoboolintro{}

  \RULEoboolelim{}
\end{mathpar}

Computation and substitution rules follow again in standard fashion, and we
define $\ltoi{(\Boolo)} := \Bool$.

Our boolean eliminator allows us to precisely annotate the resources used by
programs which branch, which constitutes a very useful refinement of linear
logic. Defining the coproduct using dependent linear types, we also have
available a dynamic version of another prominent connective of linear logic.

\begin{example}
  We define additive disjunction as
  \begin{mathpar}
    A \oplus B := \Natpair[b]{\Boolo}{(\elimBool{b}{A}{B})}
  \end{mathpar}

  and its constructors as $\mathsf{inl}~x := \pair{\true}{x}$ and
  $\mathsf{inr}~y := \pair{\false}{y}$. Using the eliminator for linear booleans
  we can derive the following dynamic elimination into some linear type $C$
  depending on $A \oplus B$.
    \begin{mathpar}
    \inferrule*[]{
      \isof<\Delta_0>z:{A \oplus B} \\
      \isof<\Delta_1 \splysep \isof.x:A >.{c_1}:{\tyapp{C}{\mathsf{inl}~x}} \\
      \isof<\Delta_2 \splysep \isof.y:B >.{c_2}:{\tyapp{C}{\mathsf{inr}~y}}
    }{
      \isof<
      \Delta_0
      \splysep \expsply{\Delta_1}{\booltonat{\mathsf{fst}~z}}
        \splysep \expsply{\Delta_2}{\booltonat{\neg \mathsf{fst}~z}}
      >.{\mathsf{case}~z~\mathsf{have}~c_1~\mathsf{else}~c_2}:{\tyapp{C}{z}}
    }
  \end{mathpar}
\end{example}

Relevant for the development of inductive types will also be this application of
the coproduct which uses elimination into linear types to obtain finite linear
types.

\begin{example}
  We define $\mathsf{Fin}~n := \elimNat{ (-) \oplus \Unito }{\Emptyo}{n}$. Note
  that we perform induction on an intuitionistic term $n$ to obtain a
  \emph{linear} type with $n$ elements. We can show by induction on $n$ that we
  do not need any resources to construct some $\isof.k:{\mathsf{Fin}~n}$.

  In the following, we will write $B\ \finltyp$ if $B$ has been constructed as
  such a finite type $\mathsf{Fin}~n$.
\end{example}

\paragraph*{Inductive types}

The ability to compute with supplies gives us a lot of expressivity, which we
will now use to incorporate inductive types in our linear system. We use
Martin-Löf's W-types \citep{martin-loef84_intuit} since they capture most
inductive types used in functional programming languages and can naturally be
extended to, e.g., inductive families and coinductive types. While we expect
that our approach can be adapted for these classes of types, we leave this to
future work.

We restrict our attention to W-types whose positions are finite. This is because
our supplies are closed only under \emph{finite} join ``$\splysep$'', and we
will need to gather resources for all positions. Intuitively, having finite
positions means that any constructor has only finitely many recursive fields,
which is a relatively modest assumption. In particular, all algebraic data types
can be expressed as such finitary W-types (e.g., also rose trees as the
constructor type may be infinite).


The formation rule for our linear $\lino{\mathcal{W}}$ type is parametrised with
a dynamic multiplicity which allows us to specify how many copies of the values
present in each constructor are to be used.
\begin{mathpar}
  \RULEWoform{}
\end{mathpar}

For example, the multiplicity $m$ can be used to type lists in which every list
element can be used $m$-many times. We could have also made the $m$ dependent on
the given $\isof.x:A$ to allow varying multiplicities for different
constructors, this is a rather straightforward generalisation which would
however complicate our presentation.

Before we can formulate introduction and elimination rules, we need to extend
our syntax with an operator $\bigotimes$ which represents the join of a supply
depending on some finite type.
\begin{mathpar}
  \RULEfinsply{}
\end{mathpar}

For closed $n$, this construct simply computes the join of all supplies
depending on $\mathsf{Fin}~n$, i.e., $\finsply{y}{\mathsf{Fin}~n}{\Delta(y)}$ is
definitionally equal to $\Delta(y_1) \splysep \ldots \splysep \Delta(y_n)$.

The introduction rule for linear W-types requires an element $\isof.a:A$,
constructed using some resources $\Delta_0$, and for each position a
prescription which element of the recursive data type it points to. These
recursive elements might be made up of different resources, which we capture
with the supply $\Delta_1$ that depends on the given position. The resulting
element of our $\lino{\mathcal{W}}$ type consequently uses $m$ copies of
$\Delta_0$ as well all the join of all $\Delta_1$'s for each position.
\begin{mathpar}
  \RULEWointro{}
\end{mathpar}

When eliminating an element $p$ of $\Wo{A}[m]{B}$, we have to specify for any
constructor $\isof.x:A$ how to construct an element of the motive $C$. This
element can be constructed with $m$ copies of $x$, a supply $\Delta_1$ specific
for this constructor, and the join of all recursively constructed elements. To
define the eliminator into dependent motives, we assume that the context of the
second premise contains a function $\isof.f:{\Fun{\ltoi{\tyapp{B}{x}}}{
    \W{\ltoi{A}}{\ltoi{B}} }}$ giving us the subtrees for all positions of a
given constructor $\isof.x:{\ltoi{A}}$, and a function
$\isof.g:{\Fun{\ltoi{\tyapp{B}{x}}}{\ltoi{\tyapp{C}{\app{f}{y}}}}}$ which
computes the actual values.

\begin{mathpar}
  \hspace{-1.9em}\RULEWoelim{}
\end{mathpar}

The eliminated element requires the supply that makes up $p$, as well as some
resources recursively computed as follows: for the given constructor $x$, we
need its associated resources $\splyapp{\Delta_1}{x}$, as well as all the
recursively computed resources $\Delta$ for each position.

In the computation rule, we are given a specific constructor $a$ and a function
$f_a$
which gives us the subtree at $a$.
\begin{equation*}
  \RULEWocomp{}
\end{equation*}

In the conclusion, we need to gather all resources associated with each subtree
alongside $m$ copies of the supply that made up $a$ and a copy of the supply
associated with this kind of constructor.


The substitution rules follow in the expected fashion, we define
$\ltoi{ (\Wo[x]{A}[m]{\tyapp{B}{x} }) } :=
\W[x]{\ltoi{A}}{\ltoi{\tyapp{B}{x}}}$.

The elimination rule for general $\ltoi{\mathcal{W}}$ types is very general, and
instances of it are, e.g., the demolition principles for lists proposed by
\citet{mcbride16_i_got_plent_onutt}. It is quite a mouthful however, let us
illustrate it with a simple example.

\begin{example}
  Defining binary trees as
  \begin{mathpar}
  \BTree{A} := \Wo[c]{\Unito \oplus A}{ \elimBool{\mathsf{fst}~c}{\Boolo}{\Emptyo} }
  \end{mathpar}

  and implementing an intuitionistic function
  $\isof.{\mathsf{leafs}}:{\Fun{\ltoi{\BTree{A}}}{\Nat}}$ we can show by
  induction on the binary tree that the following recursor is admissible.
  \begin{mathpar}
      \inferrule*[]{
      \isof<\Delta_0>.t:{\BTree{A}} \\
      \isof<\isof.{y}:C \splysep \isof.{z}:C >.{n(y,z)}:C \\
      \isof<\Delta_1 \splysep \isof.x:A>.{l(x)}:C
    }{
      \isof< \Delta_0 \splysep \expsply{\Delta_1}{(\mathsf{leafs}~t)} >.
      {\mathsf{rec}_\mathsf{BTree}(n , l , t)}
      :{C}
    }
  \end{mathpar}

  To fold a binary tree $t$ using a $n$ode function which is linear and a $l$eaf
  function which takes in some resources $\Delta_1$ as well as the value of the
  leaf, we need the resources that made up $t$ and $(\mathsf{leafs}~t)$-many
  copies of $\Delta_1$. Using this recursor we can easily implement the map
  function for binary trees discussed in \autoref{sec:intro}.

  We can generalise the recursor in different ways, and all of these
  generalisations follow from $\lino{\mathcal{W}}$E: we can also have a supply
  for the recursive case, which will then be in the conclusion as many times as
  the number of nodes in the tree; we can use our multiplicity parameter for
  $\lino{\mathcal{W}}$-types to allow multiple copies of each leaf value to be
  used; and we can turn the recursor into a dependent eliminator.
\end{example}


\subsection{Completeness and Type-Checking}
\label{ssec:syntaxresults}

Before turning to a semantical study of our system, we will in this section
gather some results that we can derive on syntactic grounds.

Our translation from linear to intuitionistic types and terms allows us to
regard any term derived in the linear fragment as one in the intuitionistic
fragment (akin to changing from the 1- to the 0-fragment in QTT \citep[Lemma
2.2]{atkey18_syntax_seman_quant_type_theor}).

\begin{lemma}
  The following rule is admissible.
  \begin{mathpar}
    \inferrule*[]{
      \isof[\Gamma]{\Delta \issply} \\
      \isof<\Delta>.a:A
    }{
      \isof[\Gamma].a:{\ltoi{A}}
    }
  \end{mathpar}
\end{lemma}

As a corollary, we can conclude consistency of $\Vdash$ since the underlying
dependent intuitionistic type theory is consistent
\citep{aczel78_type_theor_inter_const_set}.

Novel about our theory is that we can go further in the other direction than any
other system---we can give precise resource annotations to many intuitionistic
terms whose resources have to be approximated in any other system.

\begin{definition}
  The \emph{clearly linear} terms are intuitionistic terms built without
  function introduction and pair and $\mathcal{W}$ elimination, i.e., all
  $\isof[\Gamma].a:A$ such that $a$ does not contain $\fun{(-)}{(-)}$,
  $\elimW{-}{-}$ and $\Pairelim[x][y]{(-)}{(-)}$.
\end{definition}

To see why we excluded certain terms from our definition of clearly linear terms
here some examples for intuitionistic terms which cannot have a linear
counterpart.

\begin{example}
  Some intuitionistic higher-order functions cannot be represented as linear
  functions since the multiplicity dependencies between different variables
  cannot be ordered, e.g., if there are mutual dependencies. Consider a
  cartesian product function for lists, which takes in two lists $xs$ and $ys$
  and returns a list of all pairs between these lists. We will need
  $\mathsf{length}~xs$ copies of $ys$ and $\mathsf{length}~ys$ copies of $xs$,
  which makes it impossible to apply $\multimap$I (note however that a cartesian
  product which is not internalised as a function in two arguments can be typed
  in $\Vdash$).
\end{example}

\begin{example} \label{exp:fst} The intuitionistic projection $\mathsf{fst}$ is
  defined using an eliminator which discards the second element, which is not
  possible with our linear pairs (in contrast, we can derive a function
  $\isof.{\mathsf{snd}}:{
    \Natfun[z]{(\Natpair{A}[0]{B})}{\tyapp{B}{\mathsf{fst}~z}}}$, which is
  desirable in a dependent system where we might only contemplate $A$.
\end{example}

\begin{example}
  We have no guarantee that an elimination principle for W-types is not dropping
  the recursively computed values, e.g., we might implement a constant function
  using $\elimW{-}{-}$. Such a program is clearly not linear.
\end{example}

With this syntactic restriction at hand, we can formulate the following
completeness result for clearly linear terms. Our result is reminiscent of the
completeness result of \citet{dal11_linear_depen_types_relat_compl}, who proved
for their dependent linear system that it can type all PCF programs. Crucially,
their linear calculus is about the runtime of programs, whereas our system is
about value linearity. As we argued with the above examples, not all programs
maintain the given values, so we believe that the below result is as far as we
can go with the level of intensionality considered here. Notably, our
completeness result incorporates eliminators for booleans, i.e., we can type
programs with arbitrary branching.


\begin{theorem}[Completeness of $\Vdash$ for clearly linear terms] \label{lemma:completeness}
  For any clearly linear term $\isof[\Gamma].a:{A}$ there exist
  $\isof[\Gamma]{\lino{A} \isltyp}$ and $\isof[\Gamma]{\Delta \issply}$ such
  that $\ltoi{(\lino{A})} = A$ and $\isof<\Delta>.a:{\lino{A}}$.

  \begin{proof}
    By structural induction on the intuitionistic term $a$, where we strengthen
    the induction hypothesis to also prove that any term of $\Unito$ and
    $\Boolo$ requires no resources. The cases are straightforward, e.g., in case
    of the eliminator for booleans our induction hypothesis gives us different
    supplies for each branch, which we can combine in the conclusion of
    $\Boolo$\textsc{E}.
  \end{proof}
\end{theorem}


The expressivity of our type system comes at a cost, however: since we can
compute with supplies using our intuitionistic theory, type-checking for
$\Vdash$ is vastly undecidable. For example, we might have to compare
$\expsply{\Delta}{m}$ with $\expsply{\Delta}{n}$ for two different open terms of
the natural numbers $m$ and $n$. \citet{dal11_linear_depen_types_relat_compl}
have dealt with this issue by parametrising their theory over the set function
symbols that are allowed in the computation of resource annotations, restricting
to simpler theories when type-checking needs to be decidable. We take a
different approach and allow the full power of DTT for the computation of
resource annotations since we can also utilise another feature of our theory: if
we can establish that some terms are propositionally equal, we can use this
witness in $\sply$\textsc{Conv} to convince the type-checker that we have used
the specified resources. E.g.,
the following rule is admissible for any 
$\isof[\Gamma].{m,n}:{\Nat}$.
  \begin{mathpar}
    \inferrule*{
      \isof<\expsply{\Delta_0}{m} \splysep \Delta_1>{\mathcal{J}} \\
      \isof[\Gamma]{\Eq{\Nat}{m}{n}} }
    { \isof<\expsply{\Delta_0}{n} \splysep \Delta_1 >{\mathcal{J}}
    }
  \end{mathpar}


\section{Modelling Dependent Linear Type Theory}
\label{sec:semantics}

We will now characterise what constitutes a model of our syntax. Our semantics
shares similarity with that of \citet{vakar15_categ_seman_linear_logic_framew,
  vakar17_in_searc_effec_depen_types}, who modelled his dependent linear type
theory using a category indexed in symmetric monoidal closed categories. Our
model will be based on a similar indexed category, however, we can use a
standard model of DTT, namely Categories with
Families~(CwF)~\citep{dybjer96_inter}, for our base category, while
\citeauthor{vakar15_categ_seman_linear_logic_framew} had to restrict context
extension to reflect that types cannot depend on linear variables. In our
system, we do not distinguish between intuitionistic (``cartesian'' in the
terminology of \citeauthor{vakar15_categ_seman_linear_logic_framew}) and linear
variables.

Our semantics has to reflect that we can compute in the host theory with
supplies and linear types, we will hence need to model these as types in the
CwF. We introduce a technique which achieves this in
\autoref{ssec:semanticsstructure} and explain how to interpret our structural
rules. Most type formers can be interpreted straightforwardly without requiring
any more structure on our model as we will see in \autoref{ssec:semanticstypes},
with only function types requiring a bit more care since they interact with the
main structural rule of the host theory, namely context extension (but again,
the structure that we need here is not very surprising and has appeared in some
variations in prior works
\citep{atkey24_polyn_time_depen_types,vakar15_categ_seman_linear_logic_framew}).

We will give some concrete models in \autoref{ssec:semanticsresults}, namely a
syntactic model and a realisability model. The latter closely resembles the
model of \citet{atkey18_syntax_seman_quant_type_theor} and underlines that we
can erase the $\vdash$ fragment for computation. We have motivated our system as
a generalisation of QTT, we give a semantic argument for this by showing that
any model for our syntax is also a model of QTT which uses the natural numbers
as resource algebra.


\subsection{Embedding Linear Logic in Type Theory}
\label{ssec:semanticsstructure}


Our syntax was based on two entailment relations, $\vdash$ for the
intuitionistic host theory and $\Vdash$ for the embedded linear logic. Since the
$\vdash$ fragment is a completely standard DTT, we can use a
standard model for this part of our language~\citep{dybjer96_inter}.

\begin{definition}
  A \emph{Category with Families (CwF)} is given by a category $\Cx$, whose objects
  are called \emph{contexts} and morphisms \emph{substitutions}, which has
  \begin{itemize}
  \item a terminal object, called the \emph{empty} context,
  \item a presheaf of \emph{types}, $\Ty \objof \psh{\Cx}$, and a presheaf of
    \emph{terms}, $\Tm \objof \psh{\catelem{\Cx}{\Ty}}$, where for some $\gamma
    : \Gamma' \to \Gamma$ we write $\subst{\_}{\gamma}$ for the actions
    $\Ty[\gamma] : \Ty[\Gamma] \to \Ty[\Gamma']$ and $\Tm[\gamma][A] :
    \Tm[\Gamma][A] \to \Tm[\Gamma'][\subst{A}{\gamma}]$,
  \item and \emph{context extension}, i.e., for any $\Gamma \objof \Cx$ and $A
    \in \Ty[\Gamma]$ we have a context $\extend{\Gamma}{A}$, substitution $\p[A]
    \objof \extend{\Gamma}{A} \to \Gamma$ and term $\q[A] \in
    \Tm[\extend{\Gamma}{A}][\subst{A}{\p[A]}]$ with the following universal
    property: for each $\gamma : \Gamma' \to \Gamma$ and $a \in
    \Tm[\Gamma][\subst{A}{\gamma}]$ there is a substitution
    $\extendsubst{\gamma}{a} : \Gamma' \to \extend{\Gamma}{A}$ such that $\p
    \circ (\extendsubst{\gamma}{a}) = \gamma$ and
    $\subst{\q}{\extendsubst{\gamma}{a}} = a$ which is unique in the sense that
    for any $\gamma' : \Gamma' \to \extend{\Gamma}{A}$ we have $\extend{(\p
      \circ \gamma')}{\subst{\q}{\gamma'}} = \gamma'$.
  \end{itemize}


  We distinguish a collection of ground types $\ATy$, i.e., $\ATy$ is a
  subpresheaf of $\Ty$ stable under reindexing. We define $\ATm$ to gather all
  terms of ground types at some context as
  \[ \ATm[\Gamma] := \amalg_{A : \ATy[\Gamma]} \Tm[\Gamma][A] . \]%
  $\ATm$ is defined componentwise on substitutions. 
\end{definition}

We motivated our supplies in \autoref{sec:syntax} as ``linear contexts''
consisting of values that can be derived in some intuitionistic context,
correspondingly, we want to send each context $\Gamma$ to a model of linear
logic. Moreover, we want to be able to eliminate into the type of supplies, for
which we embedded supplies as intuitionistic types. Let us capture this idea for
general indexed categories.

\begin{definition}
  A CwF $\Cx$ \emph{embeds} a functor $\modelfont{F} : \op{\Cx} \to
  \mathbf{Cat}$ if for any $\Gamma \objof \Cx$ we have an $\mathsf{F}_\Gamma \in
  \Ty[\Gamma]$ such that
  $$\Tm[\Gamma][\mathsf{F}_\Gamma] \cong \ob{\modelfont{F}(\Gamma)}$$
  and $\mathsf{hom} : \Tm[\Gamma][\mathsf{F}_\Gamma] \to
  \Tm[\Gamma][\mathsf{F}_\Gamma] \to \Ty[\Gamma]$ such that
    $$ \amalg_{x,y \in
      \Tm[\Gamma][\mathsf{F}]} \Tm[\Gamma][\mathsf{hom}~x~y] \cong \amalg_{x,y
      \objof \modelfont{F}(\Gamma)}\hom{\modelfont{F}(\Gamma)}{x}{y},$$ all natural
    in $\Gamma$.
\end{definition}

Note that our embedded categories do not come with universal properties in the
sense of elimination principles---we only want to construct supplies using the
host theory, but never map out of a supply or its hom-set.
We also assumed that linear types form an intuitionistic type, to model this we
will use dependent pair types.

\begin{definition}
  A CwF $\Cx$ supports dependent pairs if for any $\Gamma \objof \Cx$ we have an
  operation $\Pair{(-)}{(-)} : (\amalg_{A : \Ty[\Gamma]} \Ty[\extend{\Gamma}{A}])
  \to \Ty[\Gamma]$ and $\Tm[\Gamma][\Pair{A}{B}] \cong \amalg_{a :
    \Tm[\Gamma][A]} \Tm[\Gamma][\subst{B}{\extendsubst{\idmorphism}{a}}]$ all
  natural in $\Gamma$.
\end{definition}

We thus have everything at hand to model the structural rules of our theory
given in \autoref{ssec:syntaxstructure}.

\begin{definition}
  A \emph{linear Category with Families (lCwF)} is a CwF $\Cx$ supporting pair
  types with an embedded functor $\Sp : \op{\Cx} \to \catfont{SMCat}$ and a
  natural transformation $\iota : \ATm \to \Sp$.
\end{definition}

The $\vdash$ fragment of our syntax is interpreted as usual in the underlying
CwF \citep{hofmann97_syntax}. The linear judgements we introduced in
\autoref{ssec:syntaxstructure} are interpreted as follows.
\begin{itemize}
\item $\isof[\Gamma]{\Delta \issply}$ is modelled as $\Delta \objof
  \Sp[\Gamma]$.
\item $\isof[\Gamma]{A \isltyp}$ is modelled as a pair
  $\Pair{\Ty[\Gamma]}{\Sp[\extend{\Gamma}{A}]}$.


  Given some $\pair{A}{\Delta_A} \in
  \Pair{\Ty[\Gamma]}{\Sp[\extend{\Gamma}{A}]}$ and $a \in \Tm[\Gamma][A]$ we
  will write $\splyapp{\Delta_A}{a}$ for
  $\subst{\Delta_A}{\extendsubst{\idmorphism}{a}}$.
\item $\isof<\Delta>.a:A$, with linear type $A$ modelled with $(A,\Delta_A)$, is
  modelled as a pair
  $\Pair[a]{\Tm[\Gamma][A]}{\hom{\Sp[\Gamma]}{\Delta}{\Delta_A(a)}}$.

\end{itemize}

The rules making supplies an intuitionistic type ($\sply\textsc{Emb}$), the
substitution rule for linear terms ($\mathsf{ltm}$\textsc{Sb}) and the
conversion rule ($\sply$\textsc{Conv}) are modelled by our embedding of supplies
and their morphisms. The embedding of linear types ($\ltyp\textsc{Emb}$) is
modelled by internalising the linear types using universes and the pair type of
the host theory. Note that the translation $\ltoi{A}$, which turns any linear
type into its underlying intuitionistic type, amounts to the first projection in
our semantics. The linear variable rule ($\lino{\text{Var}}$) is modelled using
the variable of the underlying type theory $\q[A] \in
\Tm[\extend{\Gamma}{A}][\subst{A}{\p[A]}]$ with the identity morphism
$\idmorphism[\splyapp{\Delta_A}{\q[A]}]$ for a given linear type $(A,\Delta_A)$.
The inclusion of ground types as linear types ($\iota\gtyp$) utilises the
inclusion $\iota$, i.e., $A \isgtyp$ as a linear type is $(A, \iota)$.


Since the quantitative aspects of our core theory were derived and not
stipulated (\autoref{def:expsply}), we are finished interpreting the structural
rules. However, in order to interpret the type formers, we have to model supply
exponentiation since this was explicit part of the syntax for linear function,
pair and W-types. To prepare the ground for this, we note that
$\expsply{(-)}{m}$ can be considered a functor for any $\isof[\Gamma].m:\Nat$,
where the functorial action on $\delta : \Delta_0 \to \Delta_1$ is
\begin{mathpar}
  \expsply{\delta}{m} :=
  \isof.{\elimNat
  { (\_ \otimes \delta) }
  {\idmorphism[\di]}
  {m}
  }
  :{\mathsf{hom}~(\expsply{\Delta_0}{m})~(\expsply{\Delta_1}{m}) }.
\end{mathpar}

In the following, we will interpret the type formers that came equipped with
some multiplicity ``$\expsply{}{m}$'' instead as parametrised by some functor
$E$, this flexibility will be useful in \autoref{sec:bang}.

\subsection{Interpreting the Linear Type Formers}
\label{ssec:semanticstypes}

We will now see how an lCwF incorporates the linear types we introduced in
\autoref{ssec:syntaxtypes}. We do not need much additional structure since we
can utilise that our supplies live inside the host theory, more specifically,
the dependent elimination principles of the intuitionistic types underlying a
linear type former will do most---in the case of positive types all---of the
work for us.

For the pair type, we have everything at hand since the CwF underlying an lCwF
is assumed to support pairs. We interpret linear pair types equipped with some
functor $E$, which was $\expsply{(-)}{m}$ in our syntax in \autoref{sec:syntax}.

\begin{lemma} \label{lemma:pair}
  Any lCwF $(\Cx , \Sp)$ models $\ {}^E\!\otimes$.
\end{lemma}
\begin{proof}
  Given linear types $(A , \Delta_A)$ and $(B , \Delta_B)$, 
  we interpret $A\ {}^E\!\!\otimes B$ as%
  $(\Pair{A}{B} , \Pairelim[x][y]{ \q[] }{ E(\splyapp{\Delta_A}{x}) \otimes
    \splyapp{\Delta_B}{y} } )$ , where we utilise the embedding of supplies and
  write $\Pairelim[x][y]{-}{-}$ for the unique object given by the universal
  property of the intuitionistic pair type.

  In the introduction rule, we have $a \in \Tm[\Gamma][A]$, $\delta_0 : \Delta_0
  \Rightarrow \splyapp{\Delta_A}{a}$ and $b \in \Tm[\extend{\Gamma}{A}][B]$,
  $\delta_1 : \Delta_1 \Rightarrow \splyapp{\Delta_B}{b}$, which give rise to
  $\isof.{\pair{a}{b}}:{\Pair{A}{B}}$ and $E(\delta_0) \otimes \delta_1 :
  E(\Delta_0) \otimes \Delta_1 \Rightarrow E(\splyapp{\Delta_A}{a}) \otimes
  \splyapp{\Delta_B}{b}$. Similarly, we can justify the elimination rule. For
  the computation rule, the underlying intuitionistic terms will be equated by
  virtue of the CwF, and the associated morphisms are equal.
\end{proof}

For our other positive type formers we can follow a similar approach, where we
just need to assume that the model of the host theory supports the underlying
intuitionistic types (as spelled out, e.g., by
\citet{angiuli25_princ_depen_type_theor}).


\begin{lemma} \label{lemma:datatypes} An lCwF $(\Cx , \Sp)$ models $\Unito$,
  $\Emptyo$ and $\Boolo$ if $\Cx$ supports $\Unit$, $\Empty$ and $\Bool$,
  respectively.
  \begin{proof}
    Straightforward, where the dependent supply in each case is $\di$ since we
    do not consider these data types as resources.
  \end{proof}
\end{lemma}





Our linear W-types require a bit more care to model, but will also not require
any more structure than is already present in any lCwF, and that the underlying
CwF supports usual W-types
\citep{moerdijk00_wellf_trees_categ,abbott05_contain}.

\begin{definition}
  A CwF $\Cx$ supports $\mathcal{W}$-types if for any $\Gamma \objof \Cx$, $A
  \in \Ty[\Gamma]$ and $B \in \Ty[\extend{\Gamma}{A}]$, the polynomial
  endofunctor associated with $A$ and $B$ has an initial algebra.
\end{definition}

Joining together a finite number of supplies with $\mathsf{fin}\sply$ just
utilisies the finite monoidal product. We thereby have everything at hand to
model our linear W-types.

\begin{lemma} \label{lemma:w} An lCwF $(\Cx , \Sp)$ models
  ${\ltoi{\mathcal{W}}}^E$ if $\Cx$ supports $\mathcal{W}$.
  \begin{proof}
    Given linear types $(A , \Delta_A)$ and $(B , \Delta_B)$, we interpret
    ${\ltoi{\mathcal{W}}}^E(A,B)$ using the initial algebra of the corresponding
    $\mathcal{W}$-type and supply $\elimW{ x \Delta . (E(\splyapp{\Delta_A}{x})
      \otimes \finsply{y}{\tyapp{B}{x}}{ \splyapp{\Delta}{y} } )}{\q[]}$, using
    again term language to refer to the unique eliminator of $\mathcal{W}$. The
    introduction, elimination and computation rules follow in similar fashion to
    those in \autoref{lemma:pair}, using the eliminator of $\mathcal{W}$
    judiciously when justifying the elimination rule for $\ltoi{\mathcal{W}}$
    (details can be found in the artefact described in
    \autoref{sec:implementation}).
  \end{proof}
\end{lemma}

The linear function type is the only type former that requires a bit more
structure on our lCwF. First recall what it means for our host theory to support
function types.

\begin{definition}
  A CwF $\Cx$ supports dependent functions if for $\Gamma \objof \Cx$ we have an
  operation $\Fun{(-)}{(-)} : (\amalg_{A : \Ty[\Gamma]} \Ty[\extend{\Gamma}{A}])
  \to \Ty[\Gamma]$ and $\Tm[\Gamma][\Fun{A}{B}] \cong
  \Tm[\extend{\Gamma}{A}][B]$ all natural in $\Gamma$.
\end{definition}

Additionally, we need exponential objects in our indexed symmetrical monoidal
categories as well as an adjoint $\forall_A$ which allows us to turn a supply
living in some context $\extend{\Gamma}{A}$ to one living in $\Gamma$. This
allows us to bind a variable in a supply, and can be understood as an inverse
principle to context extension for supplies. This quantifier is a relatively
natural assumption---after all, we will need to bring together our linear
calculus with the structural rules of the host theory somehow---and has in fact
appeared in some form already in the literature
\citep{atkey24_polyn_time_depen_types, vakar15_categ_seman_linear_logic_framew}.

Similar to before, we again slightly generalise our considerations and consider
function types whose domain is subjected to some given functor $E$.

\begin{lemma} \label{lemma:fun}
  An lCwF $(\Cx , \Sp)$ models $\ {}^E\!\!\!\multimap$ if each $\Sp[\Gamma]$ is closed
  and $\Sp[\p[A]] : \Sp[\Gamma] \to \Sp[\extend{\Gamma}{A}]$ has a right adjoint
  $\forall_A$ for any $\Gamma \objof \Cx$ and $A \in \Ty[\Gamma]$ natural in
  $\Gamma$.
\end{lemma}

  \begin{proof}
    If all our supply categories are closed, our indexed category is in fact
    given by $\Sp : \op{\Cx} \to \catfont{SMCCat}$ and we write $[\_,\_]$ for
    the exponential of
    $\Sp[\Gamma]$. 
    The type formation rule is interpreted by taking two linear types
    $(A,\Delta_A)$ and $(B,\Delta_B)$ to the linear type $(\Fun{A}{B} ,
    \forall_{A} [E( \splyapp{\Delta_A}{\q[]} ) ,
    \splyapp{\Delta_B}{\mathsf{app}(\subst{\q}{\p[]},{\q})} ] )$, using
    $\mathsf{app}$ for the application principle derivable from $\Cx$ supporting
    function types. Intuitively, inhabitants of this linear type contain an
    intuitionistic function $f$ and an internalised morphism which witnesses
    that for any input $\isof.x:A$, $E(\splyapp{\Delta_A}{x})$ can be turned
    into $\splyapp{\Delta_B}{f~x}$.

    In the introduction rule, we are given $b \in \Tm[\extend{\Gamma}{A}][B]$
    and $\delta : \Delta \otimes E(\splyapp{\Delta_A}{x}) \Rightarrow
    \splyapp{\Delta_B }{b}$. The linear dependent function is then defined as
    $\fun{x}{b} \in \Tm[\Gamma][\Fun{A}{B}]$, and the forwards direction of the
    isomorphism between $\hom{ \Sp[\extend{\Gamma}{A}]}{ \Sp[\p[A]](\Delta_0) }{
      \Delta_1 }$ and $\hom{\Sp[\Gamma]}{ \Delta_0 }{ \forall_A(\Delta_1) }$ as
    well as currying gives us the required morphism $\Delta \Rightarrow
    \forall_{\isof.x:A}{[E(\splyapp{\Delta_A}{x}) , \splyapp{\Delta_B}{b}]}$. We
    model the application rule similarly using the inverse morphisms. Soundness
    of our interpretation follows immediately from soundness of our CwF and the
    associated supply morphisms being equal.
  \end{proof}


\subsection{Models and Relation to Quantitative CwFs}
\label{ssec:semanticsresults}

We have seen that models of our syntax can be characterised as a rather
straightforward combination of models of linear logic indexed in a model of DTT.
This also simplifies our work when we want to evince some concrete models.

\begin{example}
  The \emph{sets-and-relations} lCwF is given by an extension to the set model
  of type theory \citep[Section 3.5]{angiuli25_princ_depen_type_theor} which
  works as follows.
  \begin{itemize}
  \item $\Cx$ is the largest Grothendieck universe $\mathcal{U}$,
  \item each $\Gamma \objof \Cx$ is a set in $\mathcal{U}$, substitutions are
    functions between these sets,
  \item a type $A \in \Ty[\Gamma]$ is a function $A : \Gamma \to
    \mathcal{U}$ and context extension $\extend{\Gamma}{A}$ as $\amalg_{x \in
      \Gamma} A(x)$,
  \item a term $\Tm[\Gamma][A]$ is an indexed products $\Pi_{x \in
      \Gamma}A(x)$, and
  \item the intuitionistic type-formers are interpreted as their set-theoretic
    counterparts.
  \end{itemize}

  Since our functor $\Sp$ is embedded as a type into our theory, we also have
  that $\Sp[\Gamma]$ gives rise to some type $\sply$, which is interpreted as a
  function $\Gamma \to \mathcal{U}$, and each $\Delta \objof \Sp[\Gamma]$ gives
  rise to some $\Delta \in \Tm[\Gamma][\sply]$ which is interpreted as a product
  $\Pi_{x \in \Gamma}\sply(x)$. The embedded hom-sets are interpreted similarly.
  We can regard each set $\sply(x)$ as a SMCC in the standard way, with
  cartesian product giving us both the monoidal structure and the internalised
  hom-sets. A ground term $x\in \Gamma$ is mapped by $\iota_\Gamma$ to the
  singleton supply.

  To interpret function types, we define $\forall_A : \Sp[\extend{\Gamma}{A}]
  \to \Sp[\Gamma]$ by sending $\phi : \Pi_{f \in \amalg_{x \in \Gamma} A(x) }
  \sply (f)$ to $(x \in \Gamma) \mapsto \amalg_{a \in A(x)} \phi(x,a)$ (note
  that $\Sp[\extend{\Gamma}{A}] = \amalg_{x \in \Gamma} A(x) \to \mathcal{U}$).
\end{example}

This model gives us confidence in our system being aptly called a dependent
linear type theory as it is just an indexed version of the usual
sets-and-relations model of linear logic. As an application, we can use it to
show that there really is no projection turning $\pair{x}{y}$ into $x$ as argued
in \autoref{exp:fst} since $((x^m),y)$ will in general not be related with $x$,
where $x^m$ denotes an $m$-tuple of $x$'s.

We have motivated our setup with $\vdash$ being only relevant for specification,
while everything derived in $\Vdash$ being the programs that we actually care
about---akin to the 0- and 1-fragments of
QTT~\citep{atkey18_syntax_seman_quant_type_theor}. To substantiate this point,
we give a realisability model in which only $\Vdash$ is realised. This model is
very similar to that of \citet{atkey18_syntax_seman_quant_type_theor} for QTT,
where our remit is simplified by the fact that we \emph{derived} the
quantitative aspects of our theory instead of annotating the structural rules of
our theory with a resource algebra. We can hence just use standard Linear
Combinatory Algebras \citep{abramsky02_geomet_inter_linear_combin_algeb} and do
not have to equip its elements with a modality taking into account the resource
algebra.

\begin{definition}
  A \emph{BCI-algebra} is given by a set $\mathcal{A}$, a binary operation
  $(-)\cdot(-)$ written left-associatively and elements $B,C,I \in \mathcal{A}$
  such that $B \cdot x \cdot y \cdot z = x \cdot (y \cdot z)$, $C \cdot x \cdot
  y \cdot z = x \cdot z \cdot y$ and $I \cdot x = x$.

\end{definition}

We do need the additional structure for modelling booleans introduced by
\citet{atkey18_syntax_seman_quant_type_theor} since boolean values can effect a
program's behaviour, despite not being present in the supplies.

\begin{definition}
  A BCI-algebra $\mathcal{A}$ \emph{supports booleans} if we have elements $T,F
  \in \mathcal{A}$ and a function $E : \mathcal{A} \times \mathcal{A} \to
  \mathcal{A}$ such that $E(p,q) \cdot T = p$ and $E(p,q) \cdot F = q$
\end{definition}


Any BCI-algebra $\mathcal{A}$ gives rise to a symmetric monoidal closed category
of assemblies $\catfont{Asm}(\mathcal{A})$ \citep{hoshino07_linear} which has
objects pairs ${(X, \vDash_X)}$, where $X$ is a set of extensional meanings and
$\vDash_X : \mathcal{A} \times X$ captures when an element $a \in \mathcal{A}$
realises some $x \in X$, such that any element is realisable, i.e., for any $x
\in X$ we have $a \vDash_X x$ for some $a \in \mathcal{A}$. A morphism between
$(X,\vDash_X)$ and $(Y,\vDash_Y)$ is a realisable function $f : X \to Y$, i.e.,
there is an $a_f \in \mathcal{A}$ such that $a \vDash_X x$ implies $a_f \cdot a
\vDash f(y)$.

As pointed out by \citet{atkey24_polyn_time_depen_types}, the original
realisability model of QTT~\citep{atkey18_syntax_seman_quant_type_theor} was
faulty since also contexts which only had 0-ed variables had realisers, even
though the 0-fragment of QTT ought to be erased. The dichotomy between contexts
and supplies makes things easier for us since this we can fully erase the
$\vdash$ fragment while realising the $\Vdash$ fragment of our syntax. Note that
a supply with only 0-ed variables $\isof.x:A[0]$ is definitionally equal to the
empty supply.



\begin{example} \label{exp:real} For any BCI-algebra $\mathcal{A}$ which
  supports booleans, the \emph{realisability model} is given by a CwF $\Cx$ of
  sets and functions which is indexed in sub-categories of
  $\catfont{Asm}(\mathcal{A})$ as follows: a set $\Gamma$ in $\Cx$ is sent by
  $\Sp$ to the category of assemblies $(\Delta , \vDash_\Delta)$ such that
  $\Delta$ is a subset of $\Gamma$. The elements of $\Gamma$ can be understood
  as the ground terms, and $\iota_\Gamma$ sends each $x$ to the singleton
  assembly containing $x$. The contexts hence capture what resources might be
  contemplated, while the supplies contain the resources actually available for
  computation.
  The model supports function types if for any $a \vDash_\Delta x$ at $(\Delta ,
  \vDash_\Delta) \objof \Sp[\extend{\Gamma}{A}]$ we have
  $(\Delta',\vDash_{\Delta'}) \objof \Sp[\Gamma]$, a function $\forall : \Delta
  \to \Delta'$ and $a_x \in \mathcal{A}$ such that $a_x \cdot a \vDash_{\Delta'}
  \forall x$.
\end{example}

Our realisability model justifies that we only treat terms derived in $\Vdash$
as relevant for computation, and that the $\vdash$ fragment can be erased
similarly to the 0-fragment of QTT. Note that we can also erase open programs,
i.e., terms derived in a non-empty intuitionistic context $\Gamma$, provided
that the supply does not mention any hypotheses, as has been worked out for QTT
by~\citet{abel23_graded_modal_depen_type_theor}.

The relationship with QTT goes deeper, and at least for a resource algebra that
can be presented inductively our syntax is more general. We refer to
\citet{atkey18_syntax_seman_quant_type_theor} for a definition of Quantitative
Categories with Families (QCwF).

\begin{lemma} \label{lemma:genqtt} An lCwF $(\Cx,\Sp)$ is a $\mathbb{N}$-QCwF if
  $\Cx$ supports a natural numbers type.
\end{lemma}
\begin{proof}
  For the CwF underlying the QCwF we have $\Cx$, while we use the collection of
  all supplies for the category of contexts with resource annotations, i.e.,
  $\mathcal{L} := \amalg_{\Gamma \objof \Cx} \Sp[\Gamma]$. The functor $U :
  \mathcal{L} \to \Cx$ is given by the first projection, and the addition and
  scaling structure can be derived similarly as in \autoref{ssec:syntaxresults}.
  The resourced terms of the QCwF are precisely our linear terms, and resourced
  context extension is given by extending supplies with an exponentiated
  variable $\expsply{x}{m}$ (the required natural transformations for resourced
  context extension hold stricly in our model).
\end{proof}

\section{Re-embedding Full Intuitionistic Logic}
\label{sec:bang}

We have seen in \autoref{exp:real} that when computing with our theory, we can
erase the $\vdash$ fragment and only keep the $\Vdash$ fragment. The latter only
allows value-linear terms, however---while this is a quite large class of terms
due to our dynamic notion of multiplicity, we still cannot type all terms, e.g.,
we have no first projection function. To rectify this, we can use the standard
approach of recovering full intuitionistic logic inside linear logic by
equipping resources whose values might be used arbitrarily often with a modality
``!''. 
\begin{mathpar}
  \RULEbangsply{}
\end{mathpar}

We also add structural rules which allow us to arbitrarily duplicate and discard
supplies annotated with ``$!$'', these are entirely standard
\citep{benton93_intuit_linear_logic} and we refer the interested reader to the
artefact described in \autoref{sec:implementation} for details.

In order to abstract over variables annotated with ``!'' we introduce a ``!''-
function type whose rules are analogous to the linear function type we gave in
\autoref{ssec:syntaxtypes}. The formation rule for such a type
$\Bangfun[x]{A}{\tyapp{B}{x}}$ applies to any two linear types $A$ and $B$, and
the introduction and elimination rules are as follows.
\begin{mathpar}
  \RULEbangfunintro{}

  \RULEbangfunapp{}
\end{mathpar}

The computation and substitution rules for !-functions are standard, and by
defining ${\ltoi{(\Bangfun[x]{A}{\tyapp{B}{x}})} :=
  \Fun[x]{\ltoi{A}}{\ltoi{\tyapp{B}{x}}}}$ we recover the underlying function
type.

Equipping our linear type system with the modality ``!'' therefore gives us the
full power of intuitionistic logic.

\begin{example}
  We can construct a function of type $\Bangfun{(\Natpair{A}{B})}{A}$ in the
  $\Vdash$ fragment of our system extended with ``!''.
\end{example}




Semantically, the rules for ``!'' corresponds to requiring an exponential
comonad \citep{benton93_intuit_linear_logic} in each $\Sp[\Gamma]$. The semantic
justification for function types we gave in \autoref{lemma:fun} also applies to
our ``!''-functions using functoriality of ``!''. We thereby have established
that we have recovered intuitionistic logic in our linear fragment $\Vdash$.

\begin{lemma} \label{lemma:bang}%
  For an lCwF $(\Cx,\Sp)$ in which each $\Sp[\Gamma]$ has an exponential
  comonad, $\Sp[\Gamma]$ is cartesian closed.
\end{lemma}

\section{Implementation of the Type System in Agda}
\label{sec:implementation}

The intuitionistic fragment of our theory is a standard DTT, which makes us
hopeful that equipping existing dependently typed languages with our type system
is relatively easy. Moreover, the semantics for our theory presented in
\autoref{sec:semantics} can be formalised in DTT, suggesting an implementation
of our type system in which linear terms are represented as terms of the DTT
equipped with an inhabitant of the supply that we associated with each linear
type in \autoref{ssec:semanticstypes}. In other words, by using a DTT both to
construct programs, and as a meta-language to impose the semantics behind our
linear type system, we can obtain a prototype for our type system. While it is
not particularly practical---we have to construct the morphisms witnessing
linearity by hand---we can guarantee value linearity of programs constructed in
the linear fragment. We describe how to implement this prototype in
Agda~\citep{norell07_towar}, the construction can be carried out similarly in
other dependently typed languages such as
Rocq~\citep{barras97_coq_proof_assis_refer_manual} or Lean~\citep{moura15_lean}.
We have uploaded the artefact for the review process to the anonymous git-repo
\url{https://github.com/anonforlics26/dltt}.


To equip Agda with supplies, we introduce inductive types \agf{sply} and
\agf{\_⊩\_}, where the latter is a relation on \agf{sply}. The constructors of
these data types correspond exactly to the objects and morphisms of our
symmetric monoidal categories. Moreover, we equip \agf{sply} with a constructor
\AgdaInductiveConstructor{\ensuremath{\iota}} which embeds any ground term as a
supply.

Furthermore, we introduce a datatype \agf{ltyp} whose constructors correspond to
the formation rules of our linear types. We can then utilise the translation
$\ltoi{(-)}$ (given inductively in \autoref{sec:syntax}) as well as the
interpretations of our type formers (Lemmas~\ref{lemma:pair}---\ref{lemma:fun})
to compute from any linear type the underlying intuitionistic type $A$ and the
supply capturing how $A$ is considered a resource.%
\AGDAltypsem{}

With this interpretation we obtain a function
\agf{\_{．}}~:~\agf{ltyp}~\AgdaSymbol{→}~\agf{ityp} and use Agda's
\AgdaKeyword{syntax} feature to be able to write $a$~\agf{∶}~$A$ for the supply
corresponding to linear type $A$ at value $a$.

With this we have everything at hand to write and type-check terms in our
system. For example, the variable rule is justified by the identity morphism.%
\AGDAVar{}

We can derive the introduction, elimination and computation rules for all the
type formers exactly as described in the proofs of
Lemmas~\ref{lemma:pair}---\ref{lemma:fun}, the artefact hence provides a
formalisation of these lemmas. Moreover, we can implement all the example
programs that we gave, e.g., we can show admissible the recursor for binary
trees, where $t$, $n$ and $l$ are in the context as terms of the respective
underlying intuitionistic types.%
\AGDArecBTree{}

By equipping \agf{sply} and \agf{\_⊩\_} with a linear exponential comonad
``\AgdaInductiveConstructor{!}'' as explained in \autoref{sec:bang}, we can also
derive a first projection function in the linear fragment with the appropriate
type.%
\AGDAbangfst{}

When we execute programs derived in \agf{\_⊩\_}, we are only interested in the
intuitionistic term underlying our linear judgments, which we can recover with
an evaluation function that amounts to taking the first projection in the host
theory.%
\AGDAeval{}

While our implementation does not erase the intuitionistic fragment, as would be
justified by the realisability model (\autoref{exp:real}), we obtain a prototype
of our type system in which a witness of $\Delta$~\agf{⊩}~$a$~\agf{∶}~$A$
guarantees that $a$ is made up of the resources $\Delta$.

\section{Related Work}
\label{sec:related}

There is a large body of work on dependent linear type theories, we sketch the
main strands and how they relate to our type system.

\paragraph*{Quantitative and graded type theories}

Our work has been greatly inspired by quantitative and graded type theories
\citep{mcbride16_i_got_plent_onutt, atkey18_syntax_seman_quant_type_theor,
  orchard19_quant_progr_reason_with_graded_modal_types,abel23_graded_modal_depen_type_theor},
which also stratify their syntax into two levels, one supporting full
intuitionistic type theory and the other one a resource-aware linear calculus
(the 0-fragment and the 1-fragment, respectively, in
QTT~\citep{atkey18_syntax_seman_quant_type_theor}). Our type system can be seen
as taking McBride's approach further: we use the underlying type theory not only
to handle type dependencies, but also to compute resource annotations. This
allows us to precisely specify resource usage with dynamic multiplicities, while
the aforementioned systems only allow for static resource annotations where
multiplicities are drawn from some resource algebra. While this algebra can be
the natural numbers, in practice mostly a semiring containing 0, 1 and $\omega$
is used \citep{brady21_idris,
  orchard19_quant_progr_reason_with_graded_modal_types}, where $\omega$ allows
for arbitrary usage of some variable. While we could in principle also use an
inductive type to represent such a resource algebra, our usage of a modality
``!'' introduced in \autoref{sec:bang} is a more standard approach to recovering
full intuitionistic logic in a linear system.

Inductive types in QTT have been characterised as Quantitative Polynomial
Functors by \citet{nakov22_quant_polyn_funct}. In contrast to their approach, we
did not change the interpretation of W-types per se, but instead made them
resource-aware by reflecting their computational behaviour in the supply that is
associated with each linear type (and we did not require additional structures
on linear Categories with Families). Our semantics justifies the general
elimination rule for W-types ($\lino{\mathcal{W}}E$), which subsumes all
elimination rules that have been considered in the literature so far
\citep{mcbride16_i_got_plent_onutt, huang25_towar_quant_induc_famil}. Further
work is necessary to understand how our interpretation of linear W-types relates
to Quantitative Polynomial Functors.


\paragraph*{Dialectica}

Our system has also been influenced by the Dialectica translation for type
theory proposed by Pédrot \citep{pedrot14, pedrot24_dialec_ultim}. Gödel's
Dialectica translation \citep{goedel58_ueber_bisher_noch_nicht_erweit} provides
a very general recipe to turn an intuitionistic calculus into a linear one
\citep{paiva91_dialec, paiva90_dialec_categ, moss18_dialec}, and our system can
be seen as turning the target of the Dialectica translation into a fully fledged
type theory by equipping the indexed categories with more structure to support
function types and a linear exponential comonad.

Recently, \citet{dore25_linear_types_with_dynam_multip} has given a programming
technique which allows for deriving linear judgments in an intuitionistic type
theory, using an approach very similar to ours. They use finite multisets
encoded as higher inductive types~\citep{hott} to represent supplies in
Cubical~Agda~\citep{vezzosi21_cubic_agda}. In contrast to our system, theirs
does not support function types. Moreover, they embed \emph{all} terms into
supplies using a natural transformation $\iota$ from intuitionistic terms to
supplies, whereas we only embed ground terms and then interpret type formers
purposefully using dependent eliminators. This saves us from having to require
additional properties of $\iota$, e.g., that it is strongly monoidal with
respect to pairs of terms, significantly simplifying our model. The treatment of
inductive types by \citet{dore25_linear_types_with_dynam_multip} is also quite
ad-hoc, whereas we give a general treatment of linear W-types.

\paragraph*{Index terms in d$\ell$PCF}

The idea of non-static resource annotations has been used fruitfully by
\citet{dal11_linear_depen_types_relat_compl} to give a highly expressive type
system for PCF, called d$\ell$PCF. Their theory works, similarly to ours, by
embedding a model of linear logic---in their case history-free game semantics
\citep{abramsky00_full_abstr_pcf}---in an intuitionistic calculus. The typing
judgment of d$\ell$PCF comes equipped with an \emph{index term} which specifies
the cost of a program. \citeauthor{dal11_linear_depen_types_relat_compl} are
primarily interested in measuring the runtime of functional programs, which
means that cost is defined differently than in our setting, e.g., they consider
the cost of doubling a natural number $n$ presented in unary as $n$ since it
requires $n$ recursive calls, whereas our system treats any function between the
natural numbers as not manipulating any resources. These differences are not
crucial however, and we could also make our notion of cost more intensional by
introducing a resource-relevant diamond type as proposed by
\citet{hofmann99_linear_types_non_size_increas} and
\citet{atkey24_polyn_time_depen_types}. The calculus of
\citet{dal11_linear_depen_types_relat_compl} does not allow for internalising
the cost-annotated judgments as some sort of linear function type, which means
d$\ell$PCF is limited to a more global analysis of cost.



\paragraph*{Dependent linear type theories}

Our approach differs from prior work on dependent linear type theory as we embed
linear logic into type theory, as opposed to combining two logics as equals.
Many approaches \citep{cervesato02_linear_logic_framew,
  vakar15_categ_seman_linear_logic_framew,
  krishnaswami15_integ_linear_depen_types,
  licata17_fibrat_framew_subst_modal_logic} are based on linear-non-linear logic
\citep{benton95, barber96_dual}, distinguishing between intuitionistic and
linear variables and only allowing dependencies on intuitionistic variables.
While we also differentiate between intuitionistic and linear types in our
theory, this is not a dichotomy---rather, the linear types are more nuanced,
substructural versions of the intuitionistic types of the host theory. In
particular, we allow dependencies on variables which are treated as resources
and hence have a more expressive system.




\section{Conclusions}
\label{sec:conclusions}

Our system presents a novel way to combine intuitionistic DTT with linear logic.
We can give precise resource annotations to a large class of programs, in
particular to many higher-order, branching and recursive programs. Our system
presents both a generalisation and simplification of previous type systems
\citep{caires10_session_types_intuit_linear_propos,krishnaswami15_integ_linear_depen_types,vakar15_categ_seman_linear_logic_framew,mcbride16_i_got_plent_onutt,atkey18_syntax_seman_quant_type_theor}
since we can adequately type a larger class of programs, while reusing
well-studied structures and languages to provide semantics and an implementation
for our system.

In the future we hope to broaden the scope of our approach, and to utilise it
for different applications of linear logic.

Embedding other substructural logics apart from linear logic can be done
straightforwardly by stipulating the relevant structural rules; we also want to
explore if we can embed classical linear logic in order to obtain a classical
linear logic with predicates. So far, we have only added inductive types to our
system, but we expect that we can also incorporate coinductive types and
dependent W-types \citep{gambino04_wellf_trees_depen_polyn_funct} in our linear
system, extending the work of \citet{dore25_linear_types_with_dynam_multip}, who
considers multiplicities drawn from the conatural numbers. More work is
necessary to understand if and how our approach can be used to linearise
univalent type theories \citep{hott, bezem14_model_type_theor_cubic_sets},
possibly giving rise to a linear homotopy type
theory~\citep{riley22_bunch_homot_type_theor_synth}.

Our system captures what we called \emph{value linearity}, which is only one of
many applications of linear logic. We also want to study if our approach can be
used for complexity theory
\citep{girard94_light,hofmann99_linear_types_non_size_increas,baillot10,
  lago11_realiz_model_implic_compl}, akin to how
\citet{atkey24_polyn_time_depen_types} has used QTT to characterise certain
complexity classes. Similarly, \citet{atkey18_syntax_seman_quant_type_theor} has
suggested that making booleans resource-aware gives rise to a system which
captures reversible computations
\citep{abramsky05_struc_approac_to_rever_comput}, which points to applications
of our system for quantum computing. Lastly, we can view classical linear
propositions as \emph{sessions}, giving rise to a calculus for concurrent
computation~\citep{caires10_session_types_intuit_linear_propos,
  toninho11_depen}. We hope that our approach will prove versatile and be useful
in these applications, only requiring small changes to the interpretation of
linear types to obtain languages tailored to different notions of linearity.


\bibliographystyle{ACM-Reference-Format}
\bibliography{bibliography}


\begin{thebibliography}{54}


\ifx \showCODEN    \undefined \def \showCODEN     #1{\unskip}     \fi
\ifx \showDOI      \undefined \def \showDOI       #1{#1}\fi
\ifx \showISBNx    \undefined \def \showISBNx     #1{\unskip}     \fi
\ifx \showISBNxiii \undefined \def \showISBNxiii  #1{\unskip}     \fi
\ifx \showISSN     \undefined \def \showISSN      #1{\unskip}     \fi
\ifx \showLCCN     \undefined \def \showLCCN      #1{\unskip}     \fi
\ifx \shownote     \undefined \def \shownote      #1{#1}          \fi
\ifx \showarticletitle \undefined \def \showarticletitle #1{#1}   \fi
\ifx \showURL      \undefined \def \showURL       {\relax}        \fi
\providecommand\bibfield[2]{#2}
\providecommand\bibinfo[2]{#2}
\providecommand\natexlab[1]{#1}
\providecommand\showeprint[2][]{arXiv:#2}

\bibitem[Abbott et~al\mbox{.}(2005)]%
        {abbott05_contain}
\bibfield{author}{\bibinfo{person}{Michael Abbott}, \bibinfo{person}{Thorsten
  Altenkirch}, {and} \bibinfo{person}{Neil Ghani}.}
  \bibinfo{year}{2005}\natexlab{}.
\newblock \showarticletitle{Containers: Constructing Strictly Positive Types}.
\newblock \bibinfo{journal}{\emph{Theoretical Computer Science}}
  \bibinfo{volume}{342}, \bibinfo{number}{1} (\bibinfo{year}{2005}),
  \bibinfo{pages}{3--27}.
\newblock
\urldef\tempurl%
\url{https://doi.org/10.1016/j.tcs.2005.06.002}
\showDOI{\tempurl}
\newblock
\shownote{Applied Semantics: Selected Topics}.


\bibitem[Abel et~al\mbox{.}(2023)]%
        {abel23_graded_modal_depen_type_theor}
\bibfield{author}{\bibinfo{person}{Andreas Abel}, \bibinfo{person}{Nils~Anders
  Danielsson}, {and} \bibinfo{person}{Oskar Eriksson}.}
  \bibinfo{year}{2023}\natexlab{}.
\newblock \showarticletitle{A Graded Modal Dependent Type Theory With a
  Universe and Erasure, Formalized}.
\newblock \bibinfo{journal}{\emph{Proc. ACM Program. Lang.}}
  \bibinfo{volume}{7}, \bibinfo{number}{ICFP}, Article \bibinfo{articleno}{220}
  (\bibinfo{date}{Aug.} \bibinfo{year}{2023}), \bibinfo{numpages}{35}~pages.
\newblock
\urldef\tempurl%
\url{https://doi.org/10.1145/3607862}
\showDOI{\tempurl}


\bibitem[Abramsky(2005)]%
        {abramsky05_struc_approac_to_rever_comput}
\bibfield{author}{\bibinfo{person}{Samson Abramsky}.}
  \bibinfo{year}{2005}\natexlab{}.
\newblock \showarticletitle{A Structural Approach To Reversible Computation}.
\newblock \bibinfo{journal}{\emph{Theoretical Computer Science}}
  \bibinfo{volume}{347}, \bibinfo{number}{3} (\bibinfo{year}{2005}),
  \bibinfo{pages}{441--464}.
\newblock
\showISSN{0304-3975}
\urldef\tempurl%
\url{https://doi.org/10.1016/j.tcs.2005.07.002}
\showDOI{\tempurl}


\bibitem[Abramsky and Duncan(2006)]%
        {abramsky06_categ_quant_logic}
\bibfield{author}{\bibinfo{person}{Samson Abramsky} {and} \bibinfo{person}{Ross
  Duncan}.} \bibinfo{year}{2006}\natexlab{}.
\newblock \showarticletitle{A Categorical Quantum Logic}.
\newblock \bibinfo{journal}{\emph{Mathematical Structures in Computer Science}}
  \bibinfo{volume}{16}, \bibinfo{number}{3} (\bibinfo{year}{2006}),
  \bibinfo{pages}{469--489}.
\newblock


\bibitem[Abramsky et~al\mbox{.}(2002)]%
        {abramsky02_geomet_inter_linear_combin_algeb}
\bibfield{author}{\bibinfo{person}{Samson Abramsky}, \bibinfo{person}{Esfandiar
  Haghverdi}, {and} \bibinfo{person}{Philip Scott}.}
  \bibinfo{year}{2002}\natexlab{}.
\newblock \showarticletitle{Geometry of Interaction and Linear Combinatory
  Algebras}.
\newblock \bibinfo{journal}{\emph{Mathematical Structures in Computer Science}}
  \bibinfo{volume}{12}, \bibinfo{number}{5} (\bibinfo{year}{2002}),
  \bibinfo{pages}{625--665}.
\newblock


\bibitem[Abramsky et~al\mbox{.}(2000)]%
        {abramsky00_full_abstr_pcf}
\bibfield{author}{\bibinfo{person}{Samson Abramsky}, \bibinfo{person}{Radha
  Jagadeesan}, {and} \bibinfo{person}{Pasquale Malacaria}.}
  \bibinfo{year}{2000}\natexlab{}.
\newblock \showarticletitle{Full Abstraction for Pcf}.
\newblock \bibinfo{journal}{\emph{Information and Computation}}
  \bibinfo{volume}{163}, \bibinfo{number}{2} (\bibinfo{year}{2000}),
  \bibinfo{pages}{409--470}.
\newblock
\showISSN{0890-5401}
\urldef\tempurl%
\url{https://doi.org/10.1006/inco.2000.2930}
\showDOI{\tempurl}


\bibitem[Aczel(1978)]%
        {aczel78_type_theor_inter_const_set}
\bibfield{author}{\bibinfo{person}{Peter Aczel}.}
  \bibinfo{year}{1978}\natexlab{}.
\newblock \showarticletitle{The Type Theoretic Interpretation of Constructive
  Set Theory}.
\newblock In \bibinfo{booktitle}{\emph{Logic Colloquium '77}},
  \bibfield{editor}{\bibinfo{person}{Angus Macintyre}, \bibinfo{person}{Leszek
  Pacholski}, {and} \bibinfo{person}{Jeff Paris}} (Eds.).
  \bibinfo{series}{Studies in Logic and the Foundations of Mathematics},
  Vol.~\bibinfo{volume}{96}. \bibinfo{publisher}{Elsevier},
  \bibinfo{pages}{55--66}.
\newblock
\showISSN{0049-237X}
\urldef\tempurl%
\url{https://doi.org/10.1016/S0049-237X(08)71989-X}
\showDOI{\tempurl}


\bibitem[Angiuli and Gratzer(2025)]%
        {angiuli25_princ_depen_type_theor}
\bibfield{author}{\bibinfo{person}{Carlo Angiuli} {and} \bibinfo{person}{Daniel
  Gratzer}.} \bibinfo{year}{2025}\natexlab{}.
\newblock \bibinfo{booktitle}{\emph{Principles of Dependent Type Theory}}.
\newblock \bibinfo{publisher}{in preparation}.
\newblock
\urldef\tempurl%
\url{https://www.danielgratzer.com/papers/type-theory-book.pdf}
\showURL{%
\tempurl}


\bibitem[Atkey(2018)]%
        {atkey18_syntax_seman_quant_type_theor}
\bibfield{author}{\bibinfo{person}{Robert Atkey}.}
  \bibinfo{year}{2018}\natexlab{}.
\newblock \showarticletitle{Syntax and Semantics of Quantitative Type Theory}.
\newblock \bibinfo{journal}{\emph{Proceedings of the 33rd Annual ACM/IEEE
  Symposium on Logic in Computer Science}} \bibinfo{number}{LICS}
  (\bibinfo{year}{2018}), \bibinfo{pages}{56--65}.
\newblock
\showISBNx{9781450355834}
\urldef\tempurl%
\url{https://doi.org/10.1145/3209108.3209189}
\showDOI{\tempurl}


\bibitem[Atkey(2024)]%
        {atkey24_polyn_time_depen_types}
\bibfield{author}{\bibinfo{person}{Robert Atkey}.}
  \bibinfo{year}{2024}\natexlab{}.
\newblock \showarticletitle{Polynomial Time and Dependent Types}.
\newblock \bibinfo{journal}{\emph{Proceedings of the ACM on Programming
  Languages}} \bibinfo{volume}{8}, \bibinfo{number}{POPL}
  (\bibinfo{year}{2024}), \bibinfo{pages}{2288--2317}.
\newblock


\bibitem[Baillot et~al\mbox{.}(2010)]%
        {baillot10}
\bibfield{author}{\bibinfo{person}{Patrick Baillot}, \bibinfo{person}{Marco
  Gaboardi}, {and} \bibinfo{person}{Virgile Mogbil}.}
  \bibinfo{year}{2010}\natexlab{}.
\newblock \showarticletitle{A polytime functional language from light linear
  logic}. In \bibinfo{booktitle}{\emph{Proceedings of the 19th European
  Conference on Programming Languages and Systems}} (Paphos, Cyprus)
  \emph{(\bibinfo{series}{ESOP'10})}. \bibinfo{publisher}{Springer-Verlag},
  \bibinfo{address}{Berlin, Heidelberg}, \bibinfo{pages}{104--124}.
\newblock
\showISBNx{3642119565}
\urldef\tempurl%
\url{https://doi.org/10.1007/978-3-642-11957-6_7}
\showDOI{\tempurl}


\bibitem[Barber and Plotkin(1996)]%
        {barber96_dual}
\bibfield{author}{\bibinfo{person}{Andrew Barber} {and} \bibinfo{person}{Gordon
  Plotkin}.} \bibinfo{year}{1996}\natexlab{}.
\newblock \bibinfo{booktitle}{\emph{Dual intuitionistic linear logic}}.
\newblock \bibinfo{type}{{T}echnical {R}eport} ECS-LFCS-96-347.
\newblock


\bibitem[Barras et~al\mbox{.}(1997)]%
        {barras97_coq_proof_assis_refer_manual}
\bibfield{author}{\bibinfo{person}{Bruno Barras}, \bibinfo{person}{Samuel
  Boutin}, \bibinfo{person}{Cristina Cornes}, \bibinfo{person}{Judica{\"e}l
  Courant}, \bibinfo{person}{Jean-Christophe Filli{\^a}tre},
  \bibinfo{person}{Eduardo Gim{\'e}nez}, \bibinfo{person}{Hugo Herbelin},
  \bibinfo{person}{G{\'e}rard Huet}, \bibinfo{person}{C{\'e}sar Mu{\~n}oz},
  \bibinfo{person}{Chetan Murthy}, \bibinfo{person}{Catherine Parent},
  \bibinfo{person}{Christine Paulin-Mohring}, \bibinfo{person}{Amokrane
  Sa{\"i}bi}, {and} \bibinfo{person}{Benjamin Werner}.}
  \bibinfo{year}{1997}\natexlab{}.
\newblock \bibinfo{booktitle}{\emph{{The Coq Proof Assistant Reference Manual :
  Version 6.1}}}.
\newblock \bibinfo{type}{Research Report} RT-0203.
  \bibinfo{institution}{{INRIA}}. \bibinfo{pages}{214} pages.
\newblock
\urldef\tempurl%
\url{https://hal.inria.fr/inria-00069968}
\showURL{%
\tempurl}


\bibitem[Benton et~al\mbox{.}(1993)]%
        {benton93_intuit_linear_logic}
\bibfield{author}{\bibinfo{person}{Nick Benton}, \bibinfo{person}{Gavin
  Bierman}, \bibinfo{person}{Valeria de Paiva}, {and} \bibinfo{person}{Martin
  Hyland}.} \bibinfo{year}{1993}\natexlab{}.
\newblock \showarticletitle{A term calculus for Intuitionistic Linear Logic}.
  In \bibinfo{booktitle}{\emph{Typed Lambda Calculi and Applications}},
  \bibfield{editor}{\bibinfo{person}{Marc Bezem} {and}
  \bibinfo{person}{Jan~Friso Groote}} (Eds.). \bibinfo{publisher}{Springer
  Berlin Heidelberg}, \bibinfo{address}{Berlin, Heidelberg},
  \bibinfo{pages}{75--90}.
\newblock
\showISBNx{978-3-540-47586-6}


\bibitem[Benton(1995)]%
        {benton95}
\bibfield{author}{\bibinfo{person}{P.~N. Benton}.}
  \bibinfo{year}{1995}\natexlab{}.
\newblock \showarticletitle{A mixed linear and non-linear logic: Proofs, terms
  and models}. In \bibinfo{booktitle}{\emph{Computer Science Logic}},
  \bibfield{editor}{\bibinfo{person}{Leszek Pacholski} {and}
  \bibinfo{person}{Jerzy Tiuryn}} (Eds.). \bibinfo{publisher}{Springer Berlin
  Heidelberg}, \bibinfo{address}{Berlin, Heidelberg},
  \bibinfo{pages}{121--135}.
\newblock
\showISBNx{978-3-540-49404-1}


\bibitem[Bernardy et~al\mbox{.}(2017)]%
        {bernardy17_linear_haskel}
\bibfield{author}{\bibinfo{person}{Jean-Philippe Bernardy},
  \bibinfo{person}{Mathieu Boespflug}, \bibinfo{person}{Ryan~R. Newton},
  \bibinfo{person}{Simon Peyton~Jones}, {and} \bibinfo{person}{Arnaud
  Spiwack}.} \bibinfo{year}{2017}\natexlab{}.
\newblock \showarticletitle{Linear Haskell: practical linearity in a
  higher-order polymorphic language}.
\newblock \bibinfo{journal}{\emph{Proc. ACM Program. Lang.}}
  \bibinfo{volume}{2}, \bibinfo{number}{POPL}, Article \bibinfo{articleno}{5}
  (\bibinfo{date}{Dec.} \bibinfo{year}{2017}), \bibinfo{numpages}{29}~pages.
\newblock
\urldef\tempurl%
\url{https://doi.org/10.1145/3158093}
\showDOI{\tempurl}


\bibitem[Bezem et~al\mbox{.}(2014)]%
        {bezem14_model_type_theor_cubic_sets}
\bibfield{author}{\bibinfo{person}{Marc Bezem}, \bibinfo{person}{Thierry
  Coquand}, {and} \bibinfo{person}{Simon Huber}.}
  \bibinfo{year}{2014}\natexlab{}.
\newblock \showarticletitle{{A Model of Type Theory in Cubical Sets}}. In
  \bibinfo{booktitle}{\emph{19th International Conference on Types for Proofs
  and Programs (TYPES)}} \emph{(\bibinfo{series}{Leibniz International
  Proceedings in Informatics (LIPIcs)}, Vol.~\bibinfo{volume}{26})},
  \bibfield{editor}{\bibinfo{person}{Ralph Matthes} {and}
  \bibinfo{person}{Aleksy Schubert}} (Eds.). \bibinfo{pages}{107--128}.
\newblock
\showISBNx{978-3-939897-72-9}
\showISSN{1868-8969}
\urldef\tempurl%
\url{https://doi.org/10.4230/LIPIcs.TYPES.2013.107}
\showDOI{\tempurl}


\bibitem[Brady(2021)]%
        {brady21_idris}
\bibfield{author}{\bibinfo{person}{Edwin Brady}.}
  \bibinfo{year}{2021}\natexlab{}.
\newblock \showarticletitle{{Idris 2: Quantitative Type Theory in Practice}}.
  In \bibinfo{booktitle}{\emph{35th European Conference on Object-Oriented
  Programming (ECOOP 2021)}} \emph{(\bibinfo{series}{Leibniz International
  Proceedings in Informatics (LIPIcs)}, Vol.~\bibinfo{volume}{194})},
  \bibfield{editor}{\bibinfo{person}{Anders M{\o}ller} {and}
  \bibinfo{person}{Manu Sridharan}} (Eds.). \bibinfo{publisher}{Schloss
  Dagstuhl -- Leibniz-Zentrum f{\"u}r Informatik}, \bibinfo{address}{Dagstuhl,
  Germany}, \bibinfo{pages}{9:1--9:26}.
\newblock
\showISBNx{978-3-95977-190-0}
\showISSN{1868-8969}
\urldef\tempurl%
\url{https://doi.org/10.4230/LIPIcs.ECOOP.2021.9}
\showDOI{\tempurl}


\bibitem[Caires and Pfenning(2010)]%
        {caires10_session_types_intuit_linear_propos}
\bibfield{author}{\bibinfo{person}{Lu{\'i}s Caires} {and}
  \bibinfo{person}{Frank Pfenning}.} \bibinfo{year}{2010}\natexlab{}.
\newblock \showarticletitle{Session Types as Intuitionistic Linear
  Propositions}. In \bibinfo{booktitle}{\emph{CONCUR 2010 - Concurrency
  Theory}}, \bibfield{editor}{\bibinfo{person}{Paul Gastin} {and}
  \bibinfo{person}{Fran{\c{c}}ois Laroussinie}} (Eds.).
  \bibinfo{publisher}{Springer Berlin Heidelberg}, \bibinfo{address}{Berlin,
  Heidelberg}, \bibinfo{pages}{222--236}.
\newblock
\showISBNx{978-3-642-15375-4}
\urldef\tempurl%
\url{https://doi.org/10.1007/978-3-642-15375-4_16}
\showDOI{\tempurl}


\bibitem[Cervesato and Pfenning(2002)]%
        {cervesato02_linear_logic_framew}
\bibfield{author}{\bibinfo{person}{Iliano Cervesato} {and}
  \bibinfo{person}{Frank Pfenning}.} \bibinfo{year}{2002}\natexlab{}.
\newblock \showarticletitle{A linear logical framework}.
\newblock \bibinfo{journal}{\emph{Inforation and Computation}}
  \bibinfo{volume}{179}, \bibinfo{number}{1} (\bibinfo{date}{Nov.}
  \bibinfo{year}{2002}), \bibinfo{pages}{19–75}.
\newblock
\showISSN{0890-5401}
\urldef\tempurl%
\url{https://doi.org/10.1006/inco.2001.2951}
\showDOI{\tempurl}


\bibitem[Dal~Lago and Gaboardi(2011)]%
        {dal11_linear_depen_types_relat_compl}
\bibfield{author}{\bibinfo{person}{Ugo Dal~Lago} {and} \bibinfo{person}{Marco
  Gaboardi}.} \bibinfo{year}{2011}\natexlab{}.
\newblock \showarticletitle{Linear Dependent Types and Relative Completeness}.
\newblock \bibinfo{journal}{\emph{Proceedings of the 26th Annual ACM/IEEE
  Symposium on Logic in Computer Science}} \bibinfo{number}{LICS}
  (\bibinfo{year}{2011}), \bibinfo{pages}{133--142}.
\newblock
\urldef\tempurl%
\url{https://doi.org/10.1109/LICS.2011.22}
\showDOI{\tempurl}


\bibitem[{Dal Lago} and Hofmann(2011)]%
        {lago11_realiz_model_implic_compl}
\bibfield{author}{\bibinfo{person}{Ugo {Dal Lago}} {and}
  \bibinfo{person}{Martin Hofmann}.} \bibinfo{year}{2011}\natexlab{}.
\newblock \showarticletitle{Realizability Models and Implicit Complexity}.
\newblock \bibinfo{journal}{\emph{Theoretical Computer Science}}
  \bibinfo{volume}{412}, \bibinfo{number}{20} (\bibinfo{year}{2011}),
  \bibinfo{pages}{2029--2047}.
\newblock
\urldef\tempurl%
\url{https://doi.org/10.1016/j.tcs.2010.12.025}
\showDOI{\tempurl}
\newblock
\shownote{Girard's Festschrift}.


\bibitem[de~Moura et~al\mbox{.}(2015)]%
        {moura15_lean}
\bibfield{author}{\bibinfo{person}{Leonardo de Moura}, \bibinfo{person}{Soonho
  Kong}, \bibinfo{person}{Jeremy Avigad}, \bibinfo{person}{Floris Van~Doorn},
  {and} \bibinfo{person}{Jakob von Raumer}.} \bibinfo{year}{2015}\natexlab{}.
\newblock \showarticletitle{The Lean theorem prover (system description)}.
\newblock  \bibinfo{number}{CADE} (\bibinfo{year}{2015}),
  \bibinfo{pages}{378--388}.
\newblock
\urldef\tempurl%
\url{https://doi.org/10.1007/978-3-319-21401-6_26}
\showDOI{\tempurl}


\bibitem[de~Paiva(1990)]%
        {paiva90_dialec_categ}
\bibfield{author}{\bibinfo{person}{Valeria Correa~Vaz de Paiva}.}
  \bibinfo{year}{1990}\natexlab{}.
\newblock \emph{\bibinfo{title}{The Dialectica Categories}}.
\newblock \bibinfo{thesistype}{Ph.\,D. Dissertation}.
  \bibinfo{school}{University of Cambridge, UK}.
\newblock


\bibitem[de~Paiva(1991)]%
        {paiva91_dialec}
\bibfield{author}{\bibinfo{person}{Valeria Correa~Vaz de Paiva}.}
  \bibinfo{year}{1991}\natexlab{}.
\newblock \bibinfo{booktitle}{\emph{{The Dialectica categories}}}.
\newblock \bibinfo{type}{{T}echnical {R}eport} UCAM-CL-TR-213.
  \bibinfo{institution}{University of Cambridge, Computer Laboratory}.
\newblock
\urldef\tempurl%
\url{https://doi.org/10.48456/tr-213}
\showDOI{\tempurl}


\bibitem[Dor{\'e}(2025)]%
        {dore25_linear_types_with_dynam_multip}
\bibfield{author}{\bibinfo{person}{Maximilian Dor{\'e}}.}
  \bibinfo{year}{2025}\natexlab{}.
\newblock \showarticletitle{Linear Types With Dynamic Multiplicities in
  Dependent Type Theory (Functional Pearl)}.
\newblock \bibinfo{journal}{\emph{Proc. ACM Program. Lang.}}
  \bibinfo{volume}{9}, \bibinfo{number}{ICFP}, Article \bibinfo{articleno}{262}
  (\bibinfo{year}{2025}), \bibinfo{numpages}{21}~pages.
\newblock
\urldef\tempurl%
\url{https://doi.org/10.1145/3747531}
\showDOI{\tempurl}


\bibitem[Dybjer(1996)]%
        {dybjer96_inter}
\bibfield{author}{\bibinfo{person}{Peter Dybjer}.}
  \bibinfo{year}{1996}\natexlab{}.
\newblock \showarticletitle{Internal type theory}. In
  \bibinfo{booktitle}{\emph{Types for Proofs and Programs (TYPES) 1995}},
  \bibfield{editor}{\bibinfo{person}{Stefano Berardi} {and}
  \bibinfo{person}{Mario Coppo}} (Eds.). \bibinfo{publisher}{Springer},
  \bibinfo{pages}{120--134}.
\newblock
\urldef\tempurl%
\url{https://doi.org/10.1007/3-540-61780-9_66}
\showDOI{\tempurl}


\bibitem[Gambino and Hyland(2004)]%
        {gambino04_wellf_trees_depen_polyn_funct}
\bibfield{author}{\bibinfo{person}{Nicola Gambino} {and}
  \bibinfo{person}{Martin Hyland}.} \bibinfo{year}{2004}\natexlab{}.
\newblock \showarticletitle{Wellfounded Trees and Dependent Polynomial
  Functors}. In \bibinfo{booktitle}{\emph{Types for Proofs and Programs}},
  \bibfield{editor}{\bibinfo{person}{Stefano Berardi}, \bibinfo{person}{Mario
  Coppo}, {and} \bibinfo{person}{Ferruccio Damiani}} (Eds.).
  \bibinfo{publisher}{Springer Berlin Heidelberg}, \bibinfo{address}{Berlin,
  Heidelberg}, \bibinfo{pages}{210--225}.
\newblock
\urldef\tempurl%
\url{https://doi.org/10.1007/978-3-540-24849-1_14}
\showDOI{\tempurl}


\bibitem[Girard(1987)]%
        {girard87_linear_logic}
\bibfield{author}{\bibinfo{person}{Jean-Yves Girard}.}
  \bibinfo{year}{1987}\natexlab{}.
\newblock \showarticletitle{Linear Logic}.
\newblock \bibinfo{journal}{\emph{Theoretical computer science}}
  \bibinfo{volume}{50}, \bibinfo{number}{1} (\bibinfo{year}{1987}),
  \bibinfo{pages}{1--101}.
\newblock


\bibitem[Girard(1994)]%
        {girard94_light}
\bibfield{author}{\bibinfo{person}{Jean-Yves Girard}.}
  \bibinfo{year}{1994}\natexlab{}.
\newblock \showarticletitle{Light linear logic}. In
  \bibinfo{booktitle}{\emph{International Workshop on Logic and Computational
  Complexity}}. Springer, \bibinfo{pages}{145--176}.
\newblock


\bibitem[Hofmann(1997)]%
        {hofmann97_syntax}
\bibfield{author}{\bibinfo{person}{Martin Hofmann}.}
  \bibinfo{year}{1997}\natexlab{}.
\newblock \bibinfo{booktitle}{\emph{Syntax and semantics of dependent types}}.
\newblock \bibinfo{publisher}{Springer London}, Chapter~2,
  \bibinfo{pages}{13--54}.
\newblock
\showISBNx{978-1-4471-0963-1}
\urldef\tempurl%
\url{https://doi.org/10.1007/978-1-4471-0963-1_2}
\showDOI{\tempurl}


\bibitem[Hofmann(1999)]%
        {hofmann99_linear_types_non_size_increas}
\bibfield{author}{\bibinfo{person}{Martin Hofmann}.}
  \bibinfo{year}{1999}\natexlab{}.
\newblock \showarticletitle{Linear Types and Non Size-Increasing Polynomial
  Time Computation}. In \bibinfo{booktitle}{\emph{Proceedings of the 14th
  Annual IEEE Symposium on Logic in Computer Science}}
  \emph{(\bibinfo{series}{LICS '99})}. \bibinfo{publisher}{IEEE Computer
  Society}, \bibinfo{address}{USA}, \bibinfo{pages}{464}.
\newblock
\showISBNx{0769501583}


\bibitem[Hoshino(2007)]%
        {hoshino07_linear}
\bibfield{author}{\bibinfo{person}{Naohiko Hoshino}.}
  \bibinfo{year}{2007}\natexlab{}.
\newblock \showarticletitle{Linear realizability}. In
  \bibinfo{booktitle}{\emph{Proceedings of the 21st International Conference,
  and Proceedings of the 16th Annuall Conference on Computer Science Logic}}
  (Lausanne, Switzerland) \emph{(\bibinfo{series}{CSL'07/EACSL'07})}.
  \bibinfo{publisher}{Springer-Verlag}, \bibinfo{address}{Berlin, Heidelberg},
  \bibinfo{pages}{420--434}.
\newblock
\showISBNx{3540749144}


\bibitem[Huang and Yallop(2025)]%
        {huang25_towar_quant_induc_famil}
\bibfield{author}{\bibinfo{person}{Yulong Huang} {and} \bibinfo{person}{Jeremy
  Yallop}.} \bibinfo{year}{2025}\natexlab{}.
\newblock \showarticletitle{Towards Quantitative Inductive Families}.
\newblock \bibinfo{journal}{\emph{TYPES 2025}} (\bibinfo{year}{2025}).
\newblock


\bibitem[Krishnaswami et~al\mbox{.}(2015)]%
        {krishnaswami15_integ_linear_depen_types}
\bibfield{author}{\bibinfo{person}{Neelakantan~R. Krishnaswami},
  \bibinfo{person}{Pierre Pradic}, {and} \bibinfo{person}{Nick Benton}.}
  \bibinfo{year}{2015}\natexlab{}.
\newblock \showarticletitle{Integrating Linear and Dependent Types}. In
  \bibinfo{booktitle}{\emph{Proceedings of the 42nd Annual ACM SIGPLAN-SIGACT
  Symposium on Principles of Programming Languages}} (Mumbai, India)
  \emph{(\bibinfo{series}{POPL '15})}. \bibinfo{publisher}{Association for
  Computing Machinery}, \bibinfo{address}{New York, NY, USA},
  \bibinfo{pages}{17–30}.
\newblock
\showISBNx{9781450333009}
\urldef\tempurl%
\url{https://doi.org/10.1145/2676726.2676969}
\showDOI{\tempurl}


\bibitem[Licata et~al\mbox{.}(2017)]%
        {licata17_fibrat_framew_subst_modal_logic}
\bibfield{author}{\bibinfo{person}{Daniel~R. Licata}, \bibinfo{person}{Michael
  Shulman}, {and} \bibinfo{person}{Mitchell Riley}.}
  \bibinfo{year}{2017}\natexlab{}.
\newblock \showarticletitle{{A Fibrational Framework for Substructural and
  Modal Logics}}. In \bibinfo{booktitle}{\emph{2nd International Conference on
  Formal Structures for Computation and Deduction (FSCD 2017)}}
  \emph{(\bibinfo{series}{Leibniz International Proceedings in Informatics
  (LIPIcs)}, Vol.~\bibinfo{volume}{84})},
  \bibfield{editor}{\bibinfo{person}{Dale Miller}} (Ed.).
  \bibinfo{address}{Dagstuhl, Germany}, \bibinfo{pages}{25:1--25:22}.
\newblock
\showISBNx{978-3-95977-047-7}
\showISSN{1868-8969}
\urldef\tempurl%
\url{https://doi.org/10.4230/LIPIcs.FSCD.2017.25}
\showDOI{\tempurl}


\bibitem[Martin-L{\"o}f(1982)]%
        {martin-loef82_const_mathem_comput_progr}
\bibfield{author}{\bibinfo{person}{Per Martin-L{\"o}f}.}
  \bibinfo{year}{1982}\natexlab{}.
\newblock \showarticletitle{Constructive Mathematics and Computer Programming}.
\newblock \bibinfo{journal}{\emph{Studies in Logic and the Foundations of
  Mathematics}}  \bibinfo{volume}{104} (\bibinfo{year}{1982}),
  \bibinfo{pages}{153 -- 175}.
\newblock
\urldef\tempurl%
\url{https://dl.acm.org/doi/10.5555/3721.3731}
\showURL{%
\tempurl}


\bibitem[Martin-L{\"o}f(1984)]%
        {martin-loef84_intuit}
\bibfield{author}{\bibinfo{person}{Per Martin-L{\"o}f}.}
  \bibinfo{year}{1984}\natexlab{}.
\newblock \bibinfo{booktitle}{\emph{Intuitionistic type theory}}.
\newblock \bibinfo{publisher}{Bibliopolis}.
\newblock


\bibitem[McBride(2016)]%
        {mcbride16_i_got_plent_onutt}
\bibfield{author}{\bibinfo{person}{Conor McBride}.}
  \bibinfo{year}{2016}\natexlab{}.
\newblock \showarticletitle{I Got Plenty o' Nuttin'}.
\newblock In \bibinfo{booktitle}{\emph{A List of Successes That Can Change the
  World: Essays Dedicated to Philip Wadler on the Occasion of His 60th
  Birthday}}, \bibfield{editor}{\bibinfo{person}{Sam Lindley},
  \bibinfo{person}{Conor McBride}, \bibinfo{person}{Phil Trinder}, {and}
  \bibinfo{person}{Don Sannella}} (Eds.). \bibinfo{publisher}{Springer},
  \bibinfo{pages}{207--233}.
\newblock
\showISBNx{978-3-319-30936-1}
\urldef\tempurl%
\url{https://doi.org/10.1007/978-3-319-30936-1_12}
\showDOI{\tempurl}


\bibitem[Moerdijk and Palmgren(2000)]%
        {moerdijk00_wellf_trees_categ}
\bibfield{author}{\bibinfo{person}{Ieke Moerdijk} {and} \bibinfo{person}{Erik
  Palmgren}.} \bibinfo{year}{2000}\natexlab{}.
\newblock \showarticletitle{Wellfounded Trees in Categories}.
\newblock \bibinfo{journal}{\emph{Annals of Pure and Applied Logic}}
  \bibinfo{volume}{104}, \bibinfo{number}{1} (\bibinfo{year}{2000}),
  \bibinfo{pages}{189 -- 218}.
\newblock
\showISSN{0168-0072}
\urldef\tempurl%
\url{https://doi.org/10.1016/S0168-0072(00)00012-9}
\showDOI{\tempurl}


\bibitem[Moss and von Glehn(2018)]%
        {moss18_dialec}
\bibfield{author}{\bibinfo{person}{Sean~K. Moss} {and} \bibinfo{person}{Tamara
  von Glehn}.} \bibinfo{year}{2018}\natexlab{}.
\newblock \showarticletitle{Dialectica models of type theory}. In
  \bibinfo{booktitle}{\emph{Proceedings of the 33rd Annual ACM/IEEE Symposium
  on Logic in Computer Science}} (Oxford, United Kingdom)
  \emph{(\bibinfo{series}{LICS '18})}. \bibinfo{publisher}{Association for
  Computing Machinery}, \bibinfo{address}{New York, NY, USA},
  \bibinfo{pages}{739--748}.
\newblock
\showISBNx{9781450355834}
\urldef\tempurl%
\url{https://doi.org/10.1145/3209108.3209207}
\showDOI{\tempurl}


\bibitem[Nakov and Nordvall~Forsberg(2022)]%
        {nakov22_quant_polyn_funct}
\bibfield{author}{\bibinfo{person}{Georgi Nakov} {and} \bibinfo{person}{Fredrik
  Nordvall~Forsberg}.} \bibinfo{year}{2022}\natexlab{}.
\newblock \showarticletitle{Quantitative Polynomial Functors}. In
  \bibinfo{booktitle}{\emph{27th International Conference on Types for Proofs
  and Programs (TYPES)}} \emph{(\bibinfo{series}{Leibniz International
  Proceedings in Informatics (LIPIcs)}, Vol.~\bibinfo{volume}{239})},
  \bibfield{editor}{\bibinfo{person}{Henning Basold}, \bibinfo{person}{Jesper
  Cockx}, {and} \bibinfo{person}{Silvia Ghilezan}} (Eds.).
  \bibinfo{address}{Dagstuhl, Germany}, \bibinfo{pages}{10:1--10:22}.
\newblock
\urldef\tempurl%
\url{https://doi.org/10.4230/LIPIcs.TYPES.2021.10}
\showDOI{\tempurl}


\bibitem[Norell(2007)]%
        {norell07_towar}
\bibfield{author}{\bibinfo{person}{Ulf Norell}.}
  \bibinfo{year}{2007}\natexlab{}.
\newblock \emph{\bibinfo{title}{Towards a practical programming language based
  on dependent type theory}}.
\newblock \bibinfo{thesistype}{Ph.\,D. Dissertation}. \bibinfo{school}{Chalmers
  University of Technology and G{\"o}teborg University}.
\newblock


\bibitem[Orchard et~al\mbox{.}(2019)]%
        {orchard19_quant_progr_reason_with_graded_modal_types}
\bibfield{author}{\bibinfo{person}{Dominic Orchard},
  \bibinfo{person}{Vilem-Benjamin Liepelt}, {and} \bibinfo{person}{Harley
  Eades~III}.} \bibinfo{year}{2019}\natexlab{}.
\newblock \showarticletitle{Quantitative program reasoning with graded modal
  types}.
\newblock \bibinfo{journal}{\emph{Proc. ACM Program. Lang.}}
  \bibinfo{volume}{3}, \bibinfo{number}{ICFP}, Article \bibinfo{articleno}{110}
  (\bibinfo{date}{July} \bibinfo{year}{2019}), \bibinfo{numpages}{30}~pages.
\newblock
\urldef\tempurl%
\url{https://doi.org/10.1145/3341714}
\showDOI{\tempurl}


\bibitem[P\'{e}drot(2014)]%
        {pedrot14}
\bibfield{author}{\bibinfo{person}{Pierre-Marie P\'{e}drot}.}
  \bibinfo{year}{2014}\natexlab{}.
\newblock \showarticletitle{A functional functional interpretation}.
\newblock \bibinfo{journal}{\emph{Proceedings of the Joint Meeting of the
  Twenty-Third EACSL Annual Conference on Computer Science Logic and the
  Twenty-Ninth Annual ACM/IEEE Symposium on Logic in Computer Science}}
  \bibinfo{number}{LICS/CSL} (\bibinfo{year}{2014}).
\newblock
\showISBNx{9781450328869}
\urldef\tempurl%
\url{https://doi.org/10.1145/2603088.2603094}
\showDOI{\tempurl}


\bibitem[P{\'e}drot(2024)]%
        {pedrot24_dialec_ultim}
\bibfield{author}{\bibinfo{person}{Pierre-Marie P{\'e}drot}.}
  \bibinfo{year}{2024}\natexlab{}.
\newblock \bibinfo{title}{Dialectica the Ultimate}.  (\bibinfo{year}{2024}).
\newblock
\urldef\tempurl%
\url{https://www.p%C3%A9drot.fr/slides/tlla-07-24.pdf}
\showURL{%
\tempurl}
\newblock
\shownote{Talk at Trends in Linear Logic and Applications}.


\bibitem[Reinking et~al\mbox{.}(2021)]%
        {reinking21_perceus}
\bibfield{author}{\bibinfo{person}{Alex Reinking}, \bibinfo{person}{Ningning
  Xie}, \bibinfo{person}{Leonardo de Moura}, {and} \bibinfo{person}{Daan
  Leijen}.} \bibinfo{year}{2021}\natexlab{}.
\newblock \showarticletitle{Perceus: garbage free reference counting with
  reuse}. In \bibinfo{booktitle}{\emph{Proceedings of the 42nd ACM SIGPLAN
  International Conference on Programming Language Design and Implementation}}
  (Virtual, Canada) \emph{(\bibinfo{series}{PLDI 2021})}.
  \bibinfo{publisher}{Association for Computing Machinery},
  \bibinfo{address}{New York, NY, USA}, \bibinfo{pages}{96--111}.
\newblock
\showISBNx{9781450383912}
\urldef\tempurl%
\url{https://doi.org/10.1145/3453483.3454032}
\showDOI{\tempurl}


\bibitem[Riley(2022)]%
        {riley22_bunch_homot_type_theor_synth}
\bibfield{author}{\bibinfo{person}{Mitchell Riley}.}
  \bibinfo{year}{2022}\natexlab{}.
\newblock \emph{\bibinfo{title}{A Bunched Homotopy Type Theory for Synthetic
  Stable Homotopy Theory}}.
\newblock \bibinfo{thesistype}{Ph.\,D. Dissertation}. \bibinfo{school}{Wesleyan
  University}.
\newblock


\bibitem[Toninho et~al\mbox{.}(2011)]%
        {toninho11_depen}
\bibfield{author}{\bibinfo{person}{Bernardo Toninho},
  \bibinfo{person}{Lu\'{\i}s Caires}, {and} \bibinfo{person}{Frank Pfenning}.}
  \bibinfo{year}{2011}\natexlab{}.
\newblock \showarticletitle{Dependent session types via intuitionistic linear
  type theory}. In \bibinfo{booktitle}{\emph{Proceedings of the 13th
  International ACM SIGPLAN Symposium on Principles and Practices of
  Declarative Programming}} (Odense, Denmark) \emph{(\bibinfo{series}{PPDP
  '11})}. \bibinfo{publisher}{Association for Computing Machinery},
  \bibinfo{address}{New York, NY, USA}, \bibinfo{pages}{161--172}.
\newblock
\showISBNx{9781450307765}
\urldef\tempurl%
\url{https://doi.org/10.1145/2003476.2003499}
\showDOI{\tempurl}


\bibitem[{Univalent Foundations Program}(2013)]%
        {hott}
\bibfield{author}{\bibinfo{person}{The {Univalent Foundations Program}}.}
  \bibinfo{year}{2013}\natexlab{}.
\newblock \bibinfo{booktitle}{\emph{Homotopy Type Theory: Univalent Foundations
  of Mathematics}}.
\newblock \bibinfo{publisher}{Available at
  \url{https://homotopytypetheory.org/book}}, \bibinfo{address}{Institute for
  Advanced Study}.
\newblock


\bibitem[V{\'a}k{\'a}r(2015)]%
        {vakar15_categ_seman_linear_logic_framew}
\bibfield{author}{\bibinfo{person}{Matthijs V{\'a}k{\'a}r}.}
  \bibinfo{year}{2015}\natexlab{}.
\newblock \showarticletitle{A Categorical Semantics for Linear Logical
  Frameworks}. In \bibinfo{booktitle}{\emph{Foundations of Software Science and
  Computation Structures}}, \bibfield{editor}{\bibinfo{person}{Andrew Pitts}}
  (Ed.). \bibinfo{publisher}{Springer Berlin Heidelberg},
  \bibinfo{address}{Berlin, Heidelberg}, \bibinfo{pages}{102--116}.
\newblock
\showISBNx{978-3-662-46678-0}
\urldef\tempurl%
\url{https://doi.org/10.1007/978-3-662-46678-0_7}
\showDOI{\tempurl}


\bibitem[V{\'a}k{\'a}r(2017)]%
        {vakar17_in_searc_effec_depen_types}
\bibfield{author}{\bibinfo{person}{Matthijs V{\'a}k{\'a}r}.}
  \bibinfo{year}{2017}\natexlab{}.
\newblock \emph{\bibinfo{title}{In Search of Effectful Dependent Types}}.
\newblock \bibinfo{thesistype}{Ph.\,D. Dissertation}.
  \bibinfo{school}{University of Oxford}.
\newblock


\bibitem[Vezzosi et~al\mbox{.}(2021)]%
        {vezzosi21_cubic_agda}
\bibfield{author}{\bibinfo{person}{Andrea Vezzosi}, \bibinfo{person}{Anders
  M{\"o}rtberg}, {and} \bibinfo{person}{Andreas Abel}.}
  \bibinfo{year}{2021}\natexlab{}.
\newblock \showarticletitle{Cubical Agda: a Dependently Typed Programming
  Language With Univalence and Higher Inductive Types}.
\newblock \bibinfo{journal}{\emph{Journal of Functional Programming}}
  \bibinfo{volume}{31} (\bibinfo{year}{2021}), \bibinfo{pages}{e8}.
\newblock


\bibitem[{Von G{\"o}del}(1958)]%
        {goedel58_ueber_bisher_noch_nicht_erweit}
\bibfield{author}{\bibinfo{person}{Kurt {Von G{\"o}del}}.}
  \bibinfo{year}{1958}\natexlab{}.
\newblock \showarticletitle{{\"U}ber Eine Bisher Noch Nicht ben{\"u}tzte
  Erweiterung Des Finiten Standpunktes}.
\newblock \bibinfo{journal}{\emph{Dialectica}} \bibinfo{volume}{12},
  \bibinfo{number}{3-4} (\bibinfo{year}{1958}), \bibinfo{pages}{280--287}.
\newblock
\urldef\tempurl%
\url{https://doi.org/10.1111/j.1746-8361.1958.tb01464.x}
\showDOI{\tempurl}


\end{thebibliography}

\end{document}